\documentclass[conference]{IEEEtran2015}
\usepackage{amsmath}
\usepackage{amssymb}
\usepackage{graphicx}
\usepackage{url}
\usepackage{pifont}
\usepackage{array}
\usepackage{booktabs}

\usepackage{graphicx,subfigure,url,cite}
\usepackage[usenames,dvipsnames]{xcolor}
\usepackage{float}
\usepackage{algorithm}
\usepackage[noend]{algpseudocode}
\usepackage{amsthm}

\newtheorem{proposition}{Proposition}

\newtheorem*{example*}{Example}

\newcommand{\head}[1]{%
  \begin{tabular}[b]{@{}c@{}}
  #1
  \end{tabular}%
}

\newcommand{\AMUSE}{{\emph{DyMo} }}

\newcommand{\AMUSENB}{{\emph{DyMo}}}

\newcommand{\DYMO}{{\emph{DyMo} }}
\newcommand{\DYMONB}{{\emph{DyMo}}}

\newcommand{\OPT}{{\emph{Optimal} }}
\newcommand{\OPTNB}{{\emph{Optimal}}}

\newcommand{\ORDERSTATS}{{\emph{Order-Statistics} }}
\newcommand{\ORDERSTATSNB}{{\emph{Order-Statistics}}}

\newcommand{\UNI}{{\emph{Uniform} }}
\newcommand{\UNINB}{{\emph{Uniform}}}

\newcommand{\ORDSTwithHIST}{{\emph{Order-Statistics w. History} }}
\newcommand{\ORDSTnoHIST}{{\emph{Order-Statistics w.o. History} }}

\newcommand{\twostep}{{\emph{Two-step estimation} }}
\newcommand{\iterative}{{\emph{Iterative estimation} }}
\newcommand{\orderstats}{{\emph{Order-Statistics estimation} }}
\newcommand{\orderstatsnb}{{\emph{Order-Statistics estimation}}}
\newcommand{\snrth}{{SNR Threshold }}
\newcommand{\snrthNB}{{SNR Threshold}}

\newcommand{\HOMOGENEOUSNBCP}{{Homogeneous}}
\newcommand{\HOMOGENEOUSNB}{{homogeneous}}
\newcommand{\HOMOGENEOUS}{{homogeneous }}

\DeclareMathOperator*{\E}{\mathbb{E}}
\DeclareMathOperator*{\Var}{\mathbb{V}ar}

\newcommand{\DRV}[2]{\frac{\partial {#1}}{\partial {#2}}}

\newcommand{\comment}[1]{}

\DeclareGraphicsExtensions{.pdf}

\newcommand{\change}[1]{{\color{black}{}#1}}
\newcommand{\ignore}[1]{{}}

\newif\ifInternalDoc    % A flag if the documnt is intented for ALU internal document.
\newif\ifShortDoc        % A flaf if the document is an extended abstrct.

\newif\ifNotInternalDoc    % A flag if the documnt is intented for ALU internal document.
\newif\ifNotShortDoc        % A flaf if the document is an extended abstrct.

\InternalDoctrue
\NotInternalDocfalse
\NotShortDoctrue
\ShortDocfalse

\def\textfloatsep{0.05in}
\begin{document}
\title{
%Counting in the Dark - Dynamic Monitoring of Very Large Wireless Systems
%Resource Efficient LTE-Multicast Video Distribution with QoS Guarantees
\emph{DyMo}: Dynamic Monitoring\\of Large Scale LTE-Multicast Systems}
\author{
{Yigal Bejerano}$^\ast$,
{Chandru Raman}$^\dag$,
{ Chun-Nam Yu}$^\ast$,
{Varun Gupta}$^\ddag$,
{Craig Gutterman}$^\ddag$,\\
{Tomas Young}$^\dag$,
{ Hugo Infante}$^\dag$,
{Yousef Abdelmalek}$^@$ and
{Gil Zussman}$^\ddag$\\ 
\phantom{0.2cm}\\
{$\ast$ Bell Labs, Nokia, Murray Hill, NJ, USA}.\\
{$\dag$ Mobile Networks, Nokia, Murray Hill, NJ, USA}.\\
{$\ddag$ Electrical Engineering, Columbia University, New York, NY, USA.}\\
{@ Verizon Wireless, Basking Ridge, NJ, USA}.  
}

%\markboth{IEEE/ACM Transactions on Networking}%
%\IEEEoverridecommandlockouts
%\IEEEpubid{\makebox[\columnwidth]{\hfill 978-1-4799-1270-4/13/\$31.00 ~\copyright~2013~IEEE} \hspace{\columnsep}\makebox[\columnwidth]{ }}
%\IEEEpubid{978-1-4799-1270-4/13/\$31.00 ~\copyright~2013~IEEE}

\maketitle

% These two commands enable page numbers in conference mode.
\thispagestyle{plain}
\pagestyle{plain}

\begin{abstract}
%%%%%%%%%%%%%%%%%%%%%%%%%%%%%%%%%%%%%%%%%%%%%%%%%%%%%%%
% File name:  abs.tex
%%%%%%%%%%%%%%%%%%%%%%%%%%%%%%%%%%%%%%%%%%%%%%%%%%%%%%%
%
LTE {\em evolved~~Multimedia~~Broadcast/Multicast~Service} (eMBMS) is an attractive solution for video delivery to very large groups in crowded venues. 
However, deployment and management of eMBMS systems is challenging, due to the lack of real-time feedback from the User Equipment (UEs). 
Therefore, we present the {\em Dynamic Monitoring} (\AMUSENB) system for low-overhead feedback collection. \AMUSE leverages eMBMS for broadcasting {\em Stochastic Group Instructions} to {\em all  UEs}. These instructions indicate the reporting rates as a function of the observed Quality of Service (QoS). 
%\comment{GZ: rates or probabilities - make sure to be consistent throughout}
This simple feedback mechanism collects very limited QoS reports from the UEs. The reports are used for network optimization, thereby ensuring high QoS to the UEs.  
We present the design aspects of \AMUSE and evaluate its performance analytically and via extensive simulations. Specifically, we show that \AMUSE infers the optimal eMBMS settings with extremely low overhead, while meeting strict QoS requirements under different UE mobility patterns and presence of network component failures. 
For instance, \AMUSE can detect the eMBMS Signal-to-Noise Ratio (SNR) experienced by the $0.1\%$ percentile of the UEs with Root Mean Square Error (RMSE) of $0.05\%$ with only 5 to 10 reports per second regardless of the number of UEs.
% under different UE mobility patters, presence of network component failures and regardless of the UE population size. 
% and with high QoS variability.

\end{abstract}

\iffalse
\begin{keywords}
Wireless Monitoring, LTE, eMBMS, Multicast, Feedback Mechanism.
\end{keywords}
\fi
%========================================================================

%%%%%%%%%%%%%%%%%%%%%%%%%%%%%%%%%%%%%%%%%%%%%%%%%%%%%%%
% File name:  intro.tex
% Changes (date, author, description):
%%%%%%%%%%%%%%%%%%%%%%%%%%%%%%%%%%%%%%%%%%%%%%%%%%%%%%%

\section{Introduction}
\label{SC:intro} 
%\comment{GZ: makes sure that all abbreviations in this and following sections are defined before they are used - don't count on definitions in the abstract}
%\comment{GZ: Varun - try to trim some of the white space at top of the figure} 

Wireless video delivery is an important service. However, unicast video streaming over LTE to a large user population in crowded venues requires a dense deployment of Base Stations (BSs)~\cite{superbowl,TYY10,Viswanathan15}. Such deployments require high capital and operational expenditure and may suffer from extensive interference between adjacent BSs. 

\begin{figure}[t]
\centering
\includegraphics[trim=0mm 0mm 0mm 0mm,width=0.45\textwidth]{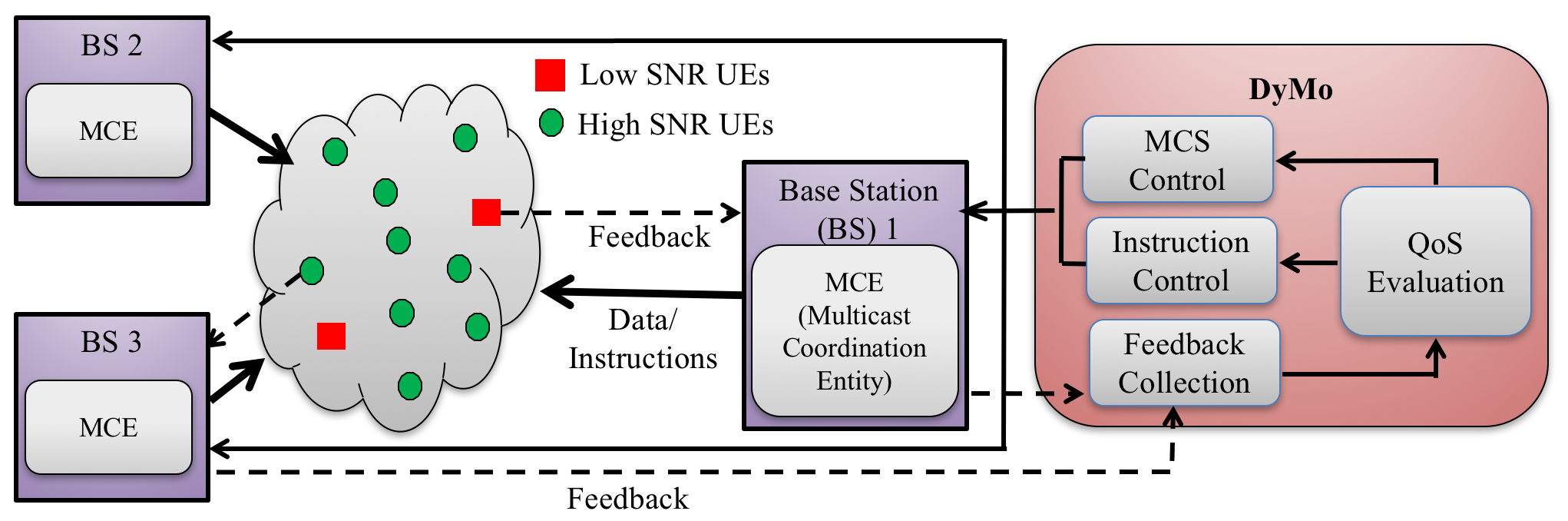}
%\includegraphics[trim=0mm 0mm 0mm 0mm,width=0.45\textwidth]{Figures/DyMoFig.png}
%\vspace*{-0.8cm}
\caption{The  \AMUSE system architecture: It exchanges control information with the Multicast Coordination Entity (MCE) of BSs which use soft signal combining for eMBMS. The Instruction Control module uses broadcast to dynamically partition the UEs into groups, each sending QoS reports at a different rate. The reports are sent to the Feedback Collection module and allow the QoS Evaluation module to identify an \snrthNB . It is used by the MCS Control module to specify the optimal MCS to the MCEs.}
\label{fig:System-Architecture}
\end{figure}

LTE-eMBMS (evolved Multimedia Broadcast/Multicast Service)~\cite{TS26.346:eMBMS-Protocols, embms-overview} provides an alternative method for content delivery in crowded venues which is based on broadcasting to a large population of User Equipment (UEs) (a.k.a. eMBMS receivers).
As illustrated in Fig.\ \ref{fig:System-Architecture}, in order to  improve the Signal-to-Noise Ratio (SNR) at the receivers, eMBMS utilizes soft signal combining techniques.\footnote{All the BSs in a particular venue transmit identical multicast signals in a time synchronized manner.} \emph{Thus, a large scale Modulation and Coding Scheme (MCS) adaptation should be conducted simultaneously for all the BSs based on the Quality of Service (QoS) at the UEs}.

\iffalse
LTE-eMBMS (evolved Multimedia Broadcast/Multicast Service)~\cite{TS26.346:eMBMS-Protocols, embms-overview} provides an alternative method for content delivery in crowded venues,
\change{by broadcasting the content to a large population of User Equipment (UEs), a.k.a. eMBMS receivers.}
Unfortunately, the eMBMS standard~\cite{TS26.346:eMBMS-Protocols} only provides a mechanism for UEs to report the experienced Quality of Service (QoS) once the communication terminates
%\comment{GZ: I reworded it. is this correct?} 
making it unsuitable for real-time traffic.
Since eMBMS utilizes soft signal combining techniques% 
\footnote{All the BSs in a particular venue transmit identical multicast signals in a time synchronized manner.},
as illustrated in Fig.\ \ref{fig:System-Architecture},
in order to  improve the Signal-to-Noise Ratio (SNR) at the receivers, service quality from unicast flows cannot be used to monitor 
the quality of eMBMS services.
This also implies that 
\emph{a large scale Modulation and Coding Scheme (MCS) adaptation should be conducted simultaneously for all the BSs based on the QoS at the UEs}.
\fi

Unfortunately, the eMBMS standard~\cite{TS26.346:eMBMS-Protocols} only provides a mechanism for UE QoS reporting once the communication terminates, thereby making it unsuitable for real-time traffic.
Recently, the Minimization of Drive Tests (MDT) protocol~\cite{TS37.320:MDT-eMBMS} was extended to provide eMBMS QoS reports periodically 
from all the UEs or when a UE joins/leaves a BS.  
However, in crowded venues with tens of thousands of UEs (e.g.,~\cite{superbowl}), even infrequent QoS reports by each UE may result in high signaling overhead and blocking of unicast traffic.\footnote{A BS can only support a limited number of connections while the minimal duration for an LTE connection is in the order of hundreds of milliseconds.}
%Further, the configuration for MDT has to be done individually at each UE %and is not straightforward.
%
Due to the limited ability to collect feedback, a deployment of an eMBMS system is very challenging. In particular, it is hindered by the following limitations:
\begin{itemize}
\item[(i)] {\em Extensive and time consuming radio frequency surveys}: 
Such surveys are conducted before each new eMBMS deployment. Yet, they provide only limited information from a few monitoring nodes. 
\item[(ii)] {\em Conservative resource allocation}: The eMBMS MCS and Forward Error Correction (FEC) codes are set 
conservatively to increase the decoding probability. 
%satisfy the disadvantaged users in the system. This leads to conservative resource allocation.
\item[(iii)] {\em Oblivious to environmental changes}: It is impossible to infer QoS degradation due to environmental changes, such as new obstacles or component failures.
\end{itemize}

Clearly, there is a need to dynamically tune the eMBMS parameters according to the feedback from UEs. 
%Therefore, the Minimization of Test Drive (MDT) protocol~\cite{TS37.320:MDT-eMBMS} was recently extended to provide 
%eMBMS QoS reports from all UEs for real-time traffic.  
%However, extensive QoS reports incur significant signaling overhead on LTE networks, 
%which are already highly subscribed in crowded areas~\cite{superbowl}. 
However, a key challenge for eMBMS parameter tuning for large scale groups is 
{\em obtaining accurate QoS reports with low overhead}. 
Schemes for efficient feedback collection in wireless multicast networks have recently received considerable attention, particulalty in the context of WiFi networks (e.g., \cite{multicastsurvey,amuseton,WWWG08,FWYL10}).
% \comment{GZ: add several references from Section II})
Yet, WiFi feedback schemes cannot be easily adapted to eMBMS since unlike WiFi, where a single Access Point transmits to a node, transmissions from multiple BSs are combined in eMBMS. Efforts for optimizing eMBMS performance focus on periodically collecting QoS reports from all UEs (e.g., \cite{Feedback-eMBMS}) but such approaches rely on extensive knowledge of the user population (for more details, see Section \ref{SSC:related}).

%For more details, see Section \ref{SSC:related}.
%\comment{GZ: please write one paragraph summary of previous work indicating that nothing relevant was done up to now... and indicate (for more details, see Section \ref{SSC:related}).... }

%\subsection{The Dynamic Monitoring System}
%\label{SSC:IntroMyMo}

%\comment{GZ: the text below in very cryptic at this stage of the paper. Very difficult to understand what the SNR threshold is - maybe combine the text with the bullets below to make things clearer? Maybe refer to Fig.\ 2?}\comment{GZ: change the paragraph start sentence based on the new related work paragraph} 
In this paper, we present the {\em Dynamic Monitoring} (\AMUSENB) system designed to support efficient LTE-eMBMS deployments in crowded and dynamic environments by providing accurate QoS reports with low overhead.
%\comment{GZ: here you already need to start using Figure 2 - it will make the explanation much clearer}
%\change{Given a {\em QoS Constaint}, $0< p\ll 1$, typically $p<1\%$,
\AMUSE identifies the maximal eMBMS \emph{SNR Threshold} 
such that 
%at most a fraction $p$} 
only a small number 
of UEs with SNR below the SNR Threshold may suffer from poor service%
\footnote{While various metrics can be used for QoS evaluation,  we consider the commonly used eMBMS SNR, referred  to as {SNR}, as a primary metric.}.
To identify the \snrth accurately, \AMUSE leverages the broadcast capabilities of eMBMS for fast dissemination of instructions to a large UE population. 
%This eliminates the need to keep track of each UE and to send individual messages.

\begin{figure}[t]
\centering
\includegraphics[trim=10mm 10mm 10mm 10mm,width=0.35\textwidth]{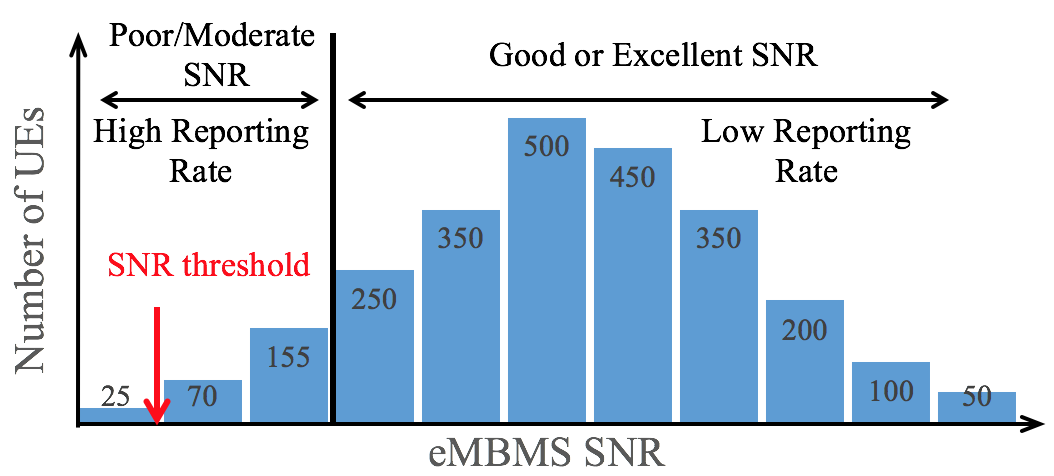}
\vspace{0.3cm}
\caption{Operation of \AMUSE for a sample UE QoS distribution: UEs are partitioned into two groups based on their SNR and each group is instructed to send QoS reports at a different rate.
%\comment{GZ: decide rate or probability and fix text and figure accordingly}.
The partitioning is dynamically adjusted based on the reports to yield more reports from UEs whose SNR is around the estimated \snrthNB . 
%\comment{the SNR thershold appears in the figure but you don't say anything about it}
}
%\vspace{0.2cm}
\label{fig:FBrate}
\end{figure}

%\comment{GZ: decide how you call things - are these ``QoS reports" or "feedback reports". once you decide make sure to unify the terminology thoroughout the paper, including in the figures and captions}
Each instruction is targeted at a sub-group of UEs that satisfies a given condition.
It instructs the UEs in the group to send a QoS report with some probability during a reporting interval.\footnote{A higher probability results in a higher reporting rate, and therefore, we will use rate and probability interchangeably.} We refer to these instructions as {\em Stochastic Group Instructions}. 
%\comment{QoS/feedback - the first time you define it use the full name. Then, you can shorten it to just ``report'' every now and them} report with some probability \comment{GZ: a recurring issue is the confusion between rate and probability - rate can be deterministic (e.g., 10 per slot while it is clear that probability is stochastic - if yo want to keep using rate below you need to indicate that high probability will result in expected high rate} during a reporting interval. 
For instance, as shown in Fig.~\ref{fig:FBrate}, \AMUSE divides UEs into two groups. UEs with poor or moderate eMBMS SNR are requested to send a report with a higher rate during the next reporting interval. 
In order to improve the accuracy of the \snrthNB, the QoS reports are analyzed and the group partitions and instructions are dynamically adapted such that the UEs whose SNR is around the \snrth report more frequently.
%estimation of the SNR threshold and configure the eMBMS parameters (e.g., eMBMS MCS) accordingly.
%It dynamically partitions the UEs into groups according to their SNR values in a way that minimizes the SNR threshold estimation error while meeting tight communication overhead budget.
The \snrth is then used for setting the eMBMS parameters, such as the MCS and FEC codes.

%\comment{GZ: better connect the following sentence  - make sure to indicate that we know that it is not a new technique but it is a cool application}
%
From a statistics perspective, \AMUSE can be viewed as a 
practical method for realizing {\em importance sampling}~\cite{mcbook} in wireless networks.
Importance sampling improves the expectation approximation of a rare event 
by sampling from a distribution that overweighs the important region.
With limited knowledge of the SNR distribution, \AMUSE leverages 
Stochastic Group Instructions to narrow down the SNR sampling 
to UEs that suffer from poor service and consequently 
obtains accurate estimation of the \snrthNB .
%\AMUSE does not require apriori knowledge about the SNR distribution %which is a prerequisite for classical importance sampling %\comment{Varun: Yigal, Chun-Nam to elaborate more here}. 
To the best of our knowledge, this is the first realization of using broadcast instructions for importance sampling in wireless networks.

The \AMUSE system architecture is illustrated in Fig.~\ref{fig:System-Architecture}. It operates on an independent server and exchanges control information with several BSs supporting eMBMS. 
%\comment{GZ: I removed cloud based since this may raise delay issues, etc.}
%\comment{GZ: make sure that the figure corresponds to the text, that all the components in the figure are described and that the way they are described here is aligned with the way they are described in the section. It is unclear from the description that there is one dymo running somewhere and many BSs}. 
%\AMUSE dynamically divides the UEs into groups according to the QoS that they experience (i.e., their eMBMS SNR %values). 
The Instruction Control module instructs the different groups to send reports at different rates. %UEs with low QoS to send reports at a high rate and the UEs with high QoS to report at a very low rate.
%\comment{again - decide on rate or probability}.
The reports are sent via unicast to the Feedback Collection module and allow the QoS Evaluation module to identify an accurate \snrthNB . 
%\comment{GZ: this sentence repeats a sentence from 2 paragraphs ago - I suggest to remove or reword} 
The \snrth is determined such that only a predefined number of UEs with SNR below the threshold, termed as \emph{outliers}, may suffer from poor service. 
The MCS Control module utilizes the \snrth to configure the eMBMS parameters (e.g., MCS) accordingly. Finally, the QoS Evaluation module continually refines group partitions based on the reports. 

%\comment{GZ: what does the MCS control module do? its in the figure but not mentioned anywhere}

%Further, \AMUSE can be easily integrated with existing eMBMS architecture as illustrated in Fig.~\ref{fig:System-Architecture}. \AMUSE includes modules for Feedback, Query, and MCS control as well as a module for QoS evaluation. 
%\AMUSE does not need any knowledge of the UE population. Consequently, \AMUSE significantly simplifies eMBMS monitoring while accurately inferring the SNR threshold with miniscule overhead.

%\subsection{Our Contributions}
%\label{SSC:contributions}

%\comment{GZ: Yigal, Varun - the following may need to be updated based on the results in the section. Yigal mention ``Our analysis provides upper bound on the achievable accuracy …" it should be mentioned here}

We focus on the QoS Evaluation module and develop a \twostep 
algorithm which can efficiently identify the \snrth 
\change{as a one time estimation.}
%when the SNR distribution of the UEs is fixed. 
We also develop an \iterative algorithm for estimating the \snrth \change{iteratively, when the distribution changes due to UE mobility or environmental changes, 
 such as network component failures}.
%We then describe and analyze the \twostep algorithm used by \AMUSE for identifying an appropriate eMBMS SNR threshold. 
%\comment{GZ: invent names for the algorithms and use them}
% such that only a \emph{small predefined number of outliers} may suffer from poor service. 
%The algorithms dynamically partition the UEs into groups according to their SNR values in such a way that minimizes the estimation error while meeting tight communication overhead budget.
Our analysis shows that the \twostep and \iterative algorithms can infer the 
\snrth with a small error and limited number of QoS reports. It is also shown that they outperform the \orderstats method, a well-known statistical method, which relies on sampling UEs with a fixed probability.
For instance, the \twostep requires only 400 reports when estimating the $1\%$ percentile to limit the error to $0.3\%$ for each re-estimation.
\change{The \iterative algorithm performs even better than the \twostep and the maximum estimation error can be bounded according to the maximum change of \snrthNB.}

%\comment{GZ: Varun, Yigal - the following simulation results require updates and more work to make the contribution clear, Yigal mentioned ``Our extensive simulations show that DyMo outperform alternative schemes, in particular order statistics   even when assuming that the average number of active receivers is known." - this should be summarized here}

We conduct extensive at-scale simulations, based on real eMBMS radio survey measurements from a stadium \change{and an urban area}. 
It is shown that \AMUSE accurately infers the \snrth and optimizes the eMBMS parameters with low overhead under different mobility patterns 
\change{and even in the event of component failures}. 
\AMUSE significantly outperforms alternative schemes based on the \orderstats method which rely on random or periodic sampling.
%\comment{use the slash something name - will avoid errors - also if its an alternative it should be mentioned above since this is the second time it appears} 

%For example, \change{our simulations show the following results. }
Our simulations show that both in a stadium-like and urban area, \AMUSE detects the {\em eMBMS SNR value of the $0.1\%$ percentile with Root Mean Square Error (RMSE) of $0.05\%$} with only $5$ messages per second \change{in total across the whole network}.
% ($60$ per reporting interval). 
This is at least $8$ times better than \orderstats based methods.
\AMUSE \change{also} infers the optimal \snrth 
%\comment{GZ: decide if threshold should be upper case or lower case and fix throughout} 
with RMSE of $0.3$~dB regardless of the UE population size, while the error of the best \orderstats method is above $1$~dB.
\AMUSE violates the outlier bound 
%\comment{GZ: these things were never defined - we have to write it assuming people did not read the paper - in the conclusions we can use definitions that were introduced in the paper} 
(of $0.1\%$) with RMSE of at most $0.35$ while the best \orderstats 
%\comment{GZ: and now its lower case} 
method incurs RMSE of over $4$ times. 
%\comment{which is what???} 
%\comment{GZ: anything about the modulation?} 
\change{The simulations also show that after a failure, \AMUSE converges instantly (i.e., in a single reporting interval) to the optimal \snrthNB.
Thus, \AMUSE is able to infer the maximum MCS while preserving QoS constraints.}
%\comment{Varun: Add outline failure scenario results and put numbers on convergence and reaction to failures} 
%We also show that \AMUSE performance significantly better (about $40\%$ improvement) with $10$ reports per second. 

%Similar to this use case, we believe that \AMUSE provides attractive monitoring and querying solutions to many large scale wireless systems.
%\end{itemize}

%\comment{GZ: Yigal, Varun - please change this paragraph but make sure that the claims here (adaptation to failures, whatever) are supported by some of the text in the 3 paragraphs above. It will be easier to rewrite it once the 3 paragraphs above are stable.}
\noindent
To summarize, the main contributions of this paper are three-fold:\\
(i) We present the concept of Stochastic Group Instructions for efficient realization of importance sampling in wireless networks.\\
(ii) We present the system architecture of \AMUSE and efficient algorithms for SNR Threshold estimation.\\
(iii) We show via extensive simulations that \AMUSE performs well in diverse scenarios.\\
The principal benefit of \AMUSE is its ability to infer the system performance based on a low number of QoS reports. 
 It converges very fast to the optimal eMBMS configuration and it reacts very fast to changes in the environment. 
Hence, it eliminates the need for service planning and extensive field trials.
% \comment{GZ: none of the results above say anything about convergence or movement - these stories should be supported by results that are summarized in the introduction or they should be removed}
Further, \AMUSE is compatible with existing LTE-eMBMS deployments and does not need any knowledge of the UE population.

The rest of the paper is organized as follows. \change{We provide background information about eMBMS and a brief review of related work in Section~\ref{SC:background}}. We introduce the model and objective in Section~\ref{SC:objectives}.
We present the \AMUSE system in Section~\ref{SC:AmuseOverview}.
\change{The algorithms for SNR threshold estimation with their analysis are given in Section~\ref{SC:Alg}.}
The numerical evaluation results appear in Section~\ref{SC:eval} before concluding in Section~\ref{SC:conclusion}.
\change{Some details of our analysis are given in the Appendix.}
%We note that, due to space constraints, the proofs and some
%simulation results are omitted and appear in~\cite{DyMoRR}.

%%%%%%%%%%%%%%%% End of file %%%%%%%%%%%%%%%%%%%%%%%%%%%%%%%%%%%%%%%%%%%%

%%%%%%%%%%%%%%%%%%%%%%%%%%%%%%%%%%%%%%%%%%%%%%%%%%%%%%%
% File name:  intro.tex
% Changes (date, author, description):
%%%%%%%%%%%%%%%%%%%%%%%%%%%%%%%%%%%%%%%%%%%%%%%%%%%%%%%
%\comment{GZ: at the end of the intro indicate that some numerical results and proofs appear in tech report [x] due to space constraints}

\change{
\section{Background information}
\label{SC:background} 

\subsection{eMBMS Background}
\label{SSC:embmsbackground}

LTE-Advanced networks provide broadcast services by using {\em evolved Multimedia Broadcast/Multicast Service} (eMBMS)~\cite{eMBMSoverview}. eMBMS is best suited to simultaneously deliver common content like video distribution to a large number of users within a contiguous region of cells. eMBMS video distribution is offered as an unidirectional service without feedback from the UE nor retransmissions of lost packets. This is enabled by all cells acting in a coordinated Single Frequency Network (SFN) arrangement, i.e., transmitting identical signals in a time synchronized manner, called {\em Multicast Broadcast Single Frequency Network} (MBSFN). The identical signals combine over the air in a non-coherent manner at each of the user locations, resulting in an improved 
%Signal-to-Interference-plus-Noise Ratio (SINR). 
Signal-Noise Ratio (SINR). 
Thus, what is normally out-of-cell interference in unicast becomes useful signal in eMBMS. For avoiding further interference from cells not transmitting the same MBSFN signal, the BSs near the boundary of the MBSFN area are used as a {\em protection tier} and they should not include eMBMS receivers in their coverage areas.
}
%--------------------------------------------------------------------------

\subsection{Related Work}
\label{SSC:related} 
%\comment{GZ: make sure here (and elsewhere that we have only UEs and nodes - get rid of receivers, stations, users, etc.} 
Wireless multicast control schemes received considerable attention in recent years (see survey in~\cite{multicastsurvey} and references therein).
Below we briefly review the most relevant papers.

\noindent
{\bf LTE-eMBMS:}
%The eMBMS standard~\cite{TS26.346:eMBMS-Protocols} provides feedback mechanism for non-real-time traffic. 
%The MDT feature in LTE \comment{GZ: add the reference, is this for general LTE or eMBMS?} enables operators to utilize UEs to collect radio \comment{GZ: channel instead of radio?} measurements and location information.
%However, the configuration of nodes \comment{GZ: replace nodes with UEs or BSs?} to support MDT needs to be done individually and can pose significant challenges. \comment{not sure that I understand the issue - if I can collect feedback, then what's the problem? - this has to be clarified. Also it is also now mentioned in the introduction - does mentioning it here add more information? it has to be the case, otherwise there is no point in repeating }
%Recently, the MDT (minimization of test drive) protocol~\cite{TS37.320:MDT-eMBMS} was extended to provide eMBMS QoS reports from all UEs also for real-time traffic. 
Most previous work on eMBMS (e.g.,~\cite{muvi,Mohapatra13:eMBMS-SVC,Militano14,embmsfair}) assumes individual feedback from all the UEs and proposes various MCS selection or resource allocation techniques. 
Yet, extensive QoS reports impose significant overhead on LTE networks, 
which are already highly congested in crowded venues~\cite{superbowl}.
An efficient feedback scheme was proposed in \cite{Feedback-eMBMS} but it relies on  knowledge of path loss (or block error) of the entire UE population to form the set of feedback nodes. 

Recently, \cite{AnonymousQuery} proposed a multicast-based 
anonymous query scheme for 
inferring the maximum MCS 
that satisfies {\em all UEs} without sending individual queries.  
%The process broadcasts a sequence of conditions, i.e., queries, 
%and all UEs that satisfy the condition transmit beacon 
%messages simultaneously. 
However, the scheme cannot be implemented in current LTE networks, since it will require UEs to
transmit simultaneous beacon messages in response to broadcast queries.

\noindent
{\bf WiFi Multicast:}
%There is a large body of literature that focuses on multicast feedback mechanisms. 
Most of the wireless multicast schemes are designed for WiFi networks. 
Some rely on {\em individual feedback} from all nodes for each packet~\cite{WWWG08,FWYL10}. %802.11aa
{\em Leader-Based Schemes}~\cite{VCOST07,DirCast ,Medusa}
collect feedback from a few selected nodes with the weakest channel quality. 
%~\cite{BSS06,VCOST07,DirCast ,Medusa}
%Pseudo-Multicast schemes~\cite{DirCast} convert the multicast feed to 
%a unicast flow and send it to one leader that acknowledges the reception 
%while the other nodes receive packets in promiscuous mode.
{\em Cluster-Based Feedback Schemes}  
in~\cite{amuseton,mudra} balance accurate reporting 
with minimization of control overhead by selecting nodes with 
the weakest channel condition in each cluster as feedback nodes.
%\comment{GZ: do we need to emphasize and define FB here - if we need the defintion its better to define in the introduction}

%%%%%%%%%%
%Most of these schemes select multicast MCS that satisfies the receiver with the worst channel condition. As shown in~\cite{amuseton,Papagiannaki2006HomeWiFi}  in crowded venues, a few unpredictable {\em outliers} may suffer  from atypical low {\em signal-to-noise-ratio} (SNR), which {\em preclude high utilization, while ensuring reliable delivery to all UEs.} The schemes in \cite{amuseton, mudra} allow a few outliers to suffer from poor service for improving the system utilization. 
%%%%%%%%%%

\iffalse
However, WiFi multicast solutions cannot easily be used in LTE-eMBMS systems. First, in WiFi, each node is associated with an Access Point, and therefore, the Access Point is aware of every node and can specify the leader or the feedback nodes. In LTE, an eMBMS UEs may stay in idle mode and the 
{\em LTE network may not be aware of UEs nor of the number of UEs.} Second, as mentioned above, LTE-eMBMS is based on soft signal combining. Thus, unlike in WiFi where the MCS is done at each Access Point independently, a large scale MCS adaptation should be conducted simultaneously at all the BSs. 
\fi

However, WiFi multicast solutions cannot easily be applied to LTE-eMBMS systems. First, in WiFi, each node is associated with an Access Point, and therefore, the Access Point is aware of every node and can specify the feedback nodes. In LTE, eMBMS UEs could be in the idle state and {\em the network may not be aware of the number of active UEs}. Second, eMBMS is based on simultaneous transmission from various BSs. Thus, unlike in WiFi where MCS adaptation is done at each Access Point independently, a common MCS adaptation should be done at all BSs.
%\comment{GZ: please fix ref 12 - search for the title in ieeexplore it has volume, page number etc., also please add a tentative link to the the tech report [9] dropbox folder and avoid the line break in the paper name}
%Recently, in ~\cite{AnonymousQuery} the authors proposed an {\em anonymous query} process for inferring the maximal MCS that satisfy {\em all UEs}.  The process broadcasts a sequence of conditions, i.e., queries, and all UEs that satisfy the condition transmit beacon messages simultaneously. While this is an interesting approach, it cannot be implemented in current LTE networks, moreover the scheme tune the MCS to the receiver with the weakest channel condition which may be over conservative when serving tens of thousands of eMBMS receivers.

%$\bullet$~It is impractical to continuously collect QoS reports from all or most UEs without affecting the network performance. Frequent QoS reports from all receivers may severely congest the uplink traffic, while infrequent reports may not be sufficient for timely detection of service quality changes. 

%\input{background-V1}

%%%%%%%%%%%%%%%%%%%%%%%%%%%%%%%%%%%%%%%%%%%%%%%%%%%%%%%
% File name:  model.tex
% Authors: Yigal Bejerano
% Version: 1.2
%----------------------------------------------------- 
% Changes (date, author, description):
%%%%%%%%%%%%%%%%%%%%%%%%%%%%%%%%%%%%%%%%%%%%%%%%%%%%%%%

%\comment{GZ: up to now in SNR threshold, t was lowercase but in some places below it is upper case and other thresholds are also uppercase - please decide on a version and search sections 1-5 to unify it. Later make sure that 6 follows} 

\change{
%\iffalse
\begin{table}[t]
\caption{Notation.}
%\vspace{0.3cm}
{\footnotesize
\centering
\begin{tabular}{|c|l|}\hline
{\bf Symbol} & ~~~~~~~~~~~~~~~~~~~~~ {\bf Semantics}\\ \hline\hline
$m$    & \change{The number of UEs in the venue, also the} \\ 
       & \change{number of active eMBMS receiver in 
		static settings.}\\ \hline
$m(t)$ & \change{The number of active eMBMS receivers at 
		time $t$.}\\ \hline
$h_v(t)$ & The individual SNR value of UE $v$ \\
	& at time interval $t$. \\ \hline
$s(t)$ & The SNR Threshold at time $t$. \\ \hline
$p$ & {\em QoS Threshold} -– The maximal portion of UEs \\
    &  with individual SNR value $h_v(t)< s(t)$.\\ \hline
$r$ & {\em Overhead Threshold} –- An upper bound on the \\
    & average number of reports in a reporting interval.\\ \hline

\end{tabular}
}
\label{TAB:notation}
%\vspace{-0.2cm}
\end{table}
%\fi 
}

\section{Model and Objective}
\label{SC:objectives} 

\subsection{Network Model}
\label{SSC:mdl}

We consider an LTE-Advanced network with multiple base stations (BSs) providing eMBMS service to a very large group of $m$ UEs in a given large venue (e.g., sports arena, transportation hub).\footnote{In this paper, we consider only the UEs subscribing to eMBMS services.} Such venues can accommodate tens of thousands of users. 
The eMBMS service is managed by a single \AMUSE server as shown in Fig.~\ref{fig:System-Architecture}
%\comment{GZ: maybe add it to the figure. Then, you can write - as shown in the figure} 
and all the BSs transmit identical multicast signals in a time synchronized manner. % for improving the eMBMS SNR of the UEs.
The multicast flows contain FEC code that allows the UEs to tolerate some level of losses $\ell$ (e.g.,  up to $5\%$ packet losses).

%\comment{GZ: change all to reporting intervals - above and below}
All UEs can detect and report the eMBMS QoS they experience. More specifically, time is divided into short {\em reporting intervals}, a few seconds each. We assume that the eMBMS SNR distribution of the UEs does not change during each reporting interval.\footnote{The SNR of each individual eMBMS packet is a random variable selected from the UE SNR distribution. We assume that this distribution does not change significantly during the reporting interval.} We define the {\em individual SNR value} $h_v(t)$, such that at least 
a given percentage $1-\ell$ (e.g., $95\%$) of the eMBMS packets received by an UE $v$ during a reporting interval $t$ have an SNR above $h_v(t)$. For a given SNR value, $h_v(t)$, there is a one-to-one mapping to an {\em eMBMS MCS} such that a UE can decode all the packets whose SNR is above $h_v(t)$~\cite{Militano14,embmsfair}.
%\comment{GZ: I rewrote this - make sure its still clear} 
The remaining packets $\ell$ can be recovered by appropriate level of FEC assuming $\ell$ is not too large.
%~\cite{Urie2013}.
\change{A summary of the main notations used throughout the paper are given in Table~\ref{TAB:notation}.} 

\subsection{Objective}
\label{SSC:objective}

We aim to design a scalable efficient eMBMS monitoring and control system for which the objective is outlined below and that satisfies the following constraints:
\begin{itemize}
\item[(i)] {\em QoS Constraint --} 
Given a {\em QoS Threshold} $p \ll 1$, at most a fraction $p$ of the UEs may suffer from packet loss of more than $\ell$. This implies that, {\em with FEC, a fraction $1-p$  of the UEs should receive all of the transmitted data}. We refer to the set UEs that suffer from packet loss after FEC as {\em outliers} and the rest are termed {\em normal} UEs.
%\comment{GZ: this terminology was used in the introduction without definition}
\item[(ii)] {\em Overhead Constraint --} The average number of UE reports during a reporting interval should be below a given {\em Overhead Threshold} $r$.
\end{itemize}

\noindent
\underline{\bf Objective:} {\em Accurately identify at any given time $t$ the maximum SNR Threshold, $s(t)$
%, as accurately as possible \comment{GZ: what does as accurately as possible means? how do we quantify it. Right now it is in the middles of the sentence. if it is important I would add another (clear) sentence for it}
 that satisfies the QoS and Overhead Constraints. } 

\noindent Namely, the calculated $s(t)$ needs to ensure that a fraction $1-p$ of the UEs have individual SNR values $h_v(t) \geq s(t)$.

%From $s(t)$, we can calculate the maximum eMBMS MCS that meets the QoS Constraint~\cite{Militano14,embmsfair} and consequently maximize the network performance, either by reducing the resource blocks allocated to the eMBMS service or improving the video quality without increasing the allocated bandwidth.

The network performance can be maximized by using $s(t)$ to calculate the maximum eMBMS MCS that meets the QoS constraint \cite{Militano14,embmsfair}.
This allows reducing the resource blocks allocated to eMBMS. Alternatively for a service such as video, the video quality can be enhanced without increasing the bandwidth allocated to the video flow. 
%\comment{GZ: we can or in Section X we show how we do it?. If we don't, what are we trying to say? This should be somwhow tied to the following sections} 
%\comment{Varun: we had these results earlier but now probably removed.. discuss with Yigal how to pitch it.}
%\noindent
%Notice that several eMBMS bearers may be configured. Conceptually, \AMUSE can optimize the performance of each bearer separately. However, this comes with the price of high reporting overhead. Therefore we recommend to use the same setting to all the eMBMS bearers and treat all the UEs as a single group.

%%%%%%%%%%%%%%%%%%%%%%%%%%%%%%%%%  End of file  %%%%%%%%%%%%%%%%%%%%%%%%%%%

%%%%%%%%%%%%%%%%%%%%%%%%%%%%%%%%%%%%%%%%%%%%%%%%%%%%%%%
% File name:  overview.tex
% Authors: Yigal Bejerano
% Version: 1.2
%------------------------------------------------------
% Changes (date, author, description):
%%%%%%%%%%%%%%%%%%%%%%%%%%%%%%%%%%%%%%%%%%%%%%%%%%%%%%%

\section{The \AMUSE System}
\label{SC:AmuseOverview} 

\change{This section introduces the \AMUSE system. 
It first presents the \AMUSE system architecture, which is 
based on the {\em Stochastic Group Instructions} concept.
Then, it provides an illustrative example of \AMUSE operations along with some technical aspects of eMBMS parameter tuning.}

\subsection{System Overview}
\label{SSC:SysOverview}

We now present the \AMUSE system architecture, shown in Fig.~\ref{fig:System-Architecture}. 
% for realizing these concepts.
%The details can be found in~\cite{DyMoRR}.

\noindent
 \textbf{Feedback Collection:} This module operates in the \AMUSE server
%\comment{GZ: what is that? be consistent with the figure and the story in the introduction} 
and in a {\em \AMUSE Mobile-Application} on each UE. 
At the beginning of each reporting interval, the Feedback Collection module 
broadcasts {\em Stochastic Group Instructions} to all the UEs. These instructions specify the QoS report probability as a function of the observed QoS (i.e., eMBMS SNR).
In response, each UE independently determines whether it should 
send a QoS report at the current reporting interval. 

\noindent
\textbf{QoS Evaluation:} The UE feedback is used to estimate the eMBMS SNR distribution, as shown in Fig.~\ref{fig:FBrate}. Since the system needs to determine the SNR Threshold, $s(t)$, the estimation of the low SNR range of the distribution has to be more accurate. To achieve this goal, the QoS Evaluation module partitions the UEs into two or more groups, according to their QoS values. This allows \AMUSE to accurately infer the optimal value of $s(t)$, by obtaining more reports from UEs with low SNR. 
We elaborate on the algorithms for $s(t)$ estimation in Section~\ref{SC:Alg}.

\noindent
\textbf{MCS Control:}  
Since the eMBMS signal is a combination of synchronized multicast transmissions from several BSs, 
the unicast SNR can be used as a lower bound on the eMBMS SNR. Therefore, the initial eMBMS MCS \change{and FEC are} 
determined from unicast SNR values reported 
by the UEs during unicast connections. 
Then, after each reporting interval, the QoS Evaluation module
%\comment{GZ: which component - be specific}
 infers the SNR Threshold, $s(t)$, and the MCS Control module determines the desired eMBMS settings, mainly the eMBMS MCS and FEC,
according to commonly used one-to-one mappings~\cite{Militano14,embmsfair}.

\begin{table}[t]
\caption{Example of the \DYMO feedback report overhead.
%\comment{reports per interval/sec is probably avg. - this is not deterministic}
}
{\footnotesize
\centering
\begin{tabular}{| c | c | c | c | c | c | c |} \hline 
	\textbf{Group} & 
      \head{\textbf{No.}\\ \textbf{of UEs}} & 
      \head{\textbf{Report} \\ \textbf{Prob. }} & 
      \head{\textbf{Avg. reports} \\ \textbf{per interval}} & 
      \head{\textbf{Avg.} \\ \textbf{per sec}} & 
      \head{\textbf{Rate} \\ \textbf{per min}} \\ \hline
H & $250$ & $20\%$ & $50$ & $5$ & $\approx 100\%$ \\ \hline
L & $2250$ & $2\%$ & $45$ & $\approx{5}$ & $\approx{12}\%$ \\ \hline
\end{tabular}
}
\label{TAB:FBoverheadExample}
\end{table}

%\comment{GZ: avoid the environment - it makes everything italics - just use bold for example and write some text afterwards. Fix this also for the other example}
%\vspace*{0.5cm}
%\begin{example*}
%\label{EX:FB}
%\noindent
%\underline{\em Example 2:}

%\vspace*{3mm}
%\noindent\textbf{Example:} 
%\begin{example*}
\subsection{Illustrative Example}
\label{SSC:example}

\change{\AMUSE operations and the Stochastic Group Instructions concept are demonstrated in the following example.}
Consider an eMBMS system that serves $2,500$ UEs with the QoS Constraint that at most $p=1\%=25$ UEs may suffer from poor service. Assume a reporting interval of $10$ seconds. 
To infer the SNR Threshold, $s(t)$, that satisfies the constraint, the UEs are divided into two groups:
% as shown in Fig.~\ref{fig:FBrate}:

\noindent
$\bullet~$ {\em High-Reporting-Rate} (H): $10\%~(250)$ of UEs that experience poor or moderate service quality report with probability of $20\%$,
i.e., an expected number of $50$ reports per interval.

\noindent
$\bullet~$ {\em Low-Reporting-Rate} (L):  $90\%~(2250)$ of the UEs that experience good or excellent service quality report with 
probability of $2\%$, implying about $45$ reports per interval.

\noindent
Table~\ref{TAB:FBoverheadExample} presents the reporting probability of each UE and the number of QoS reports per reporting interval by each group.
It also shows the number of QoS reports per second and the reporting rate per minute (i.e., the expected fraction of UEs that send QoS reports in a minute). 
Since the QoS Constraint implies that only 25 UEs may suffer from poor service, these UEs must belong to group H.
Although only $10$ QoS reports are received at each second, all the UEs in group H send QoS reports at least once a minute. Thus, the SNR Threshold can be accurately detected within one minute. 
%\comment{GZ: Varun/Yigal - I can see (perhaps with too much detail) how many reports are being  collected but I don't see how the SNR threshold is accurately detected as claimed in the last sentence - does it mean that it can be found out after a minute?}
%\BOX 
%\end{example*}

%--------------------------------------------------------------
\change{
\subsection{Dynamic eMBMS Parameter Tuning}
\label{SSC:prmTuning}

Besides the MCS,  \AMUSE can leverage the UE feedback and the calculated SNR Threshold, $s(t)$,  for optimizing other eMBMS parameters including FEC, video coding 
and protection tier.
While this aspect is not the focus of this study, we briefly discuss the challenges 
and the solutions for dynamic tuning of the eMBMS parameters. 

Once the SNR Threshold $s(t)$ is selected, \AMUSE tunes the eMBMS parameters accordingly.   
%\change{by using commonly accepted mapping~\cite{Urie2013}.} 
% In practice, the eMBMS MCS is a function of $H$~\cite{Urie2013}.
Every time \AMUSE changes the eMBMS parameters, the consumption of wireless resources for the service is affected as well. For instance,
when the eMBMS MCS index is increased, some of the wireless resources allocated for eMBMS are not needed and can be released. 
Alternatively, the service provider may prefer to improve the video quality by instructing the content server to increase the video resolution. Similarly, before the eMBMS MCS index is lowered, the wireless resources should be increased or the video resolution should be reduced to match the content bandwidth requirements to the available wireless resources.

Since the eMBMS signal is a soft combination of the signals from all BSs in the venue, any change of eMBMS parameters must be synchronized at all the BSs to avoid interruption of service. The fact that all the clocks of the BSs are synchronized can be used and a 
scheme similar to the {\em two phase commit} protocol (which is commonly used in distributed databases~\cite{Avi:databases}) can be used.
}

\iffalse
%\subsubsection{Initial eMBMS MCS Setting}
%\label{SSC:AmsueInit}
\noindent
{\bf Initial eMBMS MCS Setting:}
\AMUSE uses an iterative mechanism to tune the eMBMS parameters like the MCS index. The choice of the initial MCS setting is important. As mentioned in Section~\ref{SC:background}, in a given MBSFN area, out-of-cell interference in unicast becomes useful signal in eMBMS. In general, the unicast SINR is a lower bound to the eMBMS SINR. So, we choose the unicast SINR (which is available as a part of the already existing feedback in unicast schemes) as the initial MCS setting for the \AMUSE iterations.
\fi

%%%%%%%%%%%%%%%%%%%%%%%% End of file %%%%%%%%%%%%%%%%%%%%%%

%%%%%%%%%%%%%%%%%%%%%%%%%%%%%%%%%%%%%%%%%%%%%%%%%%%%%%%
% File name:  algorithm-V1.tex
% Authors: Chun-Nam Yu
% Version: 1.1
%------------------------------------------------------
% Changes (date, author, description):
%%%%%%%%%%%%%%%%%%%%%%%%%%%%%%%%%%%%%%%%%%%%%%%%%%%%%%%
%\comment{GZ: according to IEEE guidelines there is no need to write eq. (x) when reffering to an equation just write (x) (unless its a beginning of a sentence}

\section{Algorithms for SNR Threshold Estimation}
\label{SC:Alg}

This section describes the algorithms utilized by \AMUSE for estimating the SNR Threshold, $s(t)$, for a given QoS Constraint, $p$ and Overhead Constraint $r$.
In particular, it addresses the challenges of partitioning the UEs into groups according to their SNR distribution as well as determining the group boundaries and the 
reporting rate from the UEs in each group, such that the overall 
estimation error of $s(t)$ is minimized.  
We first consider a static setting \change{with fixed number of eMBMS receivers, $m$,} where the SNR values of UEs are fixed.
\change{Then, we extend our solution} to the case of dynamic environments and UE mobility.
%The proofs are omitted due to space constraints and can be found in~\cite{DyMoRR}.
%Due to space limitation our claims are provided without proofs.
 
\subsection{Order Statistics}

We first briefly review a known statistical method in quantile estimation, referred to as \orderstatsnb.
It provides a baseline for estimating $s(t)$ and is also used by \AMUSE for determining the initial SNR distribution 
in its first iteration assuming a single group.
Let $F(x)$ be a Cumulative Distribution Function (CDF) for a random variable $X$, the quantile function $F^{-1}(p)$ is given by, 
%\[
$\inf \{x \mid F(x) \geq p\}$. 
%\]

Let $X_1, X_2, \ldots, X_r$ be a sample from the distribution $F$, and $F_r$ its empirical distribution function. 
It is well known that the empirical quantile $F_r^{-1}(p)$ converges to the population quantile $F^{-1}(p)$ at all points $p$ where $F^{-1}$ is continuous~\cite{van2000asymptotic}. 
Moreover, the true quantile, \change{$\hat{p}=F(F_r^{-1}(p))$}, of the empirical quantile estimate $F_r^{-1}(p)$ is asymptotically normal~\cite{van2000asymptotic} 
with mean $p$ and variance 
\change{
\begin{equation}
%\frac{p(1-p)}{r}. \label{eq:variance}
\Var[\hat{p}]= \frac{p(1-p)}{r}
\label{eq:variance}
\end{equation}
}

For SNR Threshold estimation, $F$ is the SNR distribution of all UEs. 
A direct way to estimate the SNR Threshold $s(t)$ is to collect QoS reports from $r$ randomly selected UEs, and calculate the empirical quantile $F_r^{-1}(p)$ as an estimate.%. 
\footnote{Note that $F$ can have at most $m$ points of discontinuity. Therefore, we assume $p$ is a point of continuity for $F^{-1}$ to enable normal approximation. If the assumption does not hold, we can always perturb $p$ by an infinitesimal amount to make it a point of continuity for $F^{-1}$.}

\subsection{The Two-Step Estimation Algorithm}
\label{SSC:twostep}
%\subsection{Static Case}

We now present the \twostep algorithm that uses two groups for 
estimating the SNR Threshold, $s(t)$, in a static setting.
We assume a fixed number of UEs, $m$, and a bound $r$ on the number of expected reports.
By leveraging \emph{Stochastic Group Instructions}, \AMUSE is not restricted to collecting reports uniformly from all UEs and 
can use these instructions to improve the accuracy of 
%the SNR Threshold estimation, 
$s(t)$. 
One way to realize this idea is to perform a two-step estimation that
approximates the shape of the SNR distribution before focusing on the low quantile tail. 
The \twostep algorithm works as follows:

\iffalse  %%% Example
\begin{algorithm}[t]
\caption{MRA Algorithm}
\label{ALG:rateadaptationprocess}
\begin{algorithmic}[1]  
{\footnotesize
%\Procedure{RateAdaptation}{}
\State $rate \gets lowestRate$, $window \gets W_{min}$, $changeTime \gets t$, $refTime \gets t, t := \textrm{current time}$
\While {($true$)}
\State Get PDR reports from all FB nodes
\State Get Status of each FB node $i$
% \State Invoke FB node Selection if needed
\State Calc $\hat{A}_{t}$ and $\hat{M}_{t}$
% \State $rate, action, changeTime \gets GetRate(rate, window, rateChangeTime ,t)$
% \State $window, refTime \gets GetWinSize(action, window, ref, t)$
\State $rate, action, changeTime \gets GetRate(...)$
\State $window, refTime \gets GetWinSize(...)$
\State set multicast rate to $rate$
\State sleep one reporting interval 
\EndWhile
%\EndProcedure
}
\end{algorithmic}
\end{algorithm}
\fi

\vspace{0.3cm}
\noindent
{\em \underline{Algorithm 1}: Two-Step Estimation for the Static Case}

\begin{enumerate}
\item Select $p_1$ and $p_2$ such that $p_1 p_2 = p$. Use $p_1$ as the percentile boundary for defining the two groups.
\item  Select number of reports $r_1$ and $r_2$ for each step such that $r_1+r_2=r$.

\item  Instruct all UEs to send QoS reports with probability $r_1/m$ and use these reports to estimate the $p_1$ quantile $\hat{x}_1 = F_{r_1}^{-1}(p_1)$.
%\item Randomly collect $r_1$ samples from all UEs, and estimate the $p_1$ quantile $\hat{x}_1 = F_{r_1}^{-1}(p_1)$. 

\item  Instruct UEs with SNR value below $\hat{x}_1$ to send reports with probability $r_2/(p_1\cdot m)$ and calculate the $p_2$ quantile $\hat{x}_2 \!=\! G_{r_2}^{-1}(p_2)$ as an estimation for $s(t)$ ($G$ is the CDF of the subpopulation with SNR below $\hat{x}_1$).
%\item Randomly collect $r_2$ samples from UEs with SNR below $F_{r_1}^{-1}(p_1)$. Report the $p_2$ quantile $\hat{x}_2 \!=\! G_{r_2}^{-1}(p_2)$ as estimate for SNR threshold $s(t)$ 
(\change{$G_{r_2}$} is the \change{empirical CDF} of the subpopulation with SNR below $\hat{x}_1$).  
\end{enumerate}

%[do we need a pictorial illustration here?]
\vspace{0.3cm}

\noindent
{\bf  Upper Bound Analysis of the Two-Step Algorithm:} To simplify the notation, we use $r_1$ and $r_2$ to denote the expected number of reports at each step.
\change{
From (\ref{eq:variance}) we know that 
\[
\hat{p}_1 = F(\hat{x}_1)~~~and~~~\hat{p}_2 = G(\hat{x}_2)
\]
are unbiased estimators of $p_1$ and $p_2$ with variance 
\begin{equation}
\Var[\hat{p}_1] = \frac{p_1(1-p_1)}{r_1} ~~and~~
\Var[\hat{p}_2] = \frac{p_2(1-p_2)}{r_2}
\label{eq:2Svar}
\end{equation}} %% End of Change
%From (\ref{eq:variance}) we know that $\hat{p}_1 = F^{-1}(\hat{x}_1)$ %and $\hat{p}_2 = G^{-1}(\hat{x}_2)$ are unbiased estimators of $p_1$ %and $p_2$ with variance $p_1(1-p_1)/r_1$ and $p_2(1-p_2)/r_2$. 
Our estimate $\hat{x}_2$ has true quantile $\hat{p}_1 \hat{p}_2$. 
Assume $\hat{p}_1$ is less than $p_1 + \epsilon_1$ and $\hat{p}_2$ is less than $p_2 + \epsilon_2$ with high probability (for example, we can take $\epsilon_1$ and $\epsilon_2$ to be 3 times the standard deviation for $>99.8\%$ probability). 
Then, the over-estimation error is bounded by 
\change{
%$(p_1 + \epsilon_1)(p_2 + \epsilon_2) - p \approx\ \epsilon_1 p_2 + %\epsilon_2 p_1$
\begin{equation}
%\[
\begin{split}
  \epsilon =\ &  (p_1 + \epsilon_1)(p_2 + \epsilon_2)\ -\ p\\ 
           =\ & p_1 p_2 + \epsilon_1 p_2 + \epsilon_2 p_1 + 
                          \epsilon_1 \epsilon_2 - p\\
     \approx\ & \epsilon_1 p_2 + \epsilon_2 p_1 
\end{split}
%\]} %% End of Change
\label{eq:2Serror}
\end{equation}} %% End of Change

\noindent
after ignoring the small higher order term $\epsilon_1 \epsilon_2$.
%
\iffalse
\[
\begin{split}
  (p_1 + \epsilon_1)(p_2 + \epsilon_2) - p 
=\ & p_1 p_2 + \epsilon_1 p_2 + \epsilon_2 p_1 + \epsilon_1 \epsilon_2 - p\\
\approx\ & \epsilon_1 p_2 + \epsilon_2 p_1. 
\end{split}
\]
\fi
%
The case for under-estimation is similar. 
\change{As shown in the Appendix,}
%We can ignore the small higher order term $\epsilon_1 \epsilon_2$. 
%By using symmetry arguments, we show in~\cite{DyMoRR} that 
%it can be shown that 
the error is minimized by taking,
\change{
\[
p_1 \!=\! p_2 \!=\! \sqrt{p}~~~and~~~r_1 \!=\! r_2 \!=\! r/2
\]
so that 
\[
\epsilon_1 \!=\! \epsilon_2 \!=\! 3\sqrt{\sqrt{p}(1-\sqrt{p})/(r/2)}
\]}

\noindent 
This leads to proposition~\ref{PR:1}.
%the following proposition: 
\begin{proposition}
\label{PR:1}
The distance between $p$ and the quantile of the Two-Step estimator $\hat{x}_2$, $\hat{p} = F^{-1}(x_2)$, is bounded by
\[
  6\sqrt{2} \sqrt{\frac{p\sqrt{p}(1-\sqrt{p})}{r}}
\]
%with $>99.6\%$ probability by union bound. 
with probability at least $1-2(1-\Phi(3)) > 99.6\%$, where $\Phi$ is the normal CDF. 
\end{proposition}

We now compare this result against the bound of 3 standard deviations in the Order Statistics case, which is $3\sqrt{p(1-p)/r}$. 
With some simple calculations, it can be easily shown that if $p\leq 1/49 \approx 2\%$, the \twostep has smaller error than 
the \orderstats method. 
Essentially the \orderstats method has an error of order $\sqrt{p}/\sqrt{r}$, while the \twostep has an error of order $p^{3/4}/\sqrt{r}$. 
Since $p\ll 1$, the difference can be significant. 

\begin{figure}[t]
\centering
\subfigure[]{
\includegraphics[trim=15mm 0mm 5mm 5mm, width=0.22\textwidth]{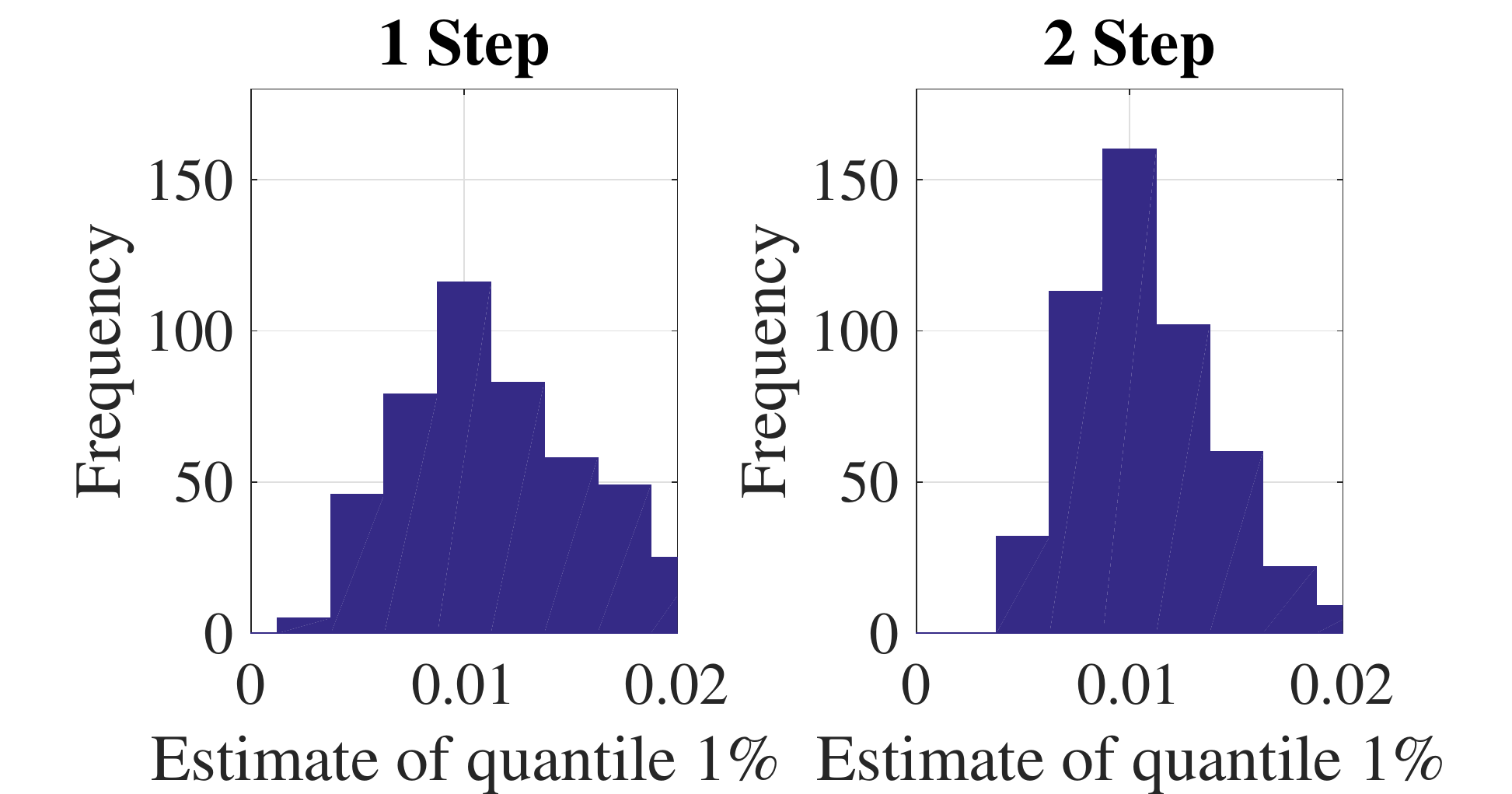}
\label{FIG:simulation} %% -(a)
}%
\subfigure[]{
\includegraphics[trim=15mm 0mm 5mm 5mm, width=0.22\textwidth]{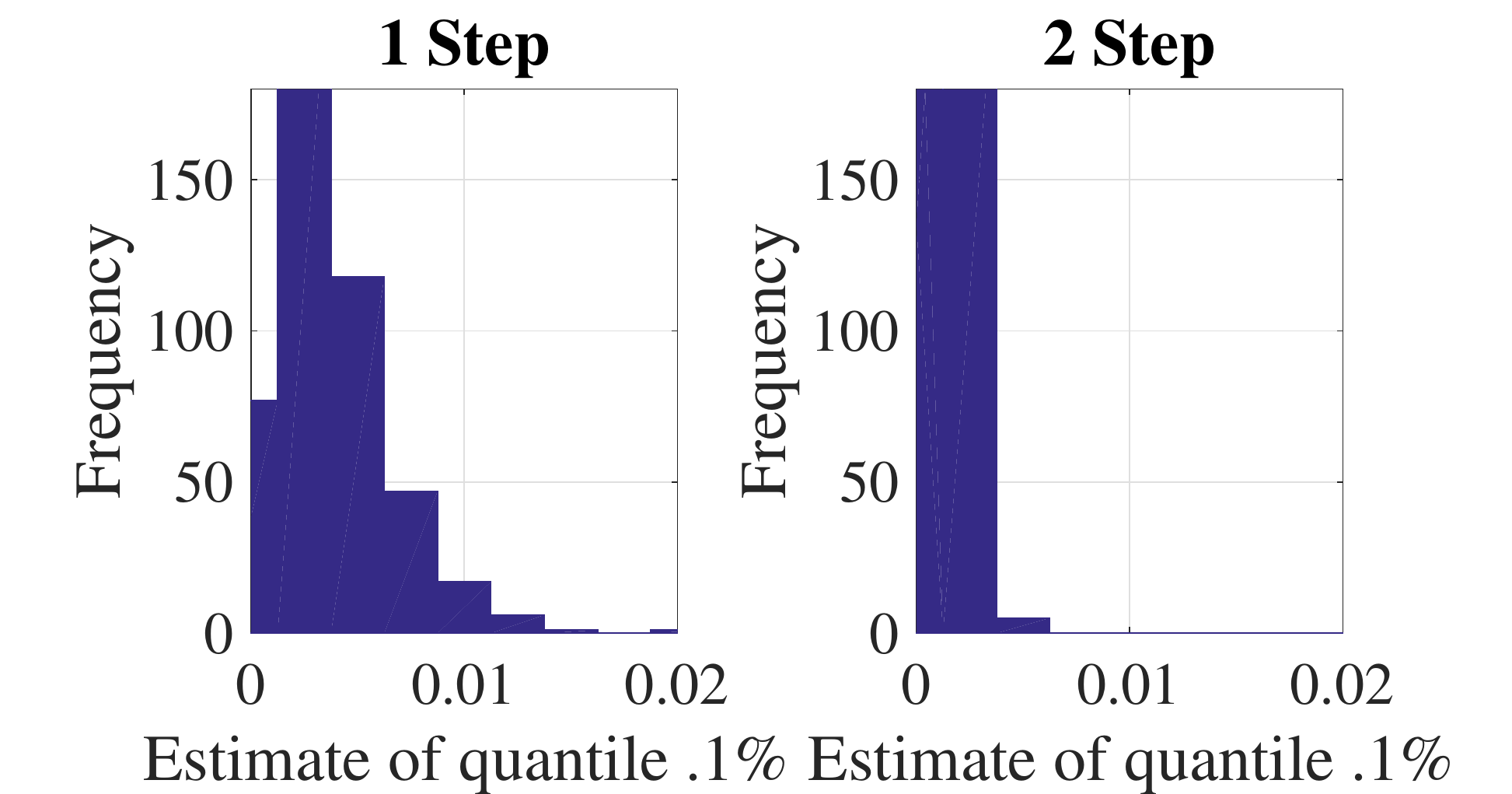}
\label{FIG:simulation_1} %% -(a)
}%
%\vspace*{-0.2cm}
\caption{Estimates of (a) $p=1\%$ and (b) $p=0.1\%$ quantiles for 500 runs for the \orderstats (1-step) method and the \twostep algorithm.}
\label{FIG:simulationQuant}
\end{figure}

\noindent
\textbf{Example:} We validated the error estimation of the \twostep algorithm and the \orderstats method by numerical analysis. 
We considered the cases of $p=1\%$ and $p=0.1\%$ of uniform distribution 
on $[0,1]$ using $r\!=\!400$ samples over population size of $10^6$.
The \twostep algorithm has smaller standard error compared to the \orderstatsnb, as shown in Fig.~\ref{FIG:simulationQuant}. 
Its accuracy is significantly better for very small $p$.
%[put in an R figure to compare two-step and direct]
%\BOX
%\end{example*}

The \twostep algorithm can be generalized to 3 or more telescoping 
group sizes, but $p$ will need to be much smaller for these sampling 
schemes in order to reduce the number of samples. 
%Therefore in this work we focus only on 2-step estimation. 

\subsection{The Iterative Estimation Algorithm}

We now turn to the dynamic case in which \AMUSE uses 
the SNR Threshold estimation  $s(t-1)$ from the previous 
reporting interval to estimate 
%the SNR Threshold 
$s(t)$ at the end of reporting interval $t$.
\change{Assume that the total number of eMBMS receivers, $m$, is fixed and it is known initially.}
%and suppose we update our estimates every unit time. 

Suppose that \AMUSE has a current estimate $\hat{x}$ of the SNR threshold, $s(t)$, and $s(t)$ changes over time. 
We assume that the change in SNR of each UE is bounded over a time period. 
Formally,
\change{
\[
|h_v(t_1) - h_v(t_2)| \leq L |t_1 - t_2| 
\]}
where $L$ is a Lipschitz constant for SNR changes. 
For example, we can assume that the UEs' SNR cannot change by more than $5$dB during a reporting interval.
\footnote{In our simulations, each reporting interval has a duration of $12$s.}
%within 10 seconds. 
This implies that within the interval, only UEs with SNR below $\hat{x} + 5$dB affect the estimation of the $p$ quantile (subject to small estimation error in $\hat{x}$). 

\AMUSE only needs to monitor UEs with SNR below $x_L = \hat{x} + L$.  
Denote the true quantile of $x_L$, defined by $F^{-1}(x_L)$, as $p_L$. 
To apply a process similar to the second step of the \twostep algorithm by focusing on UEs with SNR below $x_L$, first an estimate of $p_L$ is required. 
\AMUSE uses the previous SNR distribution to estimate $p_L$ and instructs the UEs to send reports at a rate $q=r/(p_L\cdot m)$. 
Let $Y$ be the number of reports received during the last reporting interval, then $Y/m\cdot q$ can be used as an updated estimator,
\change{$\hat{p_L}$},  for $p_L$.  
This estimator is unbiased and has variance 
\change{
\begin{equation}
\Var[\hat{p_L}]=\Var[\frac{Y}{m\cdot q}] = \frac{p_L}{m}\frac{1-q}{q} 
\label{eq:ItrVarPL}
\end{equation}}%% End of change
%
%$\frac{p_L}{m}\frac{1-q}{q}$. 

\noindent
As a result, the Iterative Estimation algorithm works as follows:
 
\vspace{0.3cm}
\noindent
{\em \underline{Algorithm 2}: Iterative Estimation for the Dynamic Case}

\begin{enumerate}
\item Instruct UEs with SNR below $\hat{x} + L$ to send reports at a rate $q$. Construct an estimator $\hat{p}_L$ of $p_L$ from the number of received reports $Y$.   
\item Set $p' = p/\hat{p}_L$. Find the $p'$ quantile $x' = G_Y^{-1}(p')$ and report it as the $p$ quantile of the whole population ($G$ is the CDF of the subpopulation with SNR below $\hat{x} + L$).  
\end{enumerate}

\vspace{0.3cm}
\noindent
{\bf  Upper Bound Analysis of the Iterative Algorithm:} Suppose the estimation error of $p_L$ is bounded by $\epsilon_1$, and the estimation error of $p' = p/\hat{p}_L$ is bounded by $\epsilon_2$ with high probability. 
Then, the estimation error is
\[
\epsilon =  (\frac{p}{\hat{p}_L} \pm \epsilon_2) p_L - p = (\frac{p}{p_L \pm \epsilon_1} \pm \epsilon_2) p_L - p. 
\]
The over-estimation error is bounded by
\begin{equation}
  \frac{p}{p_L - \epsilon_1}\epsilon_1 + p_L\epsilon_2. \label{eq:L_bound}
\end{equation}
If we assume $p_L - \epsilon_1 \geq p$ (we know $p_L \geq p$ by the Lipschitz assumption), then the bound can be simplified to $\epsilon_1 + p_L\epsilon_2$. 
The same bound also works for the under-estimation error. 
If $r$ denotes also the expected number of samples collected, 
$r = p_L\cdot m \cdot q$. 
The standard deviation of $\hat{p}_L$ can be written as:
\[
  \sqrt{\frac{p_L}{m}\frac{1-q}{q}} = \sqrt{\frac{p_L^2}{r}(1 - \frac{r}{p_L m})} \leq \frac{p_L}{\sqrt{r}}. 
\]

If we assume $\epsilon_1= 3p_L/\!\sqrt{r}$, the error of $\hat{p}_L$ is less than $\epsilon_1$ with probability at least $\Phi(3)$. 
Since we assume $p_L \!-\! \epsilon_1 \!\geq\! p$ above, this implies 
$(1\!-\!3/\sqrt{r})p_L\! \geq\! p$. 
If $r\!\geq\! 100$, then $p\!<\!0.7 p_L$ will satisfy our requirement. 

The standard deviation of estimating the $p' = p/\hat{p}_L$ quantile is %approximately
\begin{equation}
  \sqrt{\frac{1}{Y} \frac{p}{\hat{p}_L} (1- \frac{p}{\hat{p}_L})} \leq \frac{1}{2\sqrt{Y}}, \label{eq:e2_bound}
\end{equation} 
by using the fact that $x(1-x)\leq 1/4$ for $x\in [0,1]$ and $Y$ is the number of reports received (a random variable). 
If the expected number of reports ${r}$ is reasonably large 
%($\geq 50$, say), 
($\geq 100$, say), 
then $Y$ can be well approximated by a normal and $Y \geq 0.7 {r}$ with high probability $\Phi(3)=99.8\%$.   
Then, (\ref{eq:e2_bound}) is bounded by $2/(3\sqrt{{r}}) \geq 1/(2\sqrt{0.7r})$ with high probability ($\Phi(3)=99.8\%$), and we can set $\epsilon_2 = 2/\sqrt{{r}}$. Substituting these back into (\ref{eq:L_bound}), gives us the following proposition. 
\begin{proposition}
The distance between $p$ and the quantile of the estimator $x$, $\hat{p} = %F^{-1}(x')$, 
F^{-1}(x)$, is approximately bounded by
\[
  5\frac{p_L}{\sqrt{r}}
\]
with probability at least $1-2(1-\Phi(3)) > 99.6\%$, if the expected sample size ${r} \geq 100$ and $p \leq 0.7 p_L$. 
\end{proposition}

This shows that the error is of order $p_L/\sqrt{{r}}$. 
We can see that the estimation error can be smaller compared to the error of order $p^{3/4}/\sqrt{{r}}$ in the static \twostep if $p_L$ is small (i.e., the SNR of individual users does not change much during a reporting interval).
 
\vspace{0.3cm}
\noindent
{\bf  Exponential Smoothing:} \AMUSE applies exponential smoothing by weighing past and current reports to smooth the estimates of the SNR Threshold, $s(t)$, and take older reports into account. It estimates the SNR Threshold as 
\change{
\[
s(t) = \alpha \hat{x}(t) + (1-\alpha) s(t-1)
\]}
where $\hat{x}(t)$ is the new raw SNR Threshold estimate using the \iterative above and $s(t-1)$ is the SNR Threshold from the previous reporting interval.
We set $\alpha = 0.5$ to allow some re-use of past reports without letting them have too strong an effect on the estimates (e.g., samples older than 7 reporting intervals have less than 1\% weight). 
\AMUSE also uses the exponential smoothing method for estimating the SNR distribution while taking into account QoS reports from previous reporting intervals.

\iffalse
To smooth the estimates of the SNR threshold $s(t)$ and take into account older reports, \AMUSE applies exponential smoothing 
\[
  s(t) = \alpha \hat{x}(t) + (1-\alpha) s(t-1), 
\]
where $\hat{x}(t)$ is the new raw SNR threshold estimate using the \iterative above and $s(t-1)$ is the SNR threshold from the previous reporting interval. 
We set $\alpha$ equals 0.5 to allow some re-use of past reports without letting them have too strong an effect on the estimates (e.g., samples older than 7 reporting intervals have less than 1\% weight (1/128)). 
\AMUSE also uses the exponential smoothing method for estimating the SNR distribution while taking into account QoS reports from previous reporting intervals.
\fi

\vspace{0.3cm}
\noindent
\change{{\bf Dynamic and Unknown Number of eMBMS Receivers:}} If the total number of UEs, \change{$m(t)$}, is unknown or changes dynamically, 
\AMUSE can estimate \change{$m(t)$} by requiring UEs above the threshold $\hat{x} + L$ to send reports. 
These UEs can send reports at a lower rate, since \change{$m(t)$}
is not expected to change rapidly.
Similar to the \twostep algorithm, \AMUSE allocates $r_1=r_2=r/2$ reports to each group.    
The errors in estimating the total number of UEs \change{$m(t)$} will contribute to the error $\epsilon_1$ in the estimation of $p_L$ in (\ref{eq:L_bound}). 
The error analysis in this case is largely similar. 
%% Yigal - removed for the RR.
%% and is omitted here due to space limits. 
%As expected from the analysis, simulations in Section~\ref{SC:eval} show that \AMUSE provides accurate estimation of $s(t)$.

%%%%%%%%%%%%%%%%% End of File %%%%%%%%%%%%%%%%%%%

\section{Performance Evaluation}
\label{SC:eval}

\begin{figure}[t]
\centering
\subfigure[]{
\includegraphics[trim=5mm 0mm 0mm 5mm, width=0.21\textwidth]{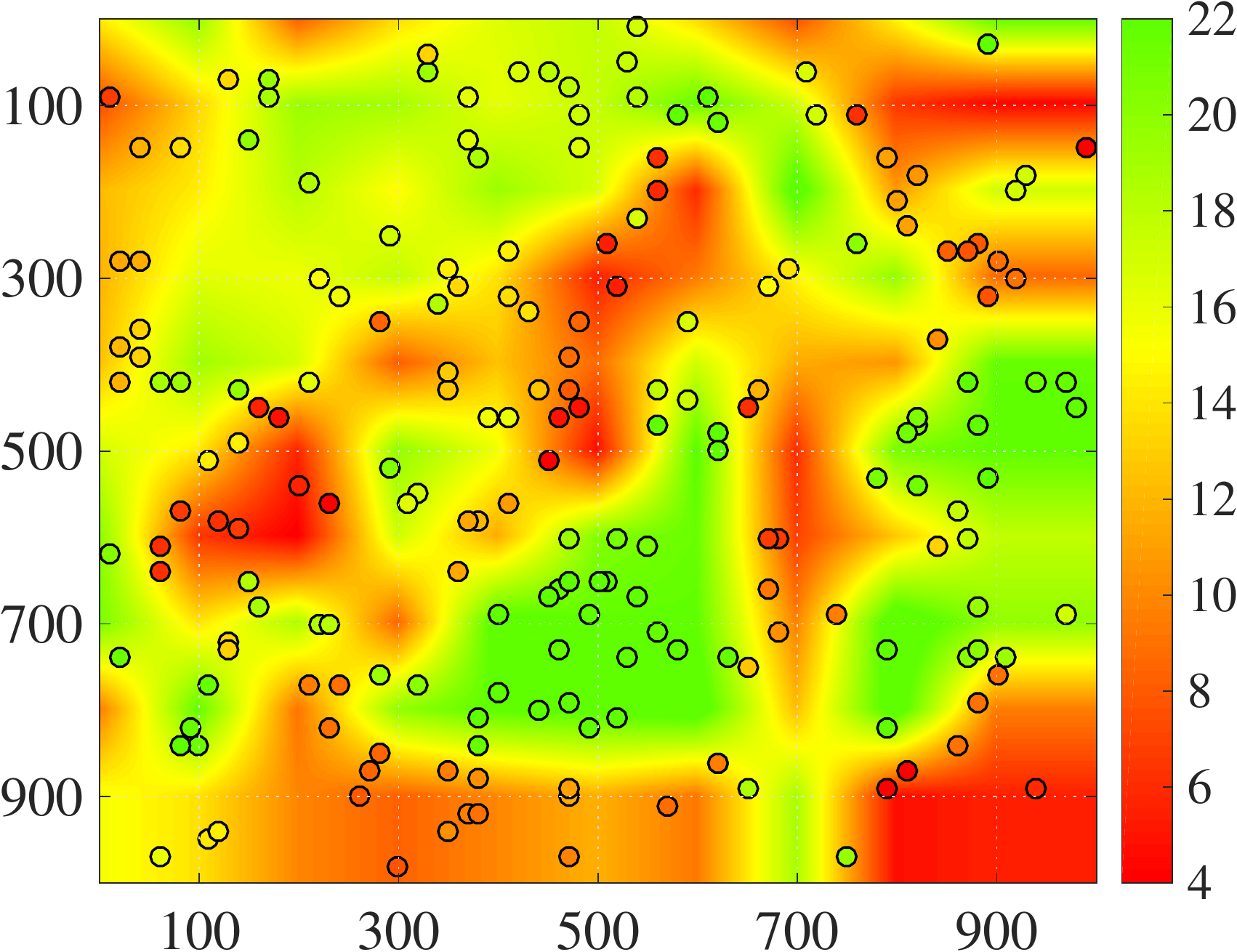}
\label{FIG:UnifHeatmap} %% -(a)
}%
\subfigure[]{
\includegraphics[trim=5mm 0mm 0mm 5mm, width=0.21\textwidth]{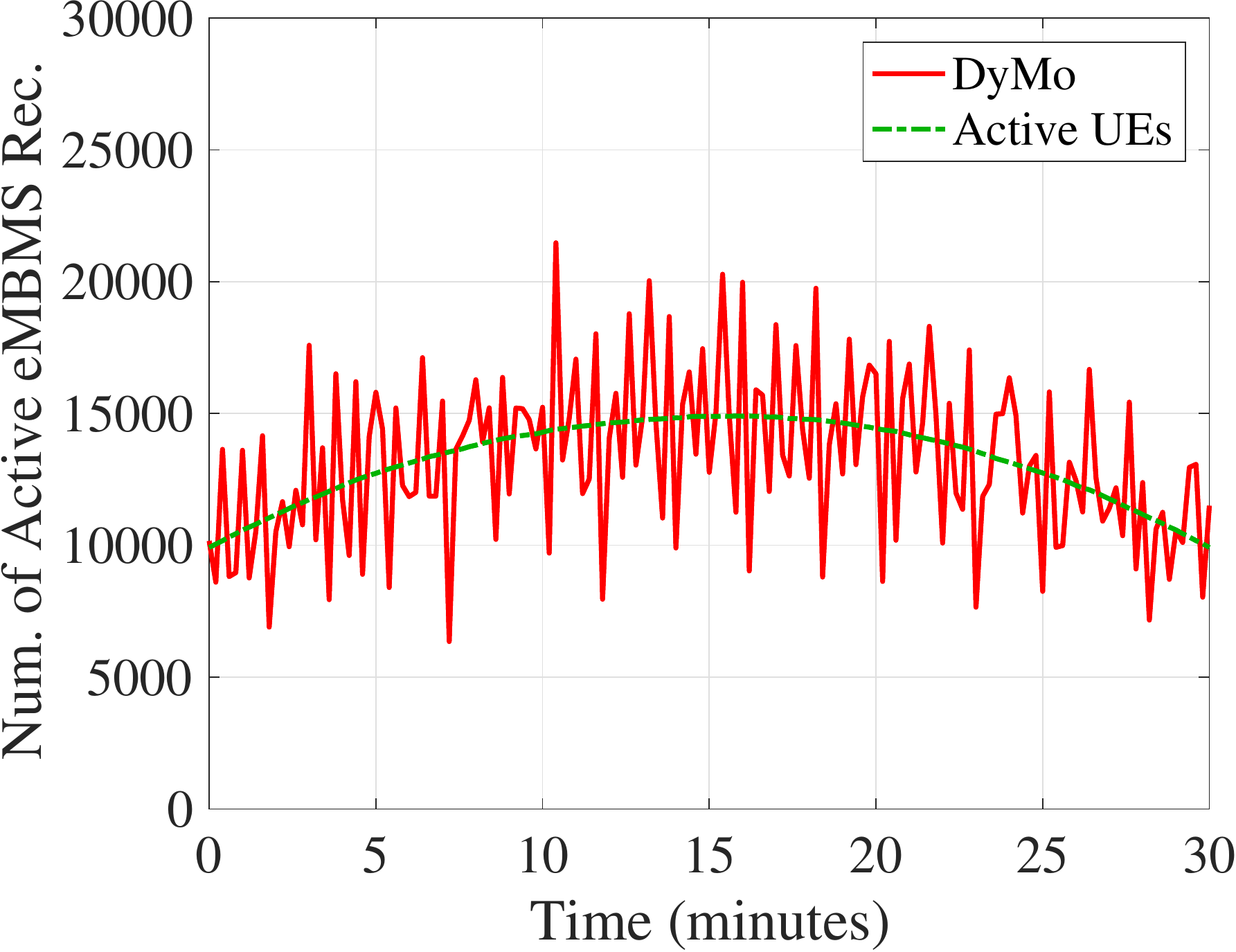}
\label{FIG:UnifActiveUsers} %% -(a)
}%
%\vspace*{-0.2cm}
\caption{(a) The heatmap of SNR distribution of UEs (b) the evolution of the number of active UEs over time compared to the number estimated by \AMUSE for a \HOMOGENEOUS environment.}
%\vspace*{-0.5cm}
\label{FIG:UnifHeatMapUEs}
\end{figure}

\begin{figure}[t]
\centering
\subfigure[]{
\includegraphics[trim=5mm 0mm 0mm 5mm, width=0.21\textwidth]{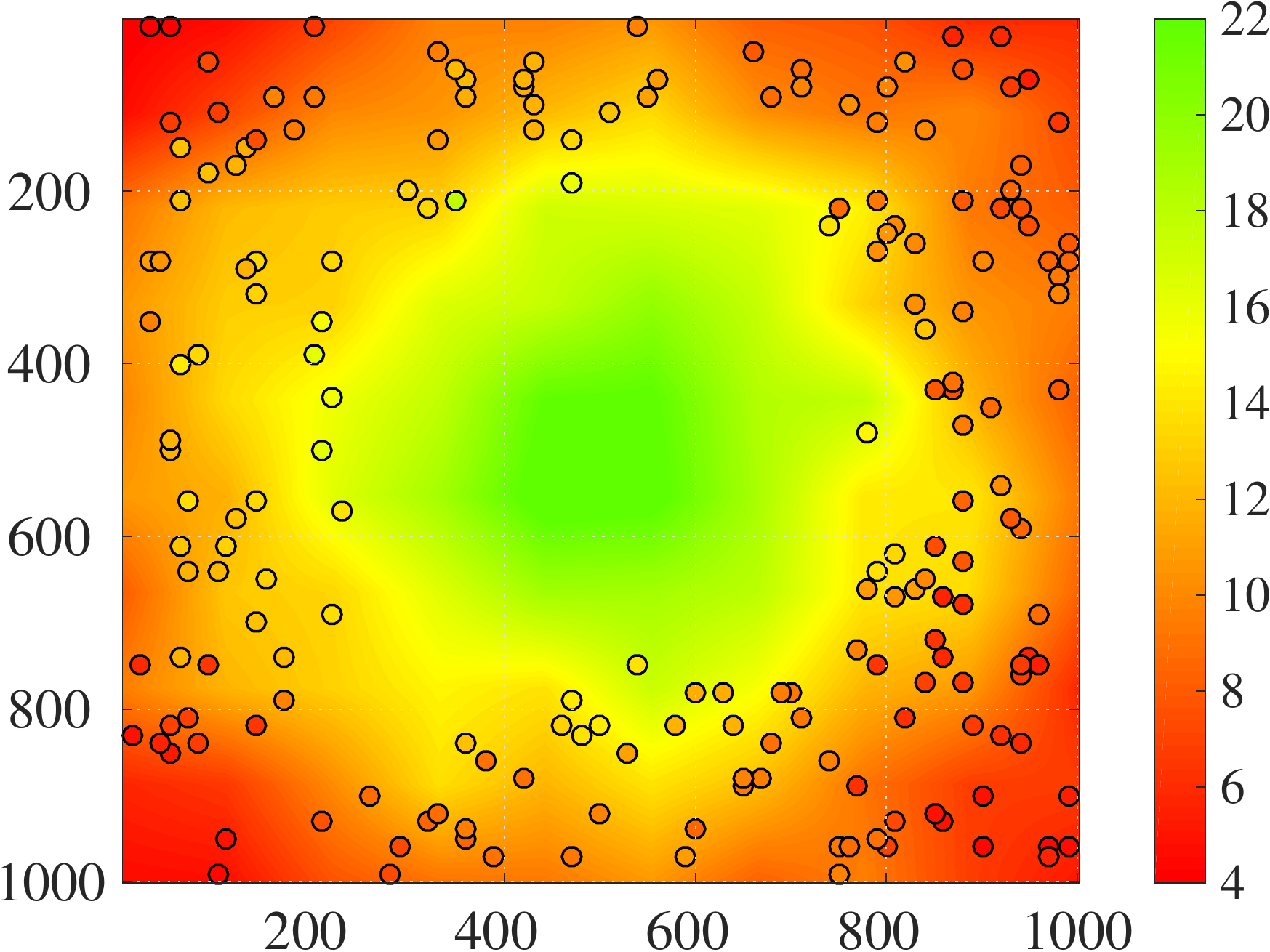}
\label{FIG:heatmap} %% -(a)
}%
\subfigure[]{
\includegraphics[trim=5mm 0mm 0mm 5mm, width=0.21\textwidth]{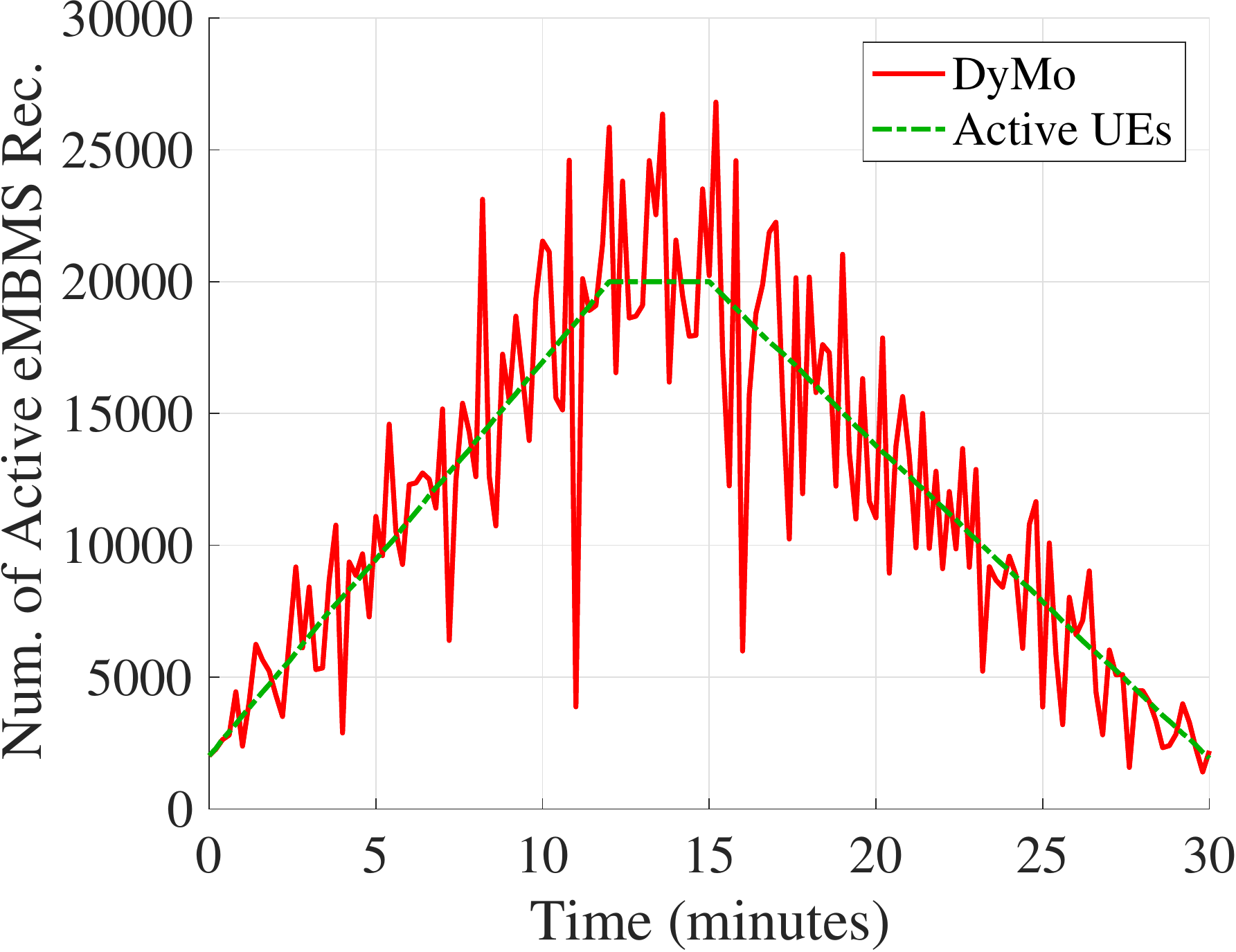}
\label{FIG:ActiveUsers} %% -(a)
}%
%\vspace*{-0.2cm}
\caption{(a) The heatmap of UE SNR distribution in a stadium area of $1000 \times 1000 m^2$ and (b) the evolution of the number of active UEs over time compared to the number estimated by \AMUSE for a stadium environment.}
%\vspace*{-0.5cm}
\label{FIG:HeatMapStadium}
\end{figure}

\begin{figure}[t]
\centering
\subfigure[]{
\includegraphics[trim=5mm 0mm 0mm 5mm, width=0.21\textwidth]{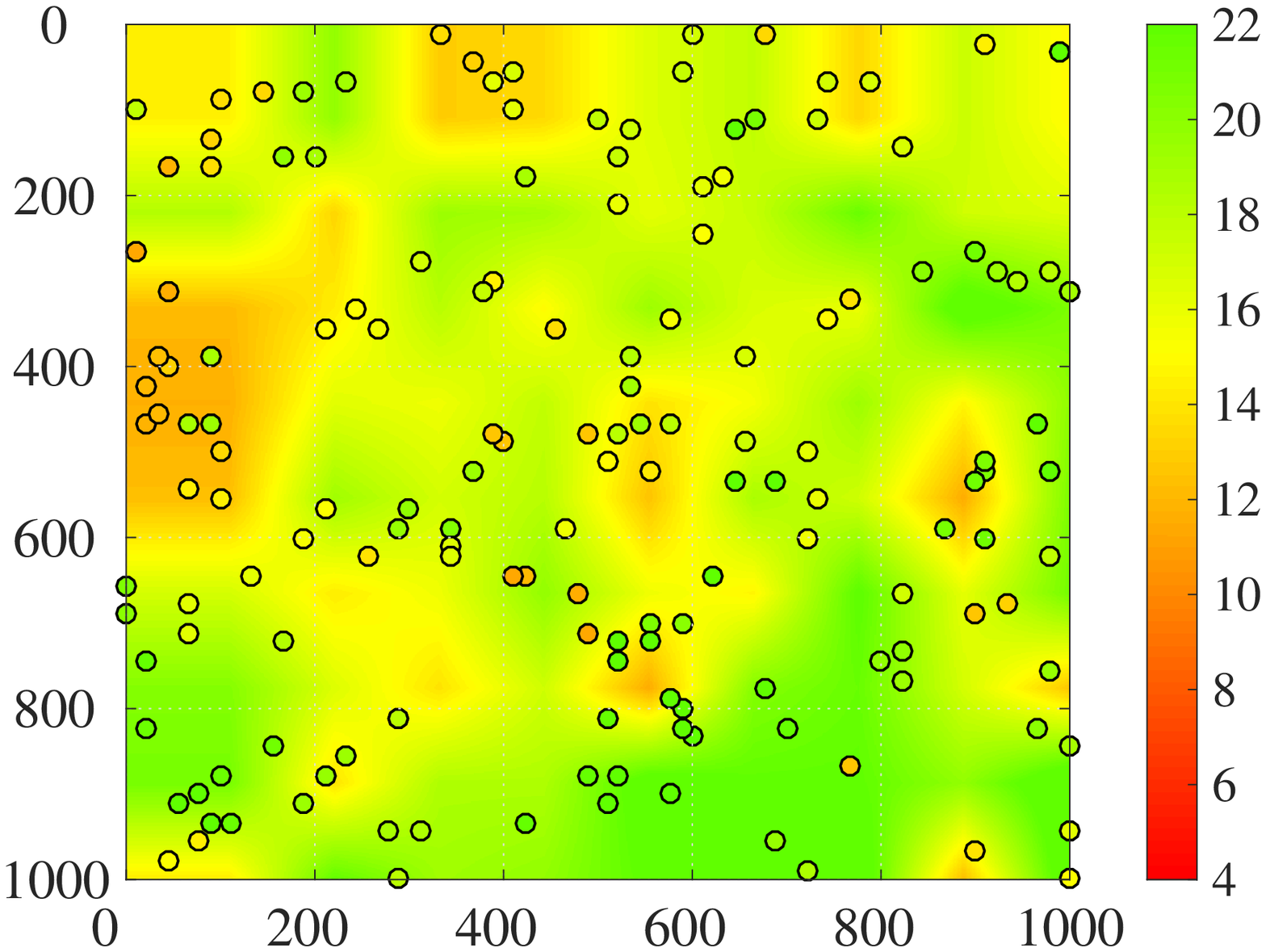}
\label{FIG:heatmap_before} %% -(a)
}%
\subfigure[]{
\includegraphics[trim=10mm 0mm 0mm 5mm, width=0.21\textwidth]{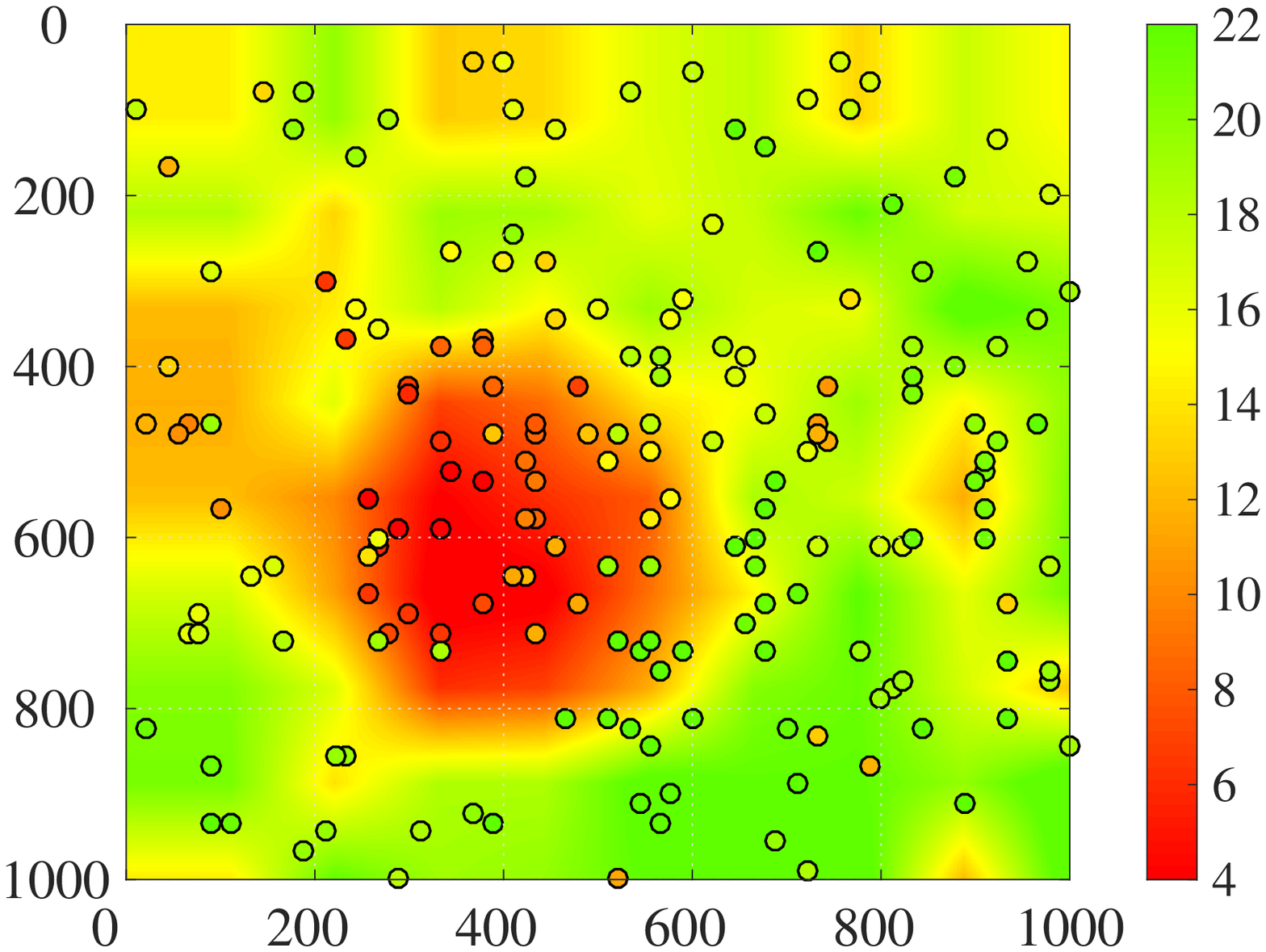}
\label{FIG:heatmap_after} %% -(a)
}%
%\vspace*{-0.5cm}
\caption{The heatmap of the SNR distribution of UEs (a) before a failure and (b) after a failure.}
\label{FIG:HeatMapFailure}
\end{figure}

%----------------------------------------------------------------
\subsection{Methodology} 
\label{SSC:EvalMethodology}

We perform extensive simulations to evaluate the performance of \DYMO with various values of QoS Constraint, $p$, Overhead Constraint, $r$, and number of UEs, $m$. 
Our evaluation considers dynamic environments with UE mobility and 
a changing number of {\em active eMBMS receivers}%
denoted by $m(t)$, dynamically selected from the given set 
of $m$ UEs \change{in the considered venue}. 
\change{In this paper, we present a few sets of simulation results, which capture various levels of variability of the SNR threshold, $s(t)$, over time.}
%in which the {\em SNR Threshold, $s(t)$, 
%changes significantly over time}. 
%Additional results can be found in~\cite{DyMoRR}.

We consider a variant of \AMUSE where the number of active UEs is unknown and is estimated from its measurements. 
We compare the performance of \AMUSE to four other schemes. 
To demonstrate the advantages of \AMUSENB, we augment each 
scheme with additional information, which is hard to obtain in practice.
%\comment{Varun: I don't understand the meaning of the following.}
%Each scheme represents the {\em most accurate estimation that can be 
%obtained by a specific SNR evaluation scheme}.
The evaluated benchmarks are the following:
 
\noindent
$~\bullet$ {\bf \OPTNB} -- Full knowledge of SNR values of the UEs at any time and consequently accurate information of the SNR distribution.
{\em This is the best possible benchmark although impractical, due to its high overhead.}

\noindent
$~\bullet$ {\bf \UNINB} -- Full knowledge of the SNR characteristics at any location while assuming uniform UE distribution and static eMBMS settings. In practice, {\em this knowledge cannot be obtained 
even with rigorous field trial measurements.} 

\noindent
$~\bullet$ {\bf \ORDERSTATSNB} -- It is based estimation of the SNR Threshold using 
random sampling. 
The active UEs send reports with a fixed probability 
of $r/\E[m(t)]$ per second, assuming that the 
expected number of active UEs, $\E[m(t)]$, is known. 
We assume that the UEs are configured with this reporting rate 
during initialization.
In practice, {\em $\E[m(t)]$ is not available.
% to \ORDERSTATS%
%\footnote{$\E[m(t)]$ is definitely unknown during the 
%configuration of the UEs}
We also ignore initial configuration overhead in our evaluation}.
\ORDERSTATS is the best possible approach when not using broadcast messages 
for UE configuration. 
We consider two variants of \ORDERSTATSNB. The first is {\bf\ORDSTnoHIST} which ignores SNR measurements from earlier reporting intervals. The second variant {\bf\ORDSTwithHIST} considers the history of reports. 

Both \AMUSE and \ORDSTwithHIST perform the same exponential 
smoothing process for assigning weights to the measurements from previous reporting intervals with a smoothing factor of $\alpha=0.5$. We use the following metrics to evaluate the performance of the schemes:

\begin{itemize}
\item[(i)] {\em Accuracy --} The accuracy of the {\em SNR Threshold} estimation, $s(t)$. After calculating $s(t)$ at each reporting interval, we check the actual {\em SNR Threshold Percentile} in the accurate SNR distribution \change{of the considered scheme}. This metric provides the percentile of active UEs with individual SNR values below $s(t)$. 
%Note that this also considers the ability of the scheme to adapt quickly to changes of number of active eMBMS receivers and their SNR values.
%\noindent 
%$~(i)$ 

\item[(ii)] {\em QoS Constraint violation --} The number of outliers above the QoS Constraint $p$. 
\change{The number of outliers of a scheme in a given reporting interval $t$ is defined as the actual SNR Threshold Percentile of the scheme times the number of active eMBMS receivers, $m(t)$, at time $t$.
}
%Recall that this metric is defined relative to the current number 
%of active eMBMS receivers and not to $m$.
%\noindent
%$~(ii)$ 

\item[(iii)] {\em Overhead Constraint violation --} The number of reports above the Overhead Threshold, $r$, \change{at each reporting interval}.
%\noindent
%$~(iii)$ 
\end{itemize}

The total simulation time for each instance is $30$mins 
with $5$ reporting intervals per minute (each is $12$s).
During each reporting interval, an active UE may 
send its SNR value at most once.
%For \AMUSE and \ORDERSTATS variants, some active eMBMS receivers send 
%their SNR values at each reporting interval.
The accuracy of each SNR report is $0.1$dB. 
%This means that the SNR distribution of UEs is a discrete function.

%--------------------------------------------------------------
\subsection{Simulated Environments}
%\subsection{Stadium Environment}

We simulated a variety of environments with different SNR distributions and UE mobility patterns.
Although the simulated environments are artificial, their SNR distributions mimic those of real eMBMS networks obtained through field trial measurements. 
To capture the SNR characteristics of an environment, 
we divide its geographical area into rectangles of $10m \times 10m$.
For each reporting interval, each UE draws its individual SNR value, $h_v(t)$, from a Gaussian-like distribution which is a characteristic of the rectangle in which its located. The rectangles have different mean SNR, but the same standard deviation of 
roughly $5$dB (as observed in real measurements).
Thus, the SNR characteristics of each environment are determined by 
the mean SNR values of the rectangles at any reporting interval.
%
\iffalse
In some environments, typically where the SNR variations are small, 
the SNR Threshold, $s(t)$, barely changes over time.
For such scenarios, the \UNI scheme, based on rigorous field trial 
measurements, is an appropriate solution.
In such a situation, \AMUSE can efficiently infer the SNR Threshold 
and reduce the need for expansive field trails. 
%The results for these simulations are in~\cite{DyMoRR}.
In this paper, we discuss two types of environments in which $s(t)$
changes significantly over time.
\fi
%
\change{
To demonstrate the performance of the different schemes, we discuss three types of environments.

\noindent
$~\bullet$ {\bf \HOMOGENEOUSNBCP:} 
In the \HOMOGENEOUSNB%
\footnote{We use the term \HOMOGENEOUS since the term uniform is already used to denote the \UNI scheme.}
setting the mean SNR value of each rectangle is fixed 
and it is uniformly selected in the range of $5-25$dB.
Fig.~\ref{FIG:UnifHeatmap} provides an example of the mean SNR 
values of such a venue as well as typical UE location distribution. 
In such instances, we assume random mobility pattern, in which each 
UE moves back and forth between two uniformly selected points. 
During the simulation, $50\%$ of the UEs are always active, while the other $50\%$ join and leave at some random time, as illustrated 
by Fig.~\ref{FIG:UnifActiveUsers}. 
As we show later in such setting $s(t)$ barely change over time.
} %% End of change

\noindent
$~\bullet$ {\bf Stadiums:} 
In a stadium, the eMBMS service quality is typically significantly better inside the stadium than in the surrounding vicinity (e.g., the parking lots).
To capture this, we simulate several stadium-like environments, 
in which the stadium, in the center of the venue, has high eMBMS SNR with 
mean values in the range of $15-25$dB. On the other hand, the vicinity has significantly 
lower SNR with means values of $5-10$dB.
%An example of such an environment is shown in Fig.~\ref{FIG:heatmap}. 
An example of a stadium is shown in Fig.~\ref{FIG:heatmap}. 
%An example of the mean SNR values of an given simulated stadium in an 
%area scaled by $1000m \times 1000m$ is shown in Fig.~\ref{FIG:heatmap}. 

We assume a mobility pattern in which, the UEs move from the edges to the inside of the stadium in $12$mins, stay there for $3$mins, and then go back to the edges.\footnote{While significant effort has been dedicated to modeling mobility (e.g., \cite{levy, temporalrobust} and references therein), we use a \emph{simplistic mobility model} since our focus is on the multicast aspects rather than the specific mobility patterns.}
As shown in Fig.~\ref{FIG:ActiveUsers}, as the UEs move toward the center, 
the number of active UEs gradually increases from $10\%$ of the UEs to $100\%$, and then declines again as they move away.

\noindent
$~\bullet$ {\bf Failures:} 
\change{Such an environment is similar to the \HOMOGENEOUS 
setting with a sudden event of a component failure.} %% End of Change
In the case of a malfunctioning component, the QoS 
in some parts of a venue can degrade significantly.
To simulate failures, we consider cases in which the 
eMBMS SNR is high with a mean between $15-25$dB. 
During the simulation, (around the $10^{th}$ minute),
we mimic a failure by reducing the mean SNR values 
of some of the rectangles by over $10$dB to the range 
of $5-10$dB. The mean SNR values are restored to their original 
values after a few minutes.
Figs.~\ref{FIG:heatmap_before}~and~\ref{FIG:heatmap_after} provide an example of the mean SNR values of such a venue before and after a failure, respectively. % at the $7^{th}$ minute.
\change{We assume the same mobility pattern like the \HOMOGENEOUS 
setting, as shown by Fig.~\ref{FIG:UnifActiveUsers}.} 
%% End of \change

\iffalse
In such instances, we assume random mobility pattern, in which each 
UE moves back and forth between two uniformly selected points. 
During the simulation, $50\%$ of the UEs are always active, while the other $50\%$ join and leave at some random time. 
%\comment{Varun: perhaps add a reference to a mobility study here}
\fi

%--------------------------------------------------------
%Uniform Results
\begin{figure*}[t]
\centering
\subfigure[]{
\includegraphics[trim=10mm 0mm 5mm 5mm, width=0.22\textwidth]{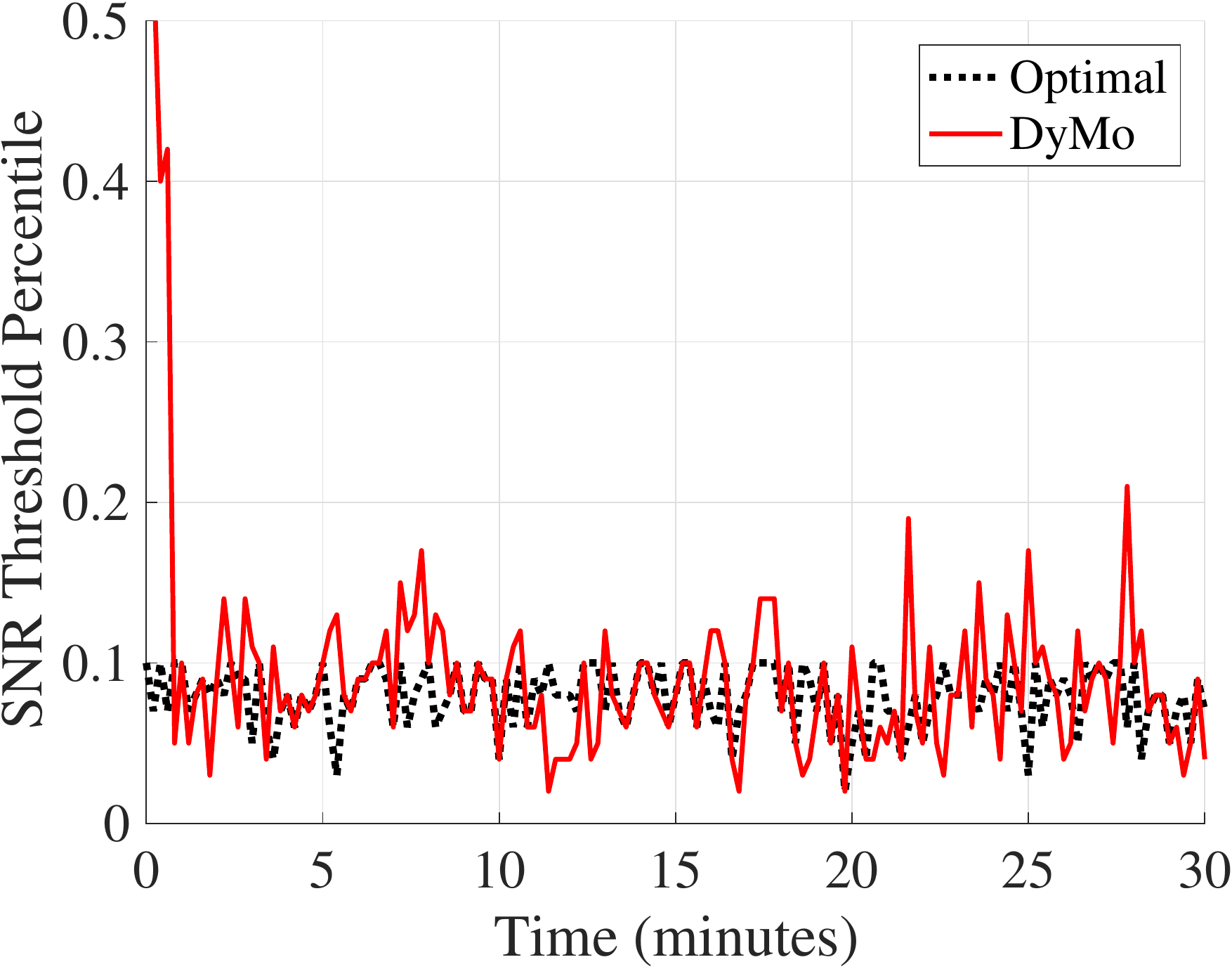}
%\caption{The actual percentile of the SNR threshold estimated by \DYMO.}
\label{FIG:UnifSNRthpDymo} %% -(b)
}
\subfigure[]{
\includegraphics[trim=10mm 0mm 5mm 5mm, width=0.22\textwidth]{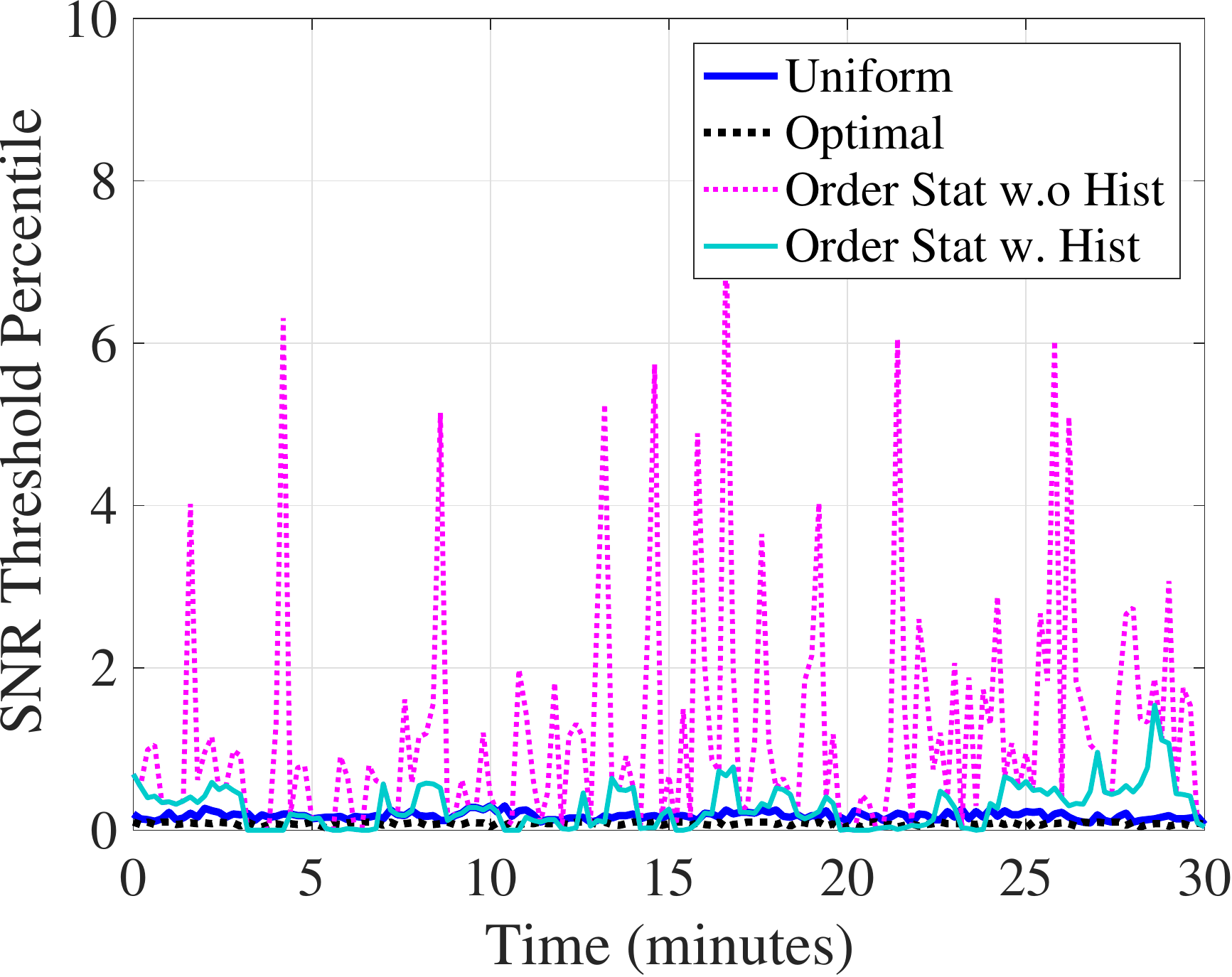}
\label{FIG:UnifSNRthpOrder} %% -(a)
%\caption{The actual percentile of the SNR threshold estimated by \ORDERSTATS.}
}
\subfigure[]{
\includegraphics[trim=10mm 0mm 5mm 5mm, width=0.22\textwidth]{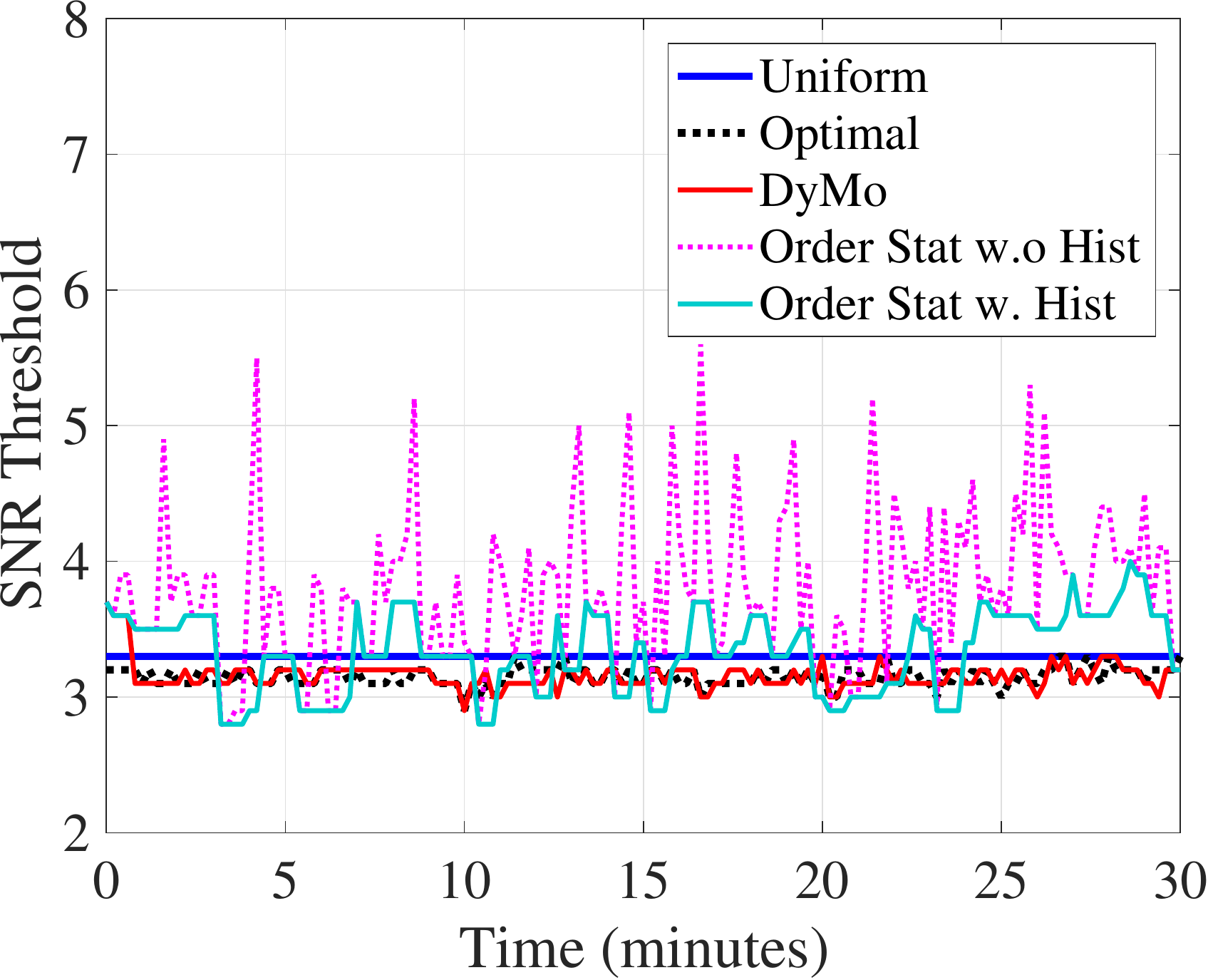}
\label{FIG:UnifSNRth} %% -(b)
%\caption{The SNR threshold estimation of \DYMO.}
}
\subfigure[]{
\includegraphics[trim=10mm 0mm 5mm 5mm, width=0.22\textwidth]{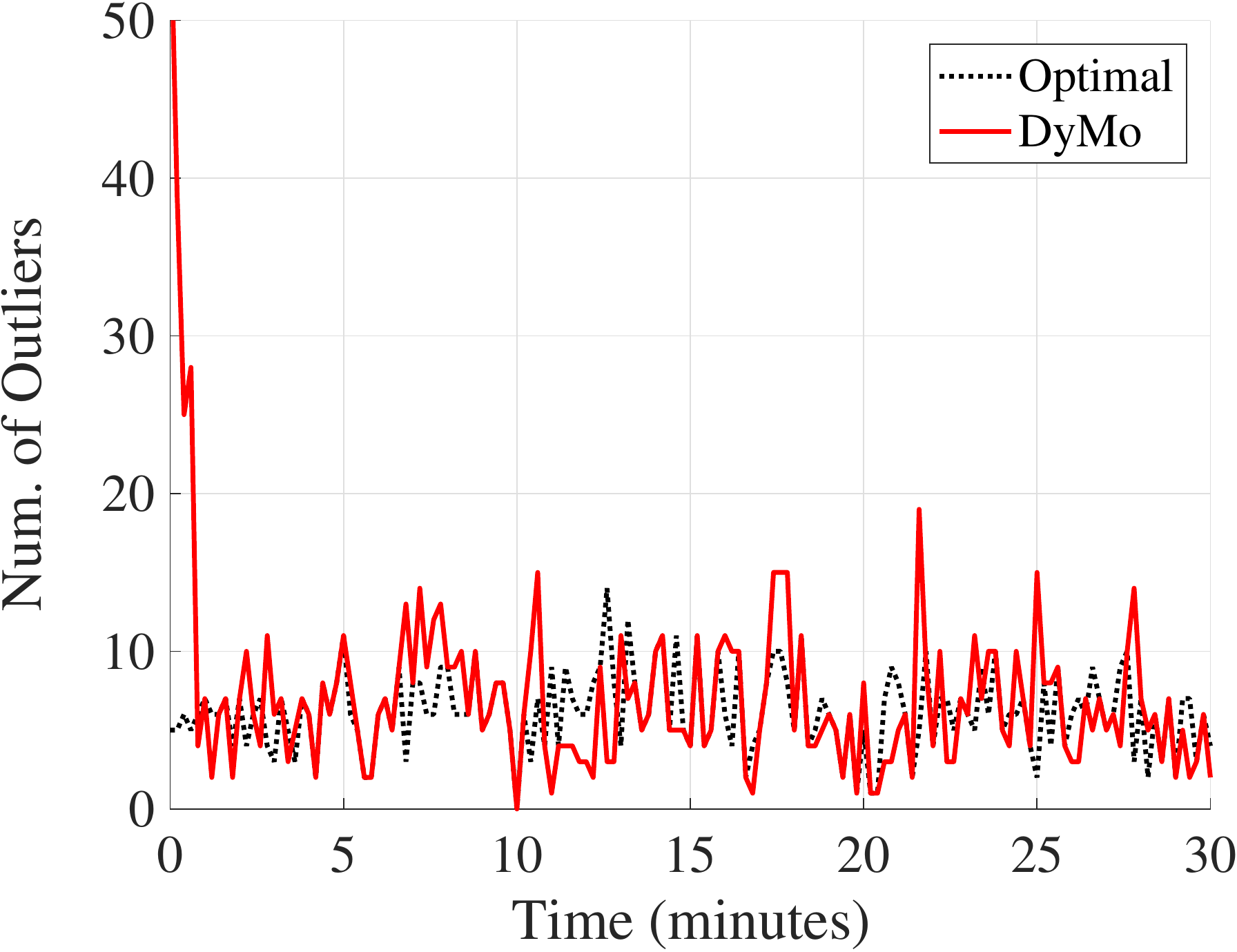}
\label{FIG:UnifOutliersDymo} %% -(c)
%\caption{The number of Outliers by using \DYMO.}
}
\\
\subfigure[]{
\includegraphics[trim=10mm 0mm 5mm 5mm, width=0.22\textwidth]{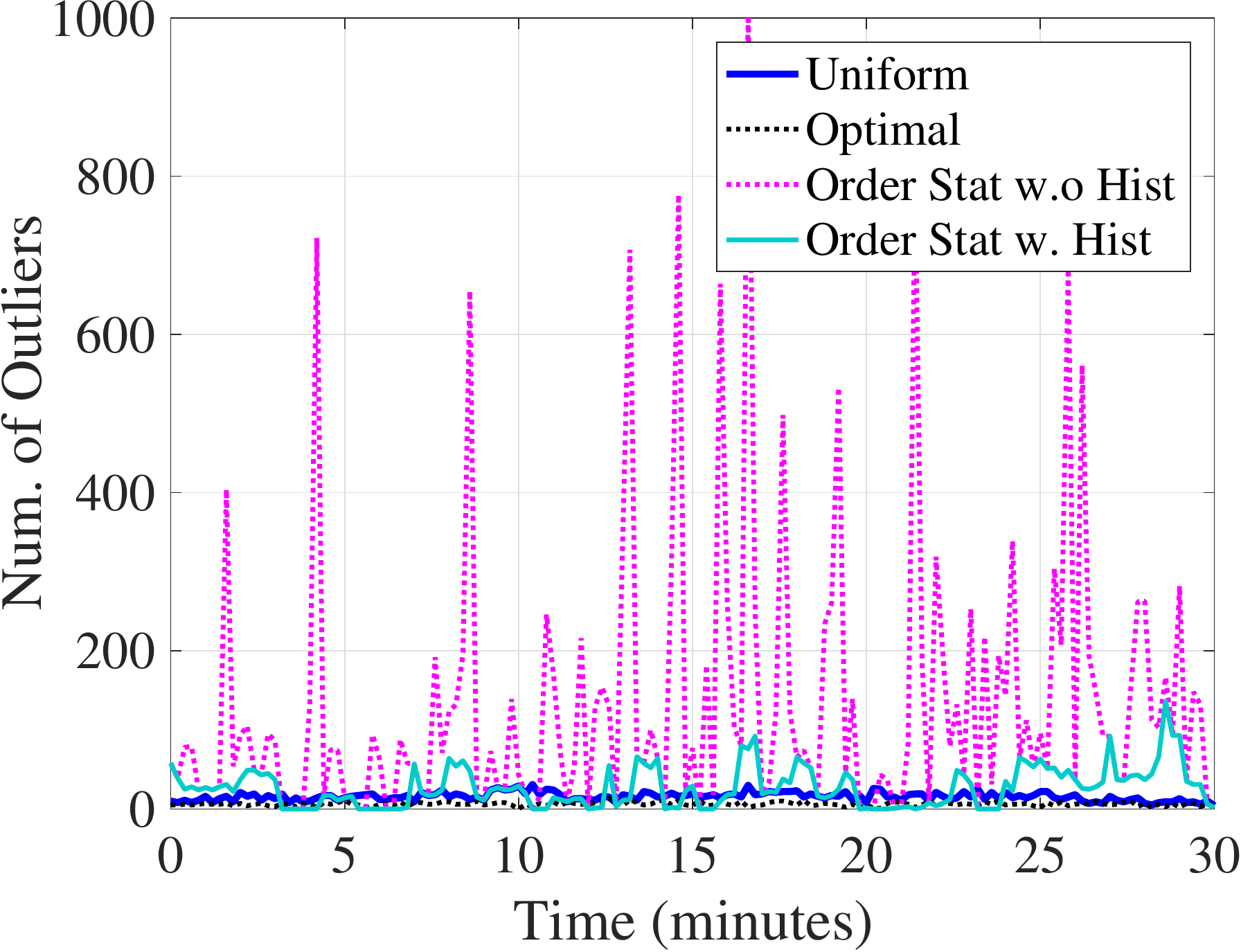}
\label{FIG:UnifOutliersOrder} %% -(d)
%\caption{The number of Outliers by using \UNI and \ORDERSTATS.}
}
\subfigure[]{
\includegraphics[trim=10mm 0mm 5mm 5mm, width=0.22\textwidth]{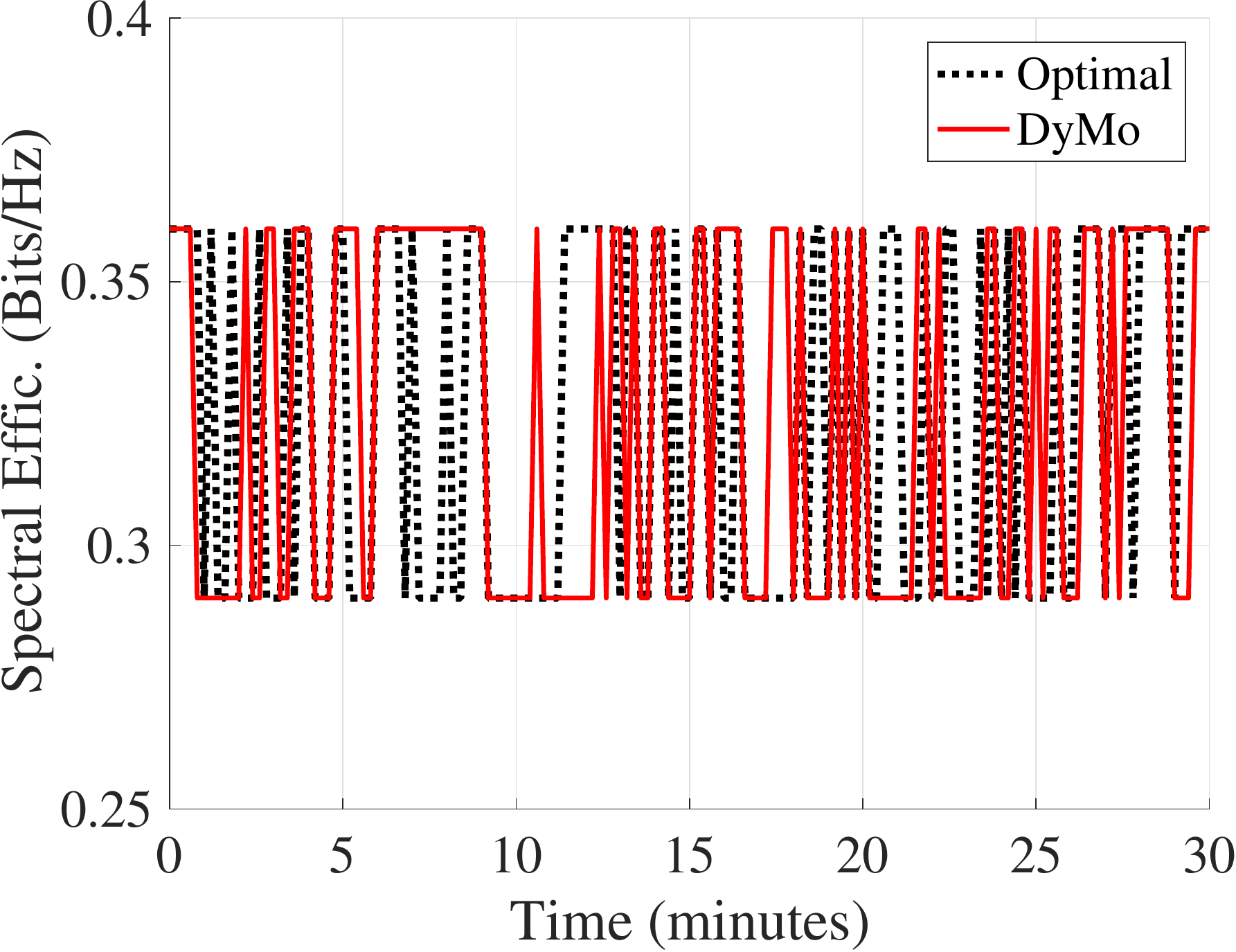}
\label{FIG:UnifSpectralEfficiencyDymo} %% -(c)
%\caption{The optimal Spectral Efficiency vs. the one achieved by \DYMO.}
}
\subfigure[]{
\includegraphics[trim=10mm 0mm 5mm 5mm, width=0.22\textwidth]{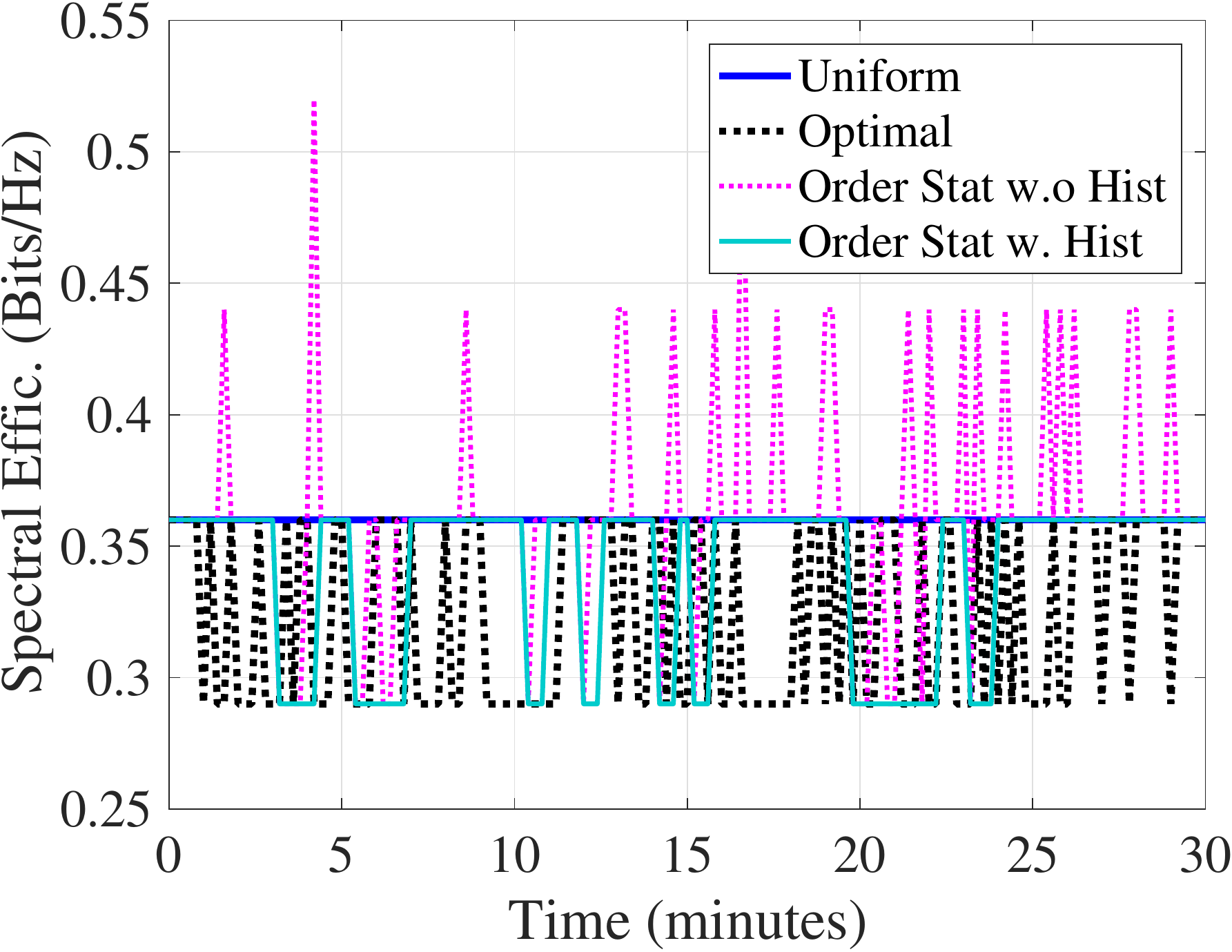}
\label{FIG:UnifSpectralEfficiencyOrder} %% -(c)
%\caption{The optimal Spectral Efficiency vs. the one achieved by \ORDERSTATS.}
}
\subfigure[]{
\includegraphics[trim=10mm 0mm 5mm 5mm, width=0.22\textwidth]{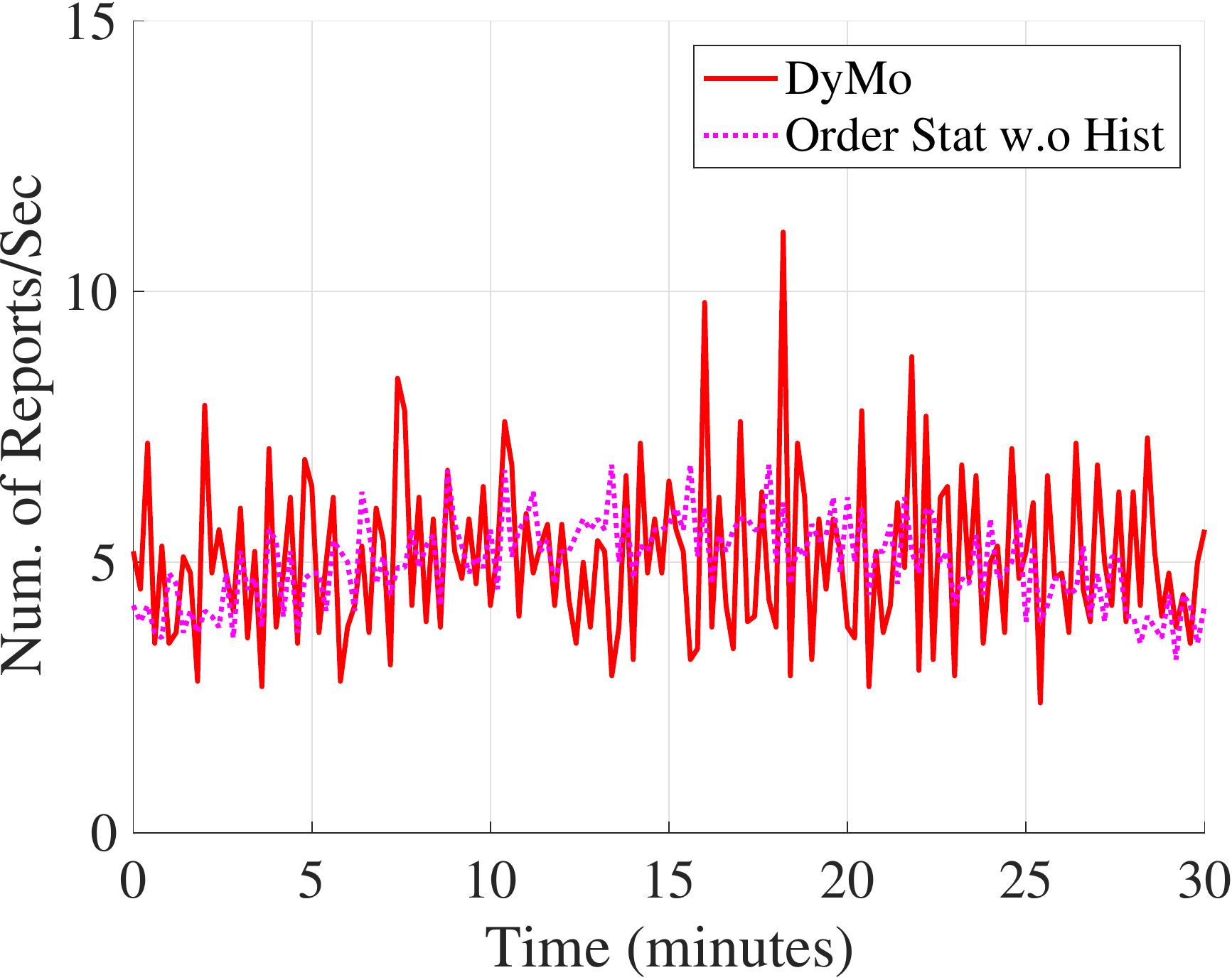}
\label{FIG:UnifOverhead} %% -(d)
%\caption{The QoS report overhead.}
}
%\vspace*{-0.5cm}
\caption[]{Simulation results from a single simulation instance lasting for $30$mins in a component \HOMOGENEOUS environment with $20,000$ UEs moving side to side between two random points, with $p=0.1$ and $r=5$ messages/sec. (a) The actual percentile of the SNR Threshold estimated by \DYMONB, (b) the actual percentile of the SNR Threshold estimated by \ORDERSTATSNB, (c) the SNR Threshold estimation, (d) spectral Efficiency of \OPT vs. \DYMONB, (e) spectral Efficiency of \OPT vs. \ORDERSTATSNB, (f) the number of Outliers by using \DYMONB, (g) the number of outliers by using \UNI and \ORDERSTATSNB, and (h) the QoS report overhead.
%,
} 
\label{FIG:vstime0}
\end{figure*}

%-------------------------------------------------------------
\begin{figure*}[t]
\centering
\subfigure[]{
\includegraphics[trim=10mm 0mm 5mm 5mm, width=0.22\textwidth]{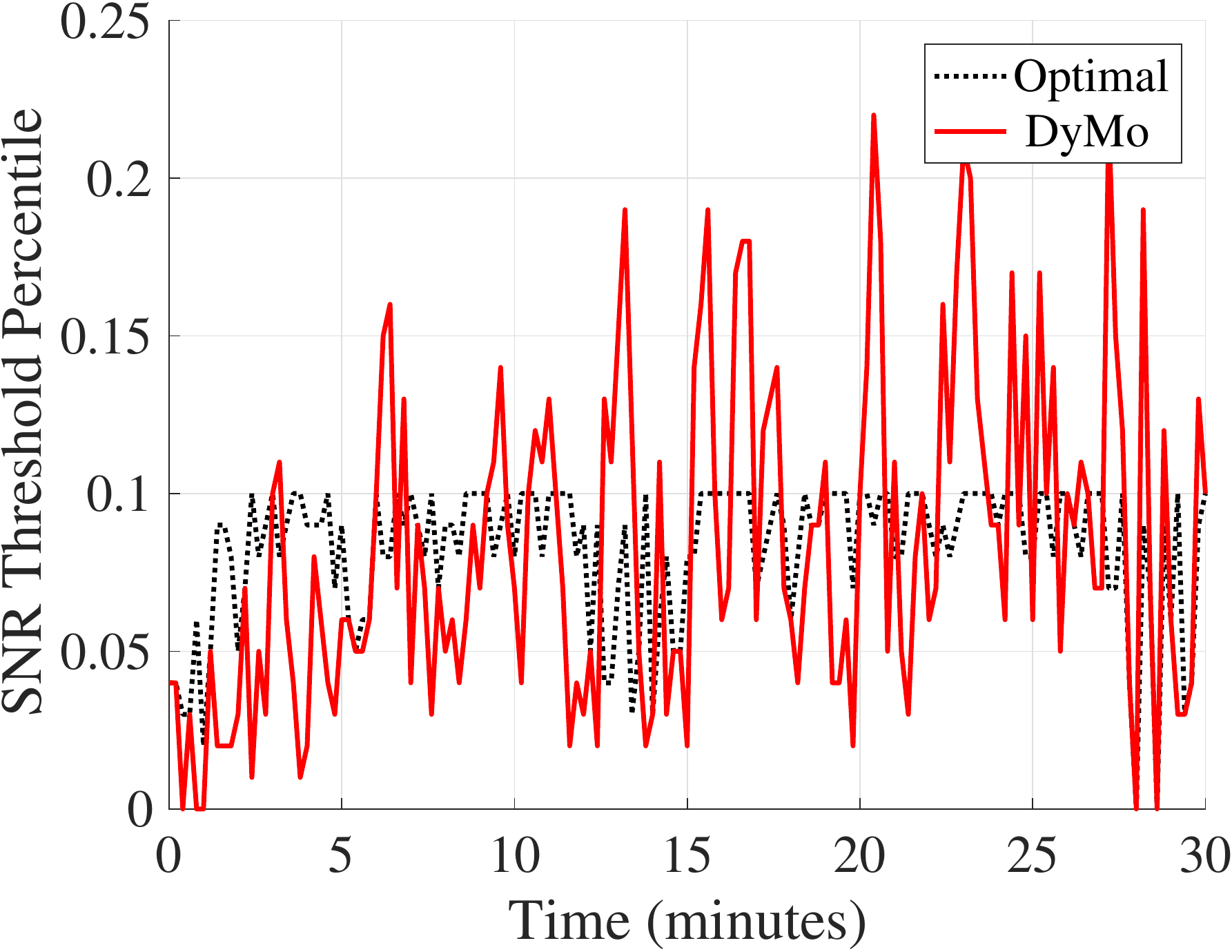}
%\caption{The actual percentile of the SNR threshold estimated by \DYMO.}
\label{FIG:SNRthpDymo} %% -(b)
}
\subfigure[]{
\includegraphics[trim=10mm 0mm 5mm 5mm, width=0.22\textwidth]{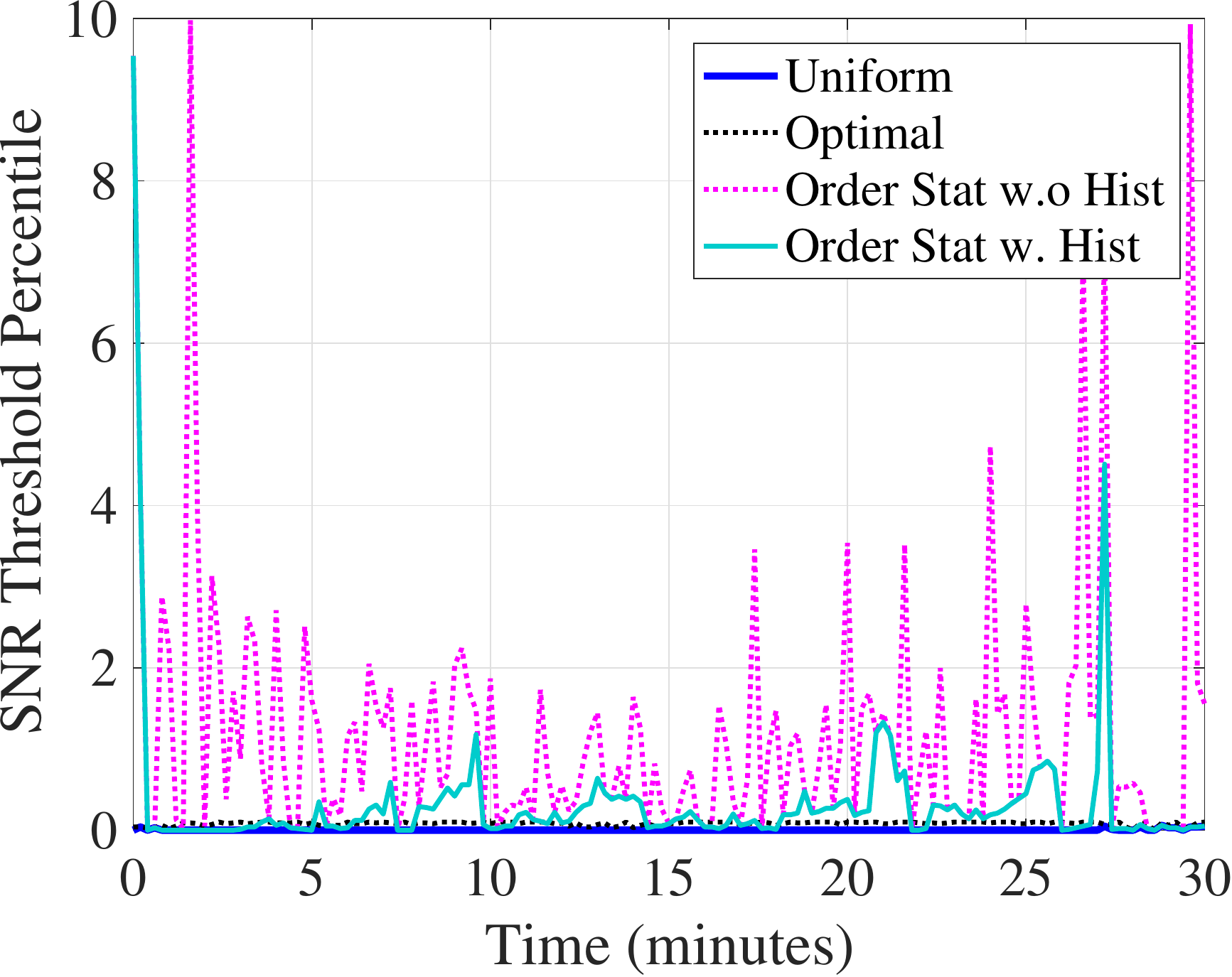}
\label{FIG:SNRthpOrder} %% -(a)
%\caption{The actual percentile of the SNR threshold estimated by \ORDERSTATS.}
}
\subfigure[]{
\includegraphics[trim=10mm 0mm 5mm 5mm, width=0.22\textwidth]{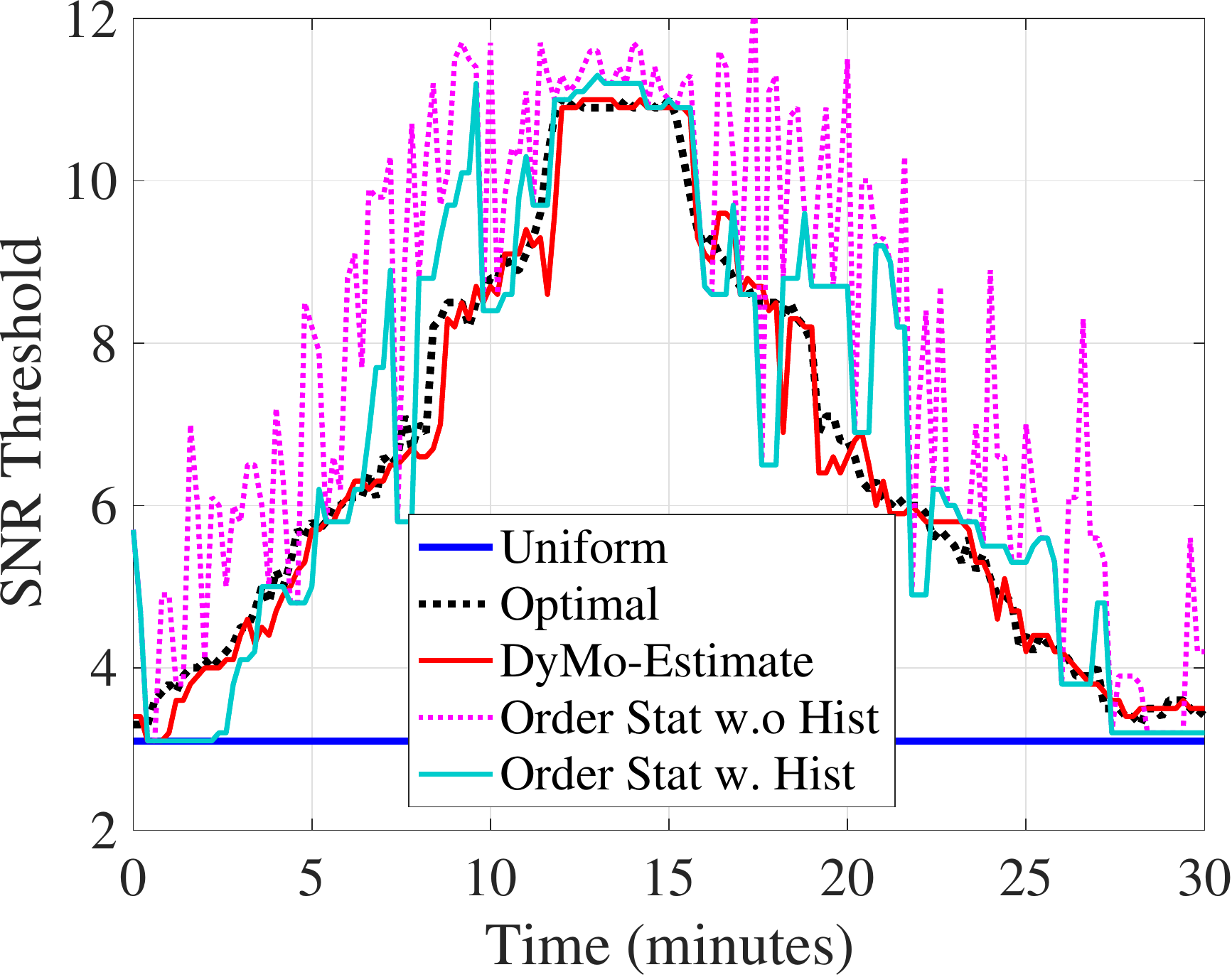}
\label{FIG:SNRth} %% -(b)
%\caption{The SNR threshold estimation of \DYMO.}
}
\subfigure[]{
\includegraphics[trim=10mm 0mm 5mm 5mm, width=0.22\textwidth]{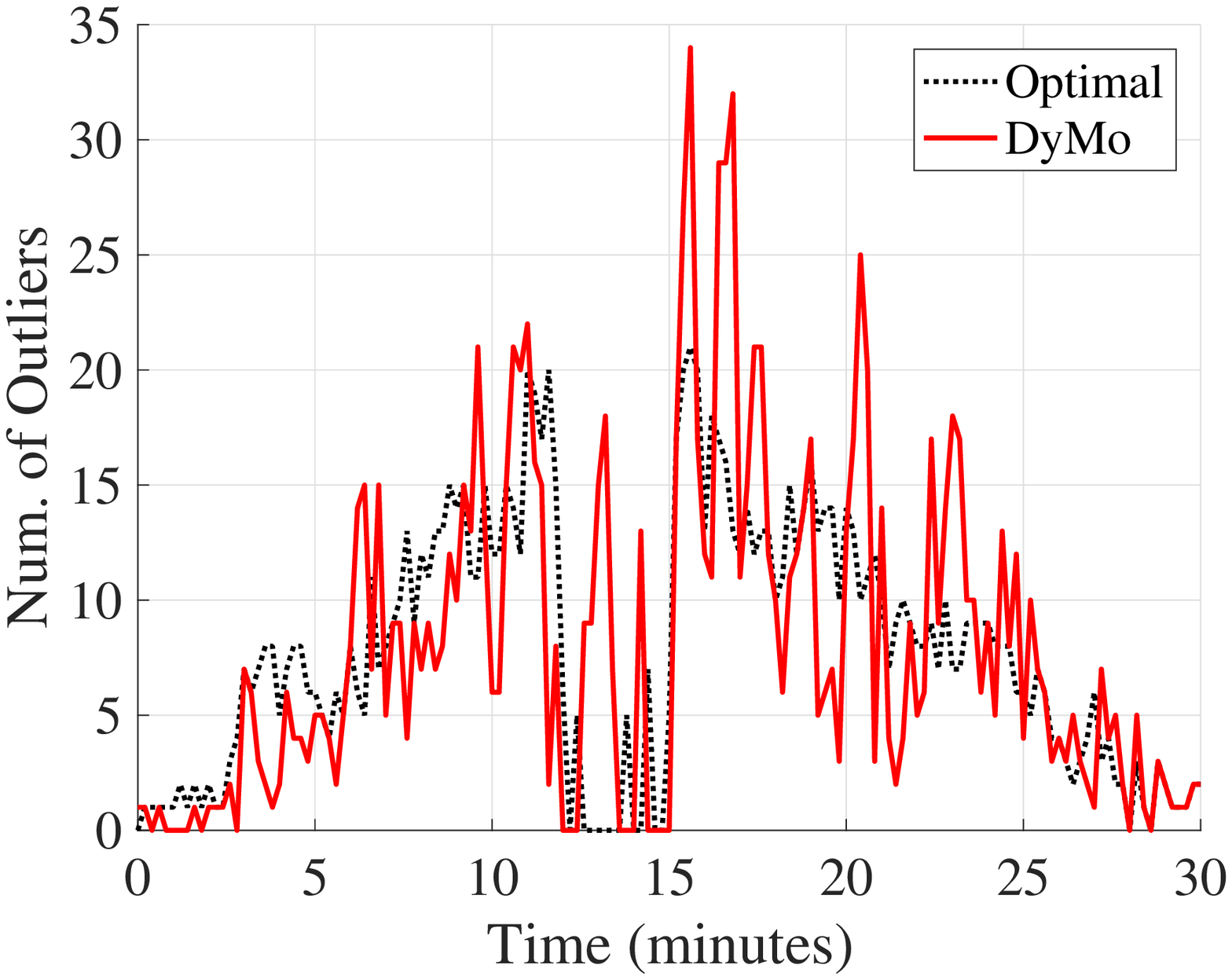}
\label{FIG:OutliersDymo} %% -(c)
%\caption{The number of Outliers by using \DYMO.}
}
\\
\subfigure[]{
\includegraphics[trim=10mm 0mm 3mm 5mm, width=0.22\textwidth]{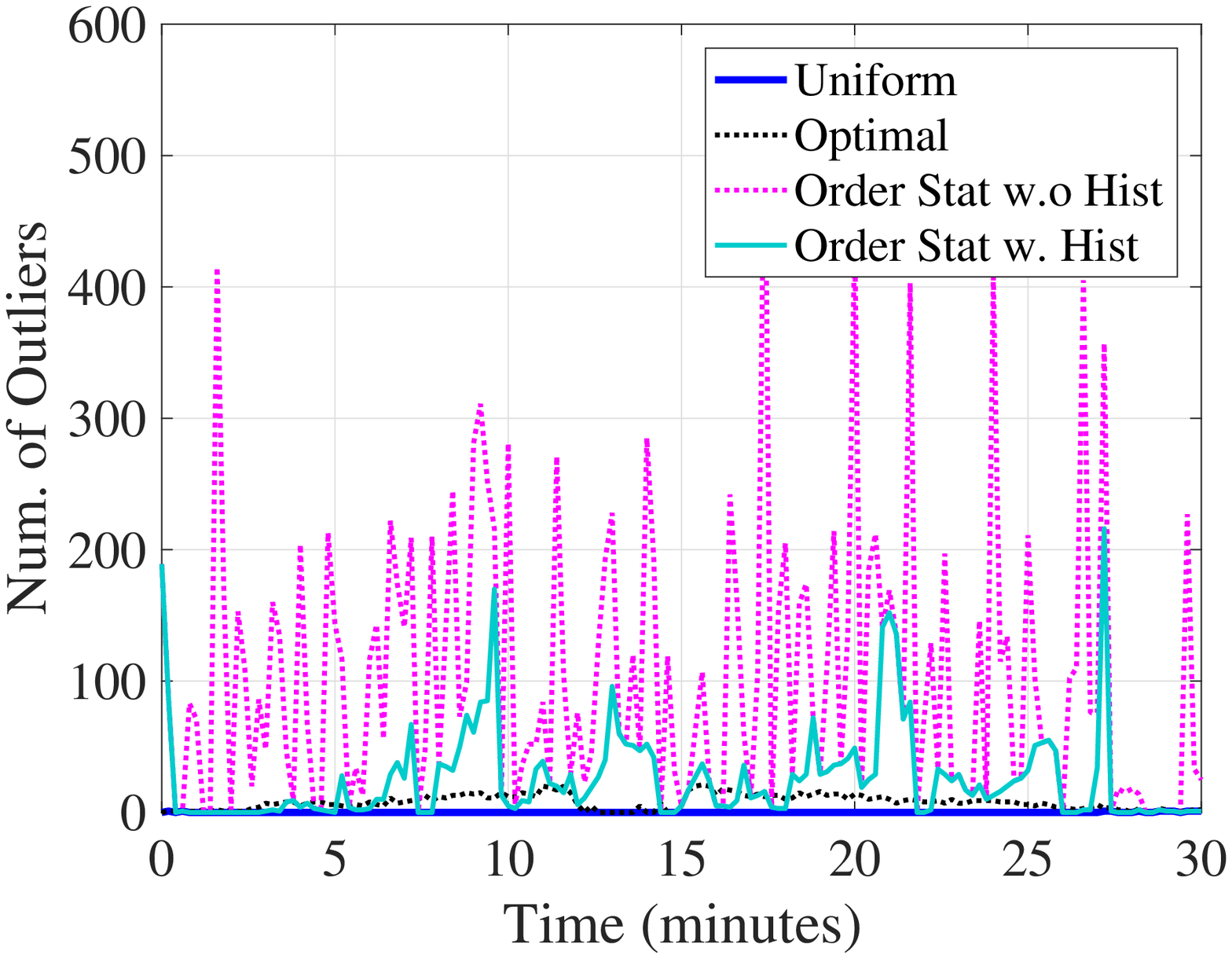}
\label{FIG:OutliersOrder} %% -(d)
%\caption{The number of Outliers by using \UNI and \ORDERSTATS.}
}
\subfigure[]{
\includegraphics[trim=10mm 0mm 5mm 5mm, width=0.22\textwidth]{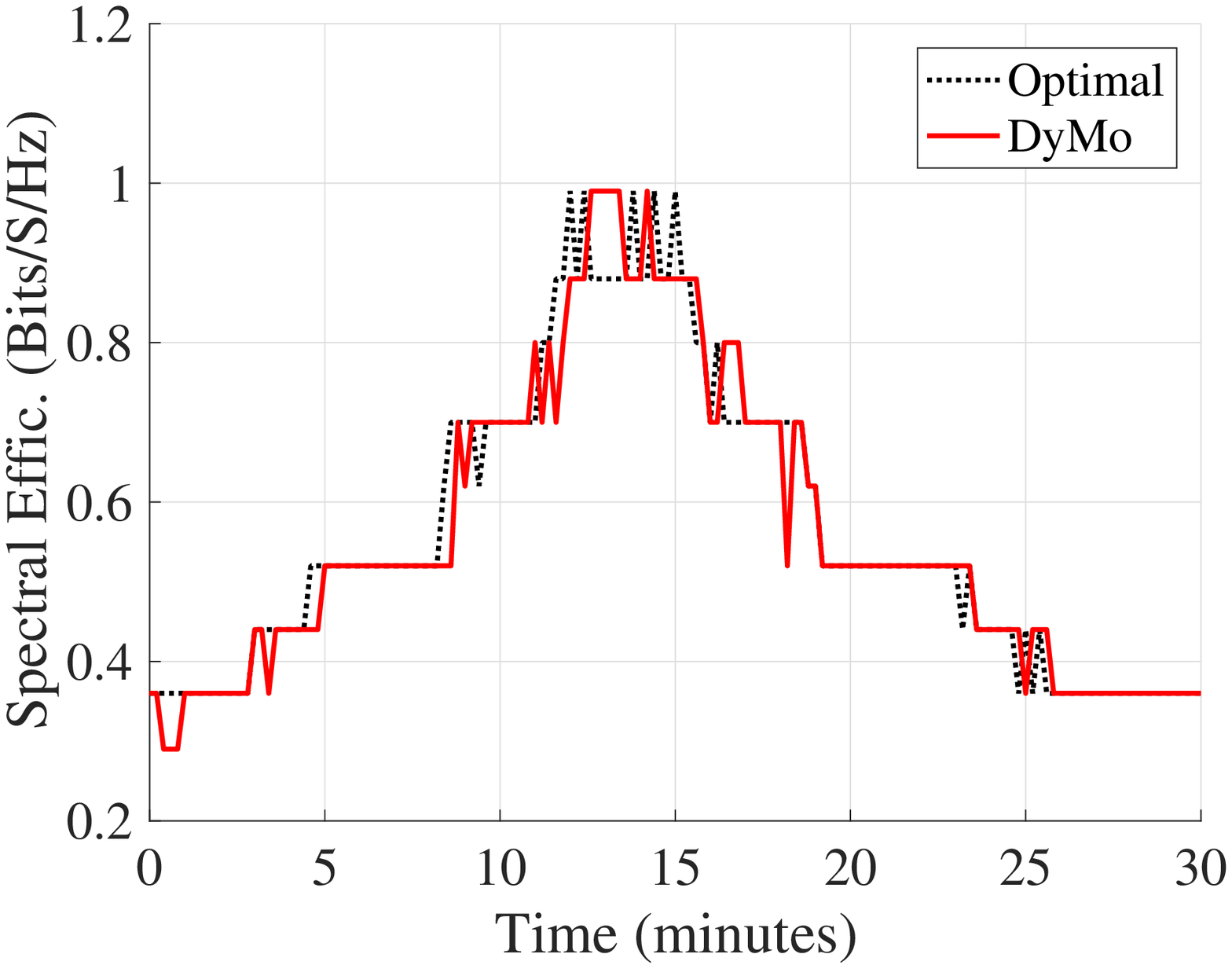}
\label{FIG:SpectralEfficiencyDymo} %% -(c)
%\caption{The optimal Spectral Efficiency vs. the  one achieved by \DYMO.}
}
\subfigure[]{
\includegraphics[trim=10mm 0mm 5mm 5mm, width=0.22\textwidth]{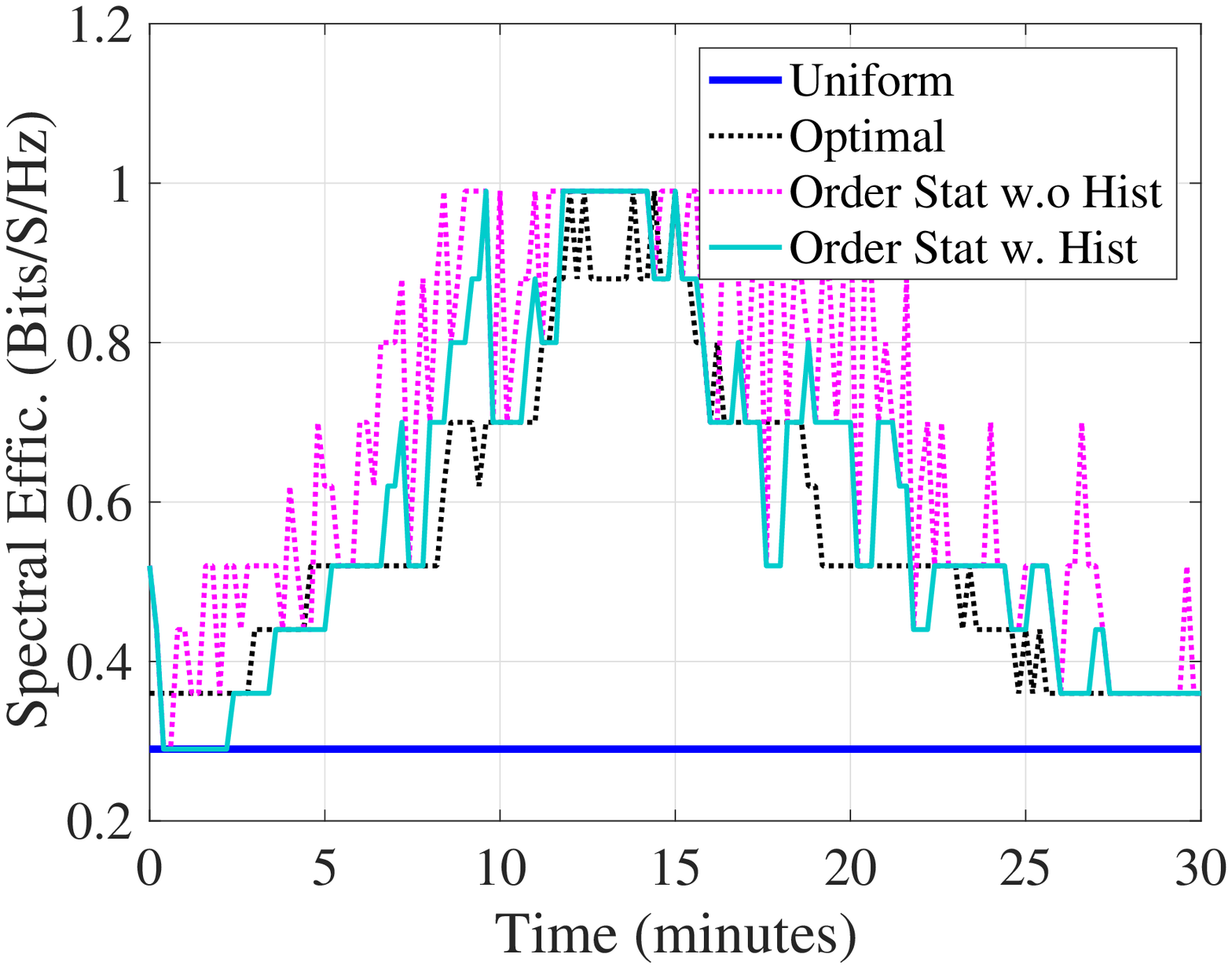}
\label{FIG:SpectralEfficiencyOrder} %% -(c)
%\caption{The optimal Spectral Efficiency vs. the one achieved by \ORDERSTATS.}
}
\subfigure[]{
\includegraphics[trim=10mm 0mm 5mm 5mm, width=0.22\textwidth]{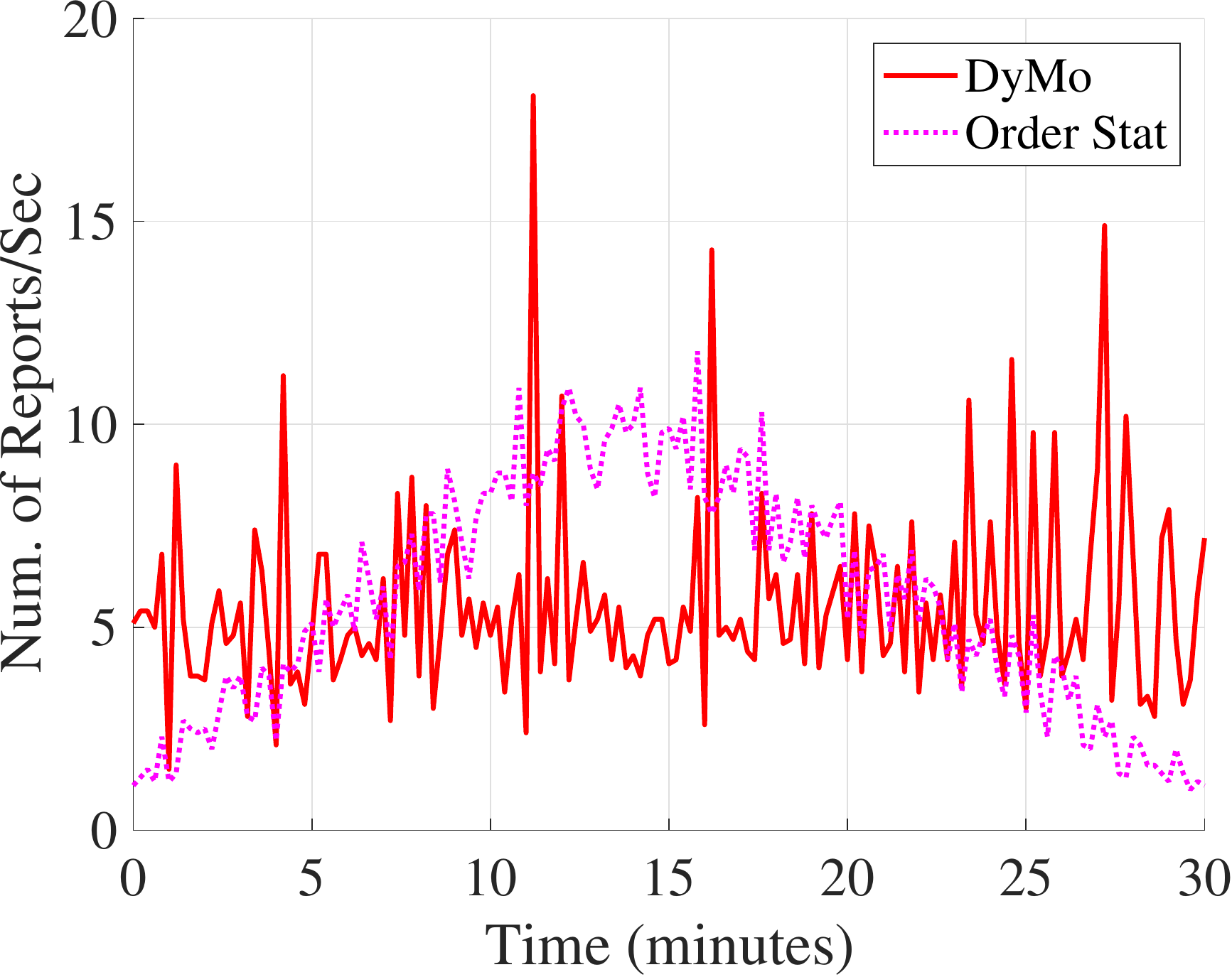}
\label{FIG:Overhead} %% -(d)
%\caption{The QoS report overhead.}
}
\caption[]{Simulation results from a single simulation instance lasting for $30$mins in a stadium environment with $20,000$ UEs moving from the edges to the center and back, with $p=0.1$ and $r=5$ messages/sec. (a) The actual percentile of the SNR Threshold estimated by \DYMONB, (b) the actual percentile of the SNR Threshold estimated by \ORDERSTATSNB, (c) the SNR Threshold estimation, (d) spectral efficiency of \OPT vs. \DYMONB, (e) spectral efficiency of \OPT vs.  \ORDERSTATSNB, (f) the number of Outliers by using \DYMONB, (g) the number of Outliers by using \UNI and \ORDERSTATSNB, and (h) the QoS report overhead.
%, 
} 
\label{FIG:vstime1}
\end{figure*}

%----------------------------------------------------------------
\begin{figure*}[t]
\centering
\subfigure[]{
\includegraphics[trim=10mm 0mm 5mm 5mm, width=0.22\textwidth]{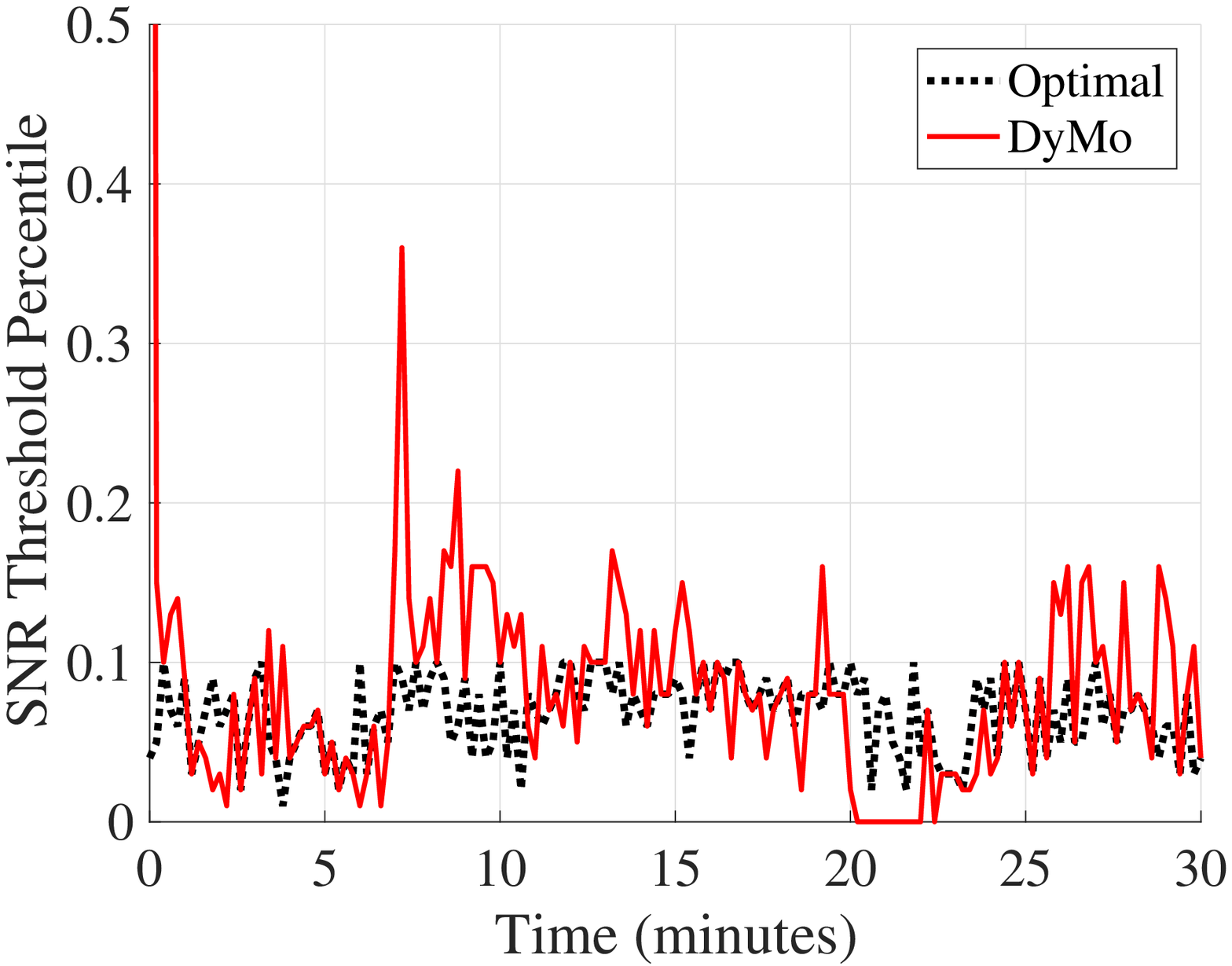}
%\caption{The actual percentile of the SNR threshold estimated by \DYMO.}
\label{FIG:FailSNRthpDymo} %% -(b)
}
\subfigure[]{
\includegraphics[trim=10mm 0mm 5mm 5mm, width=0.22\textwidth]{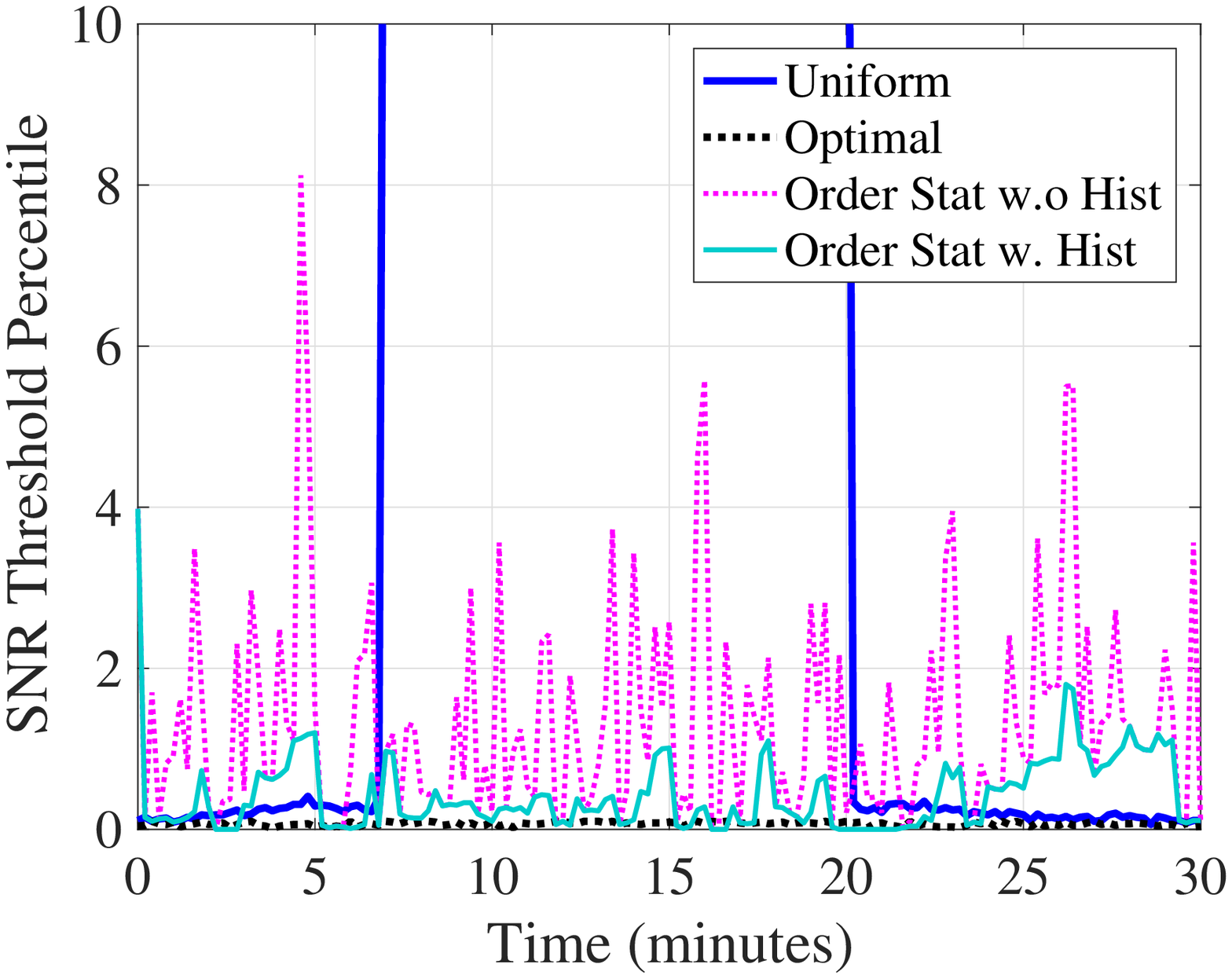}
\label{FIG:FailSNRthpOrder} %% -(a)
%\caption{The actual percentile of the SNR threshold estimated by \ORDERSTATS.}
}
\subfigure[]{
\includegraphics[trim=10mm 0mm 5mm 5mm, width=0.22\textwidth]{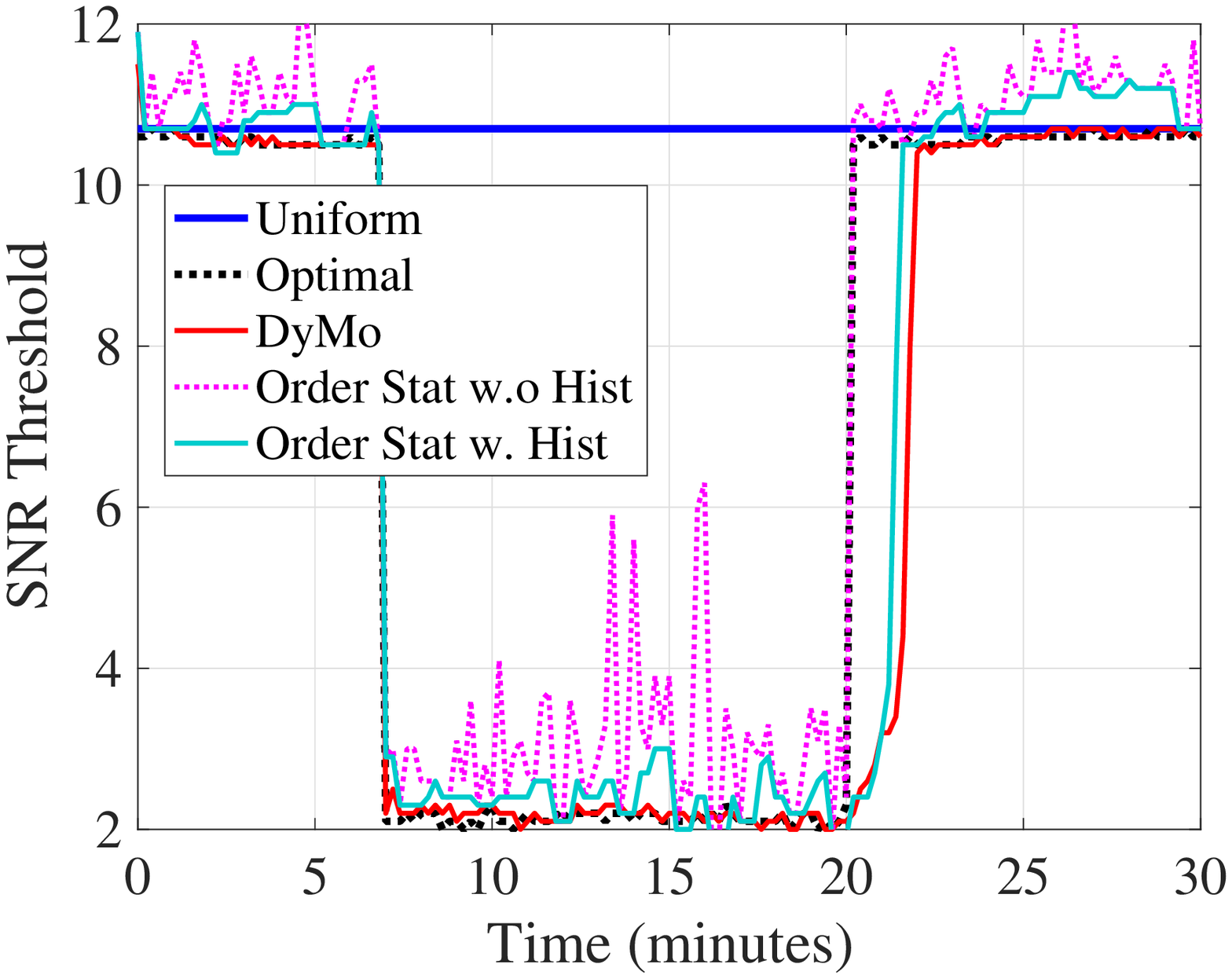}
\label{FIG:FailSNRth} %% -(b)
%\caption{The SNR threshold estimation of \DYMO.}
}
\subfigure[]{
\includegraphics[trim=10mm 0mm 5mm 5mm, width=0.22\textwidth]{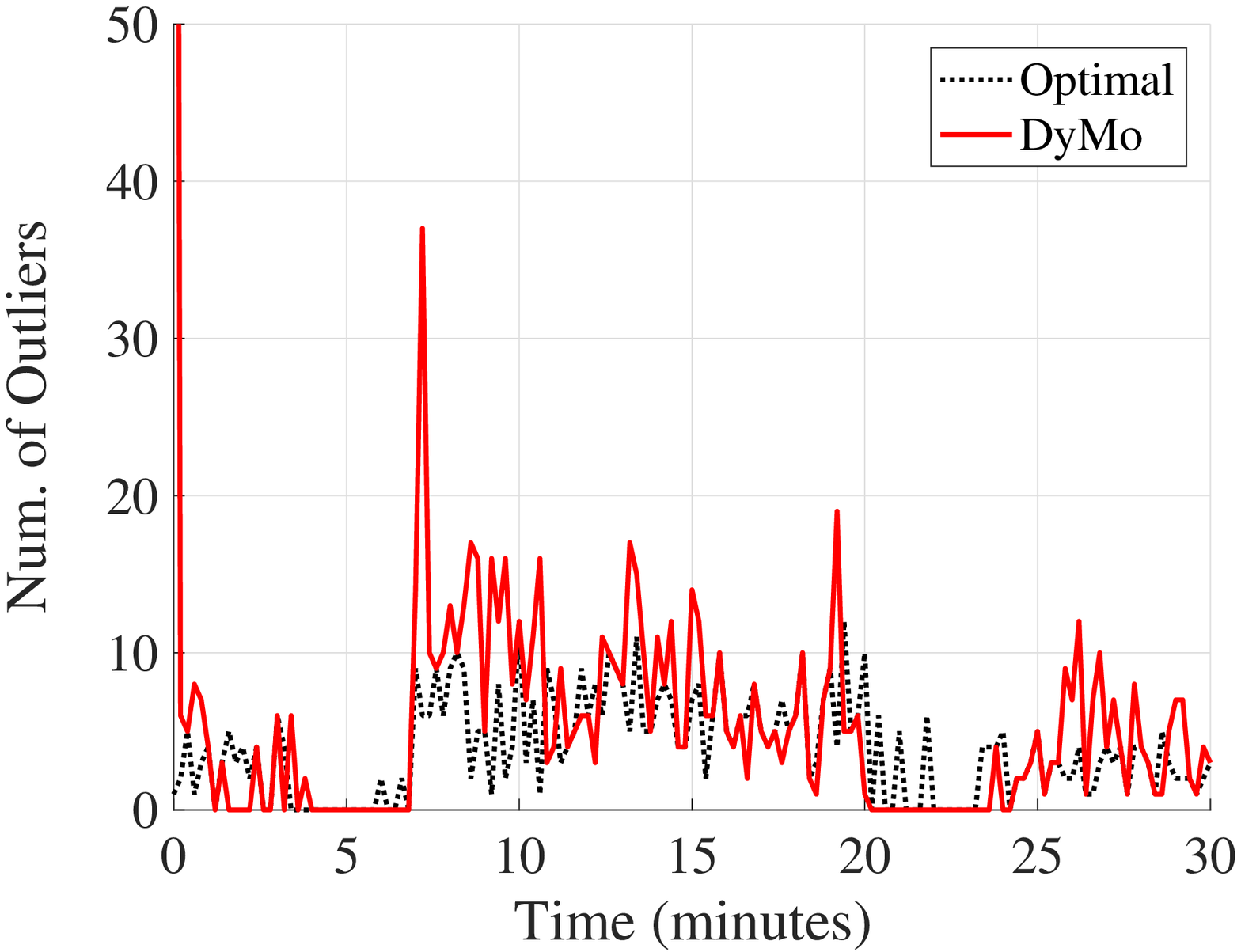}
\label{FIG:FailOutliersDymo} %% -(c)
%\caption{The number of Outliers by using \DYMO.}
}
\\
\subfigure[]{
\includegraphics[trim=10mm 0mm 5mm 5mm, width=0.22\textwidth]{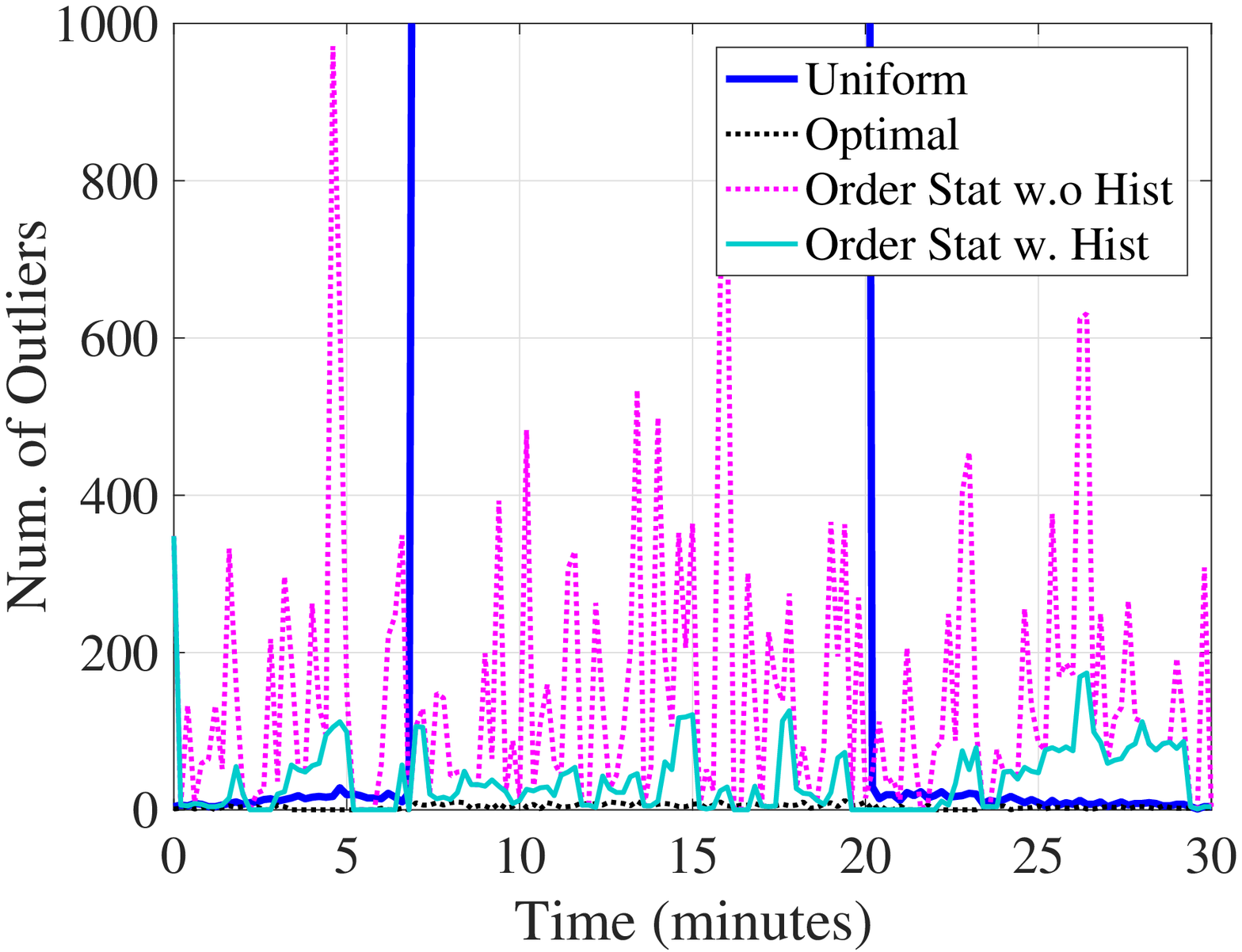}
\label{FIG:FailOutliersOrder} %% -(d)
%\caption{The number of Outliers by using \UNI and \ORDERSTATS.}
}
\subfigure[]{
\includegraphics[trim=10mm 0mm 5mm 5mm, width=0.22\textwidth]{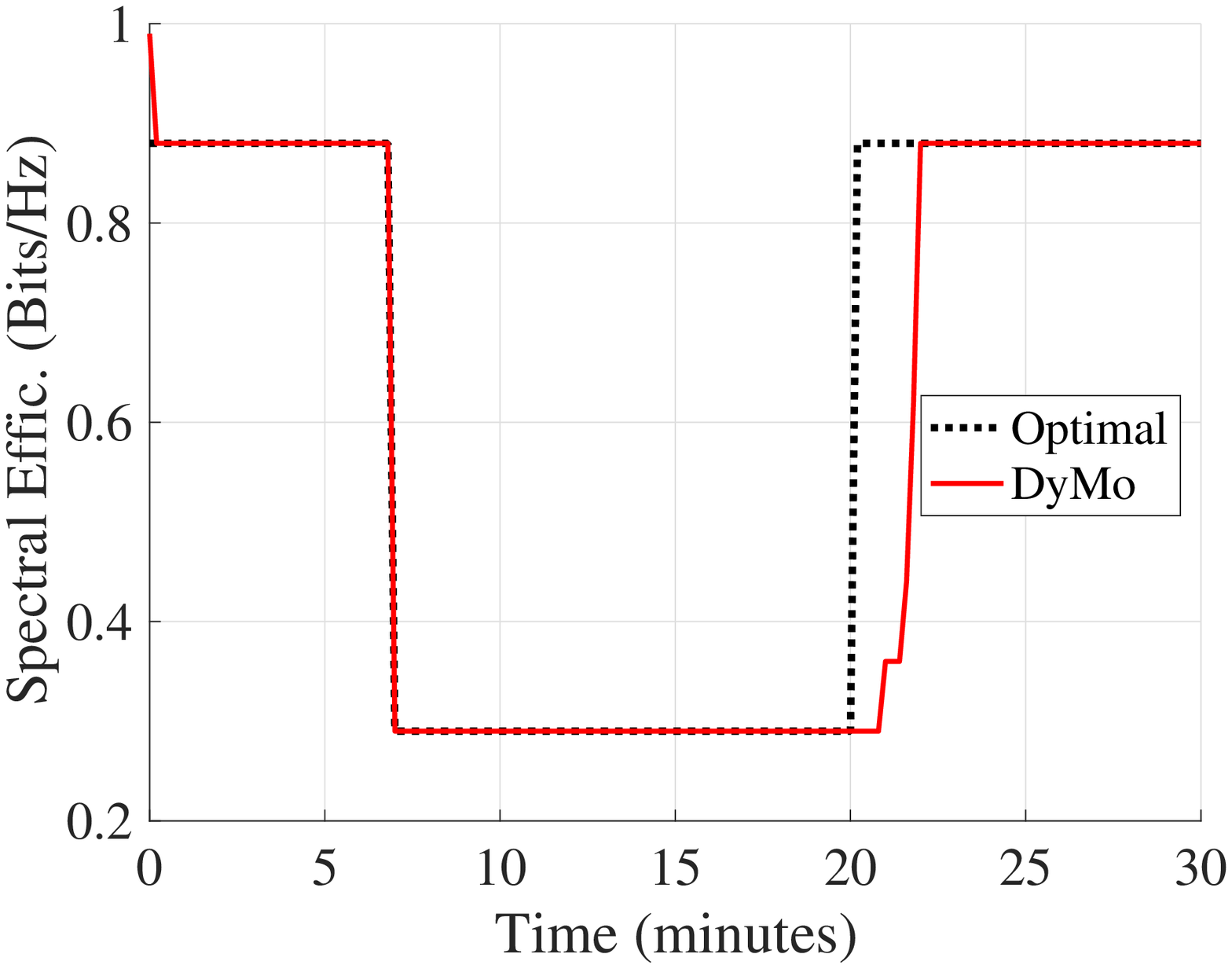}
\label{FIG:FailSpectralEfficiencyDymo} %% -(c)
%\caption{The optimal Spectral Efficiency vs. the  one achieved by \DYMO.}
}
\subfigure[]{
\includegraphics[trim=10mm 0mm 5mm 5mm, width=0.22\textwidth]{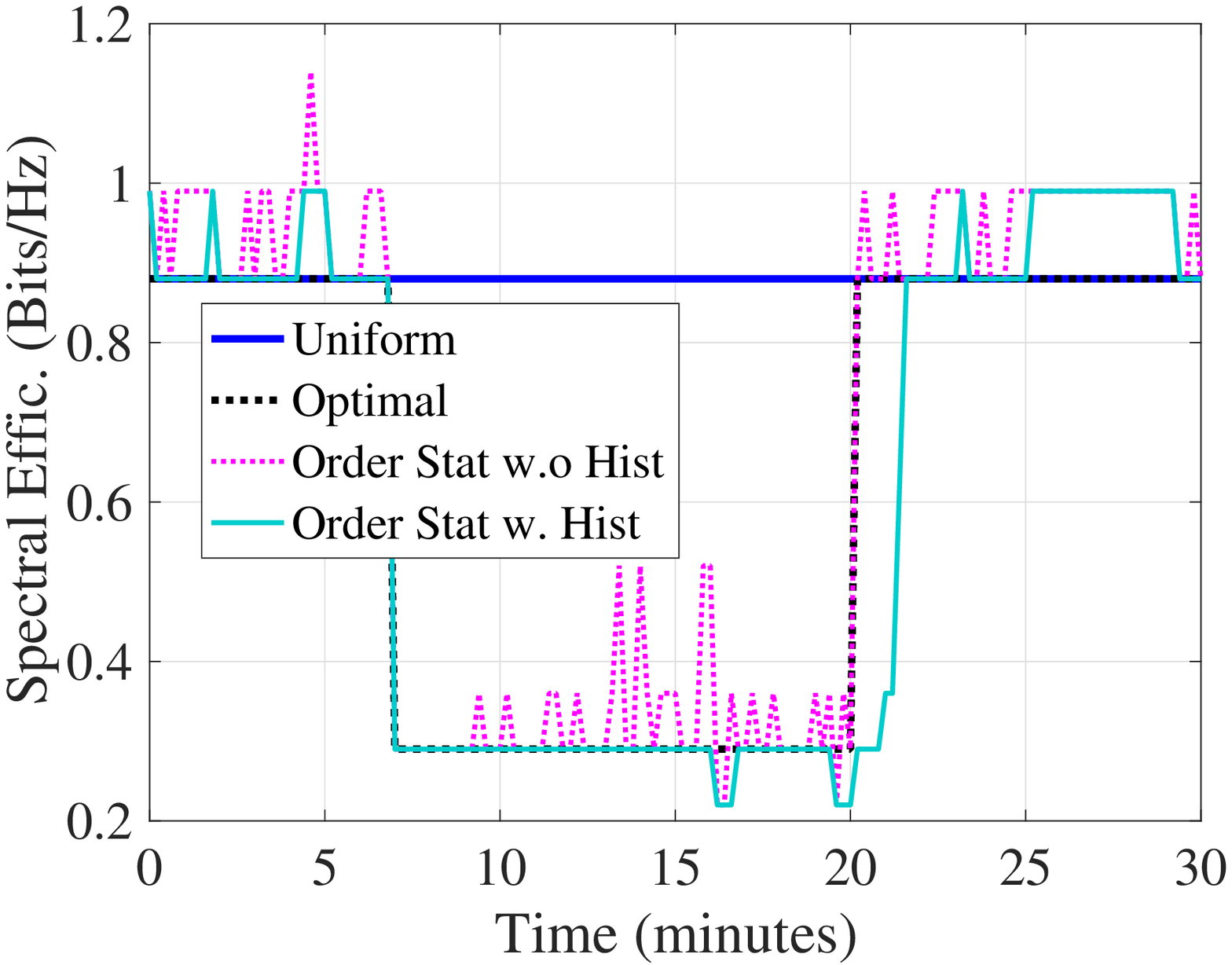}
\label{FIG:FailSpectralEfficiencyOrder} %% -(c)
%\caption{The optimal Spectral Efficiency vs. the one achieved by \ORDERSTATS.}
}
\subfigure[]{
\includegraphics[trim=10mm 0mm 5mm 5mm, width=0.22\textwidth]{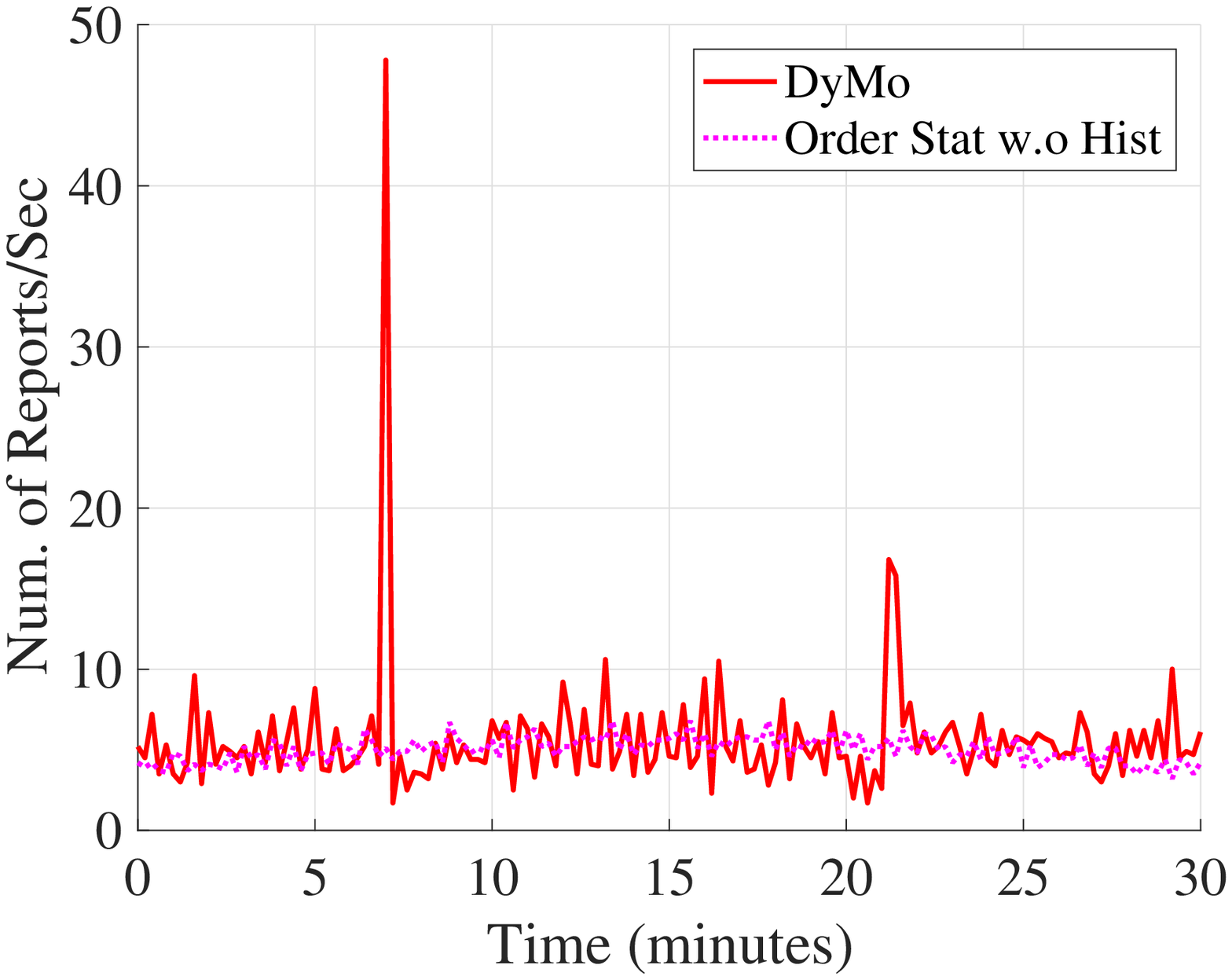}
\label{FIG:FailOverhead} %% -(d)
%\caption{The QoS report overhead.}
}
%\vspace*{-0.5cm}
\caption[]{Simulation results from a single simulation instance lasting for $30$mins in a component failure environment with $20,000$ UEs moving side to side between two random points, with $p=0.1$ and $r=5$ messages/sec. (a) The actual percentile of the SNR Threshold estimated by \DYMONB, (b) the actual percentile of the SNR Threshold estimated by \ORDERSTATSNB, (c) the SNR Threshold estimation, (d) spectral Efficiency of \OPT vs. \DYMONB, (e) spectral Efficiency of \OPT vs. \ORDERSTATSNB, (f) the number of Outliers by using \DYMONB, (g) the number of outliers by using \UNI and \ORDERSTATSNB, and (h) the QoS report overhead.
%,
} 
\label{FIG:vstime2}
\end{figure*}

%----------------------------------------------------------------
\begin{figure*}[t]
\centering
\subfigure[]{
\includegraphics[trim=10mm 0mm 3mm 5mm, width=0.27\textwidth]{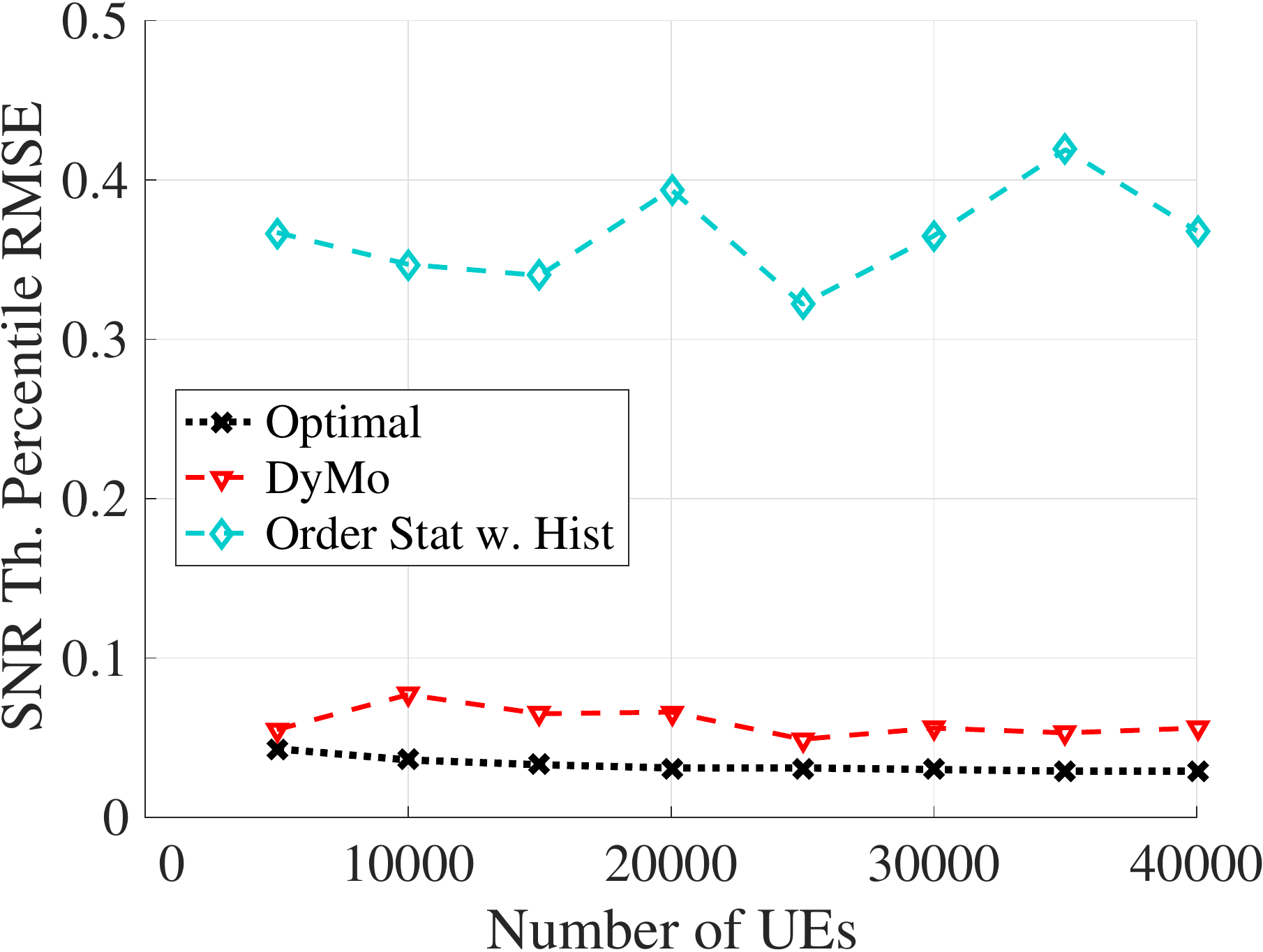}
\label{FIG:UnifSNRthpvsUE} %% -(a)
}%
\subfigure[]{
\includegraphics[trim=5mm 0mm 3mm 5mm, width=0.27\textwidth]{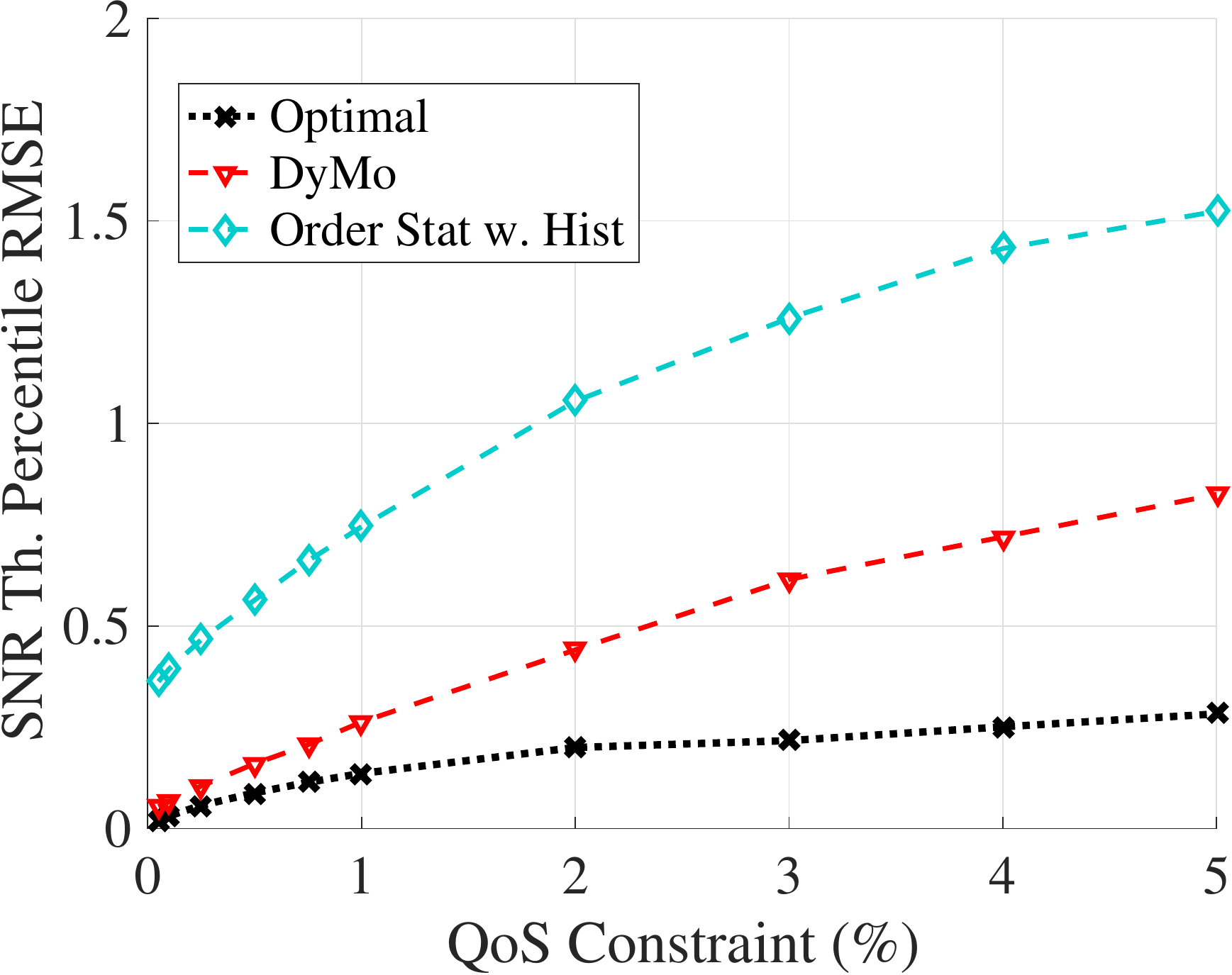}
\label{FIG:UnifSNRthpvsP} %% -(a)
}%
\subfigure[]{
\includegraphics[trim=5mm 0mm 3mm 5mm, width=0.27\textwidth]{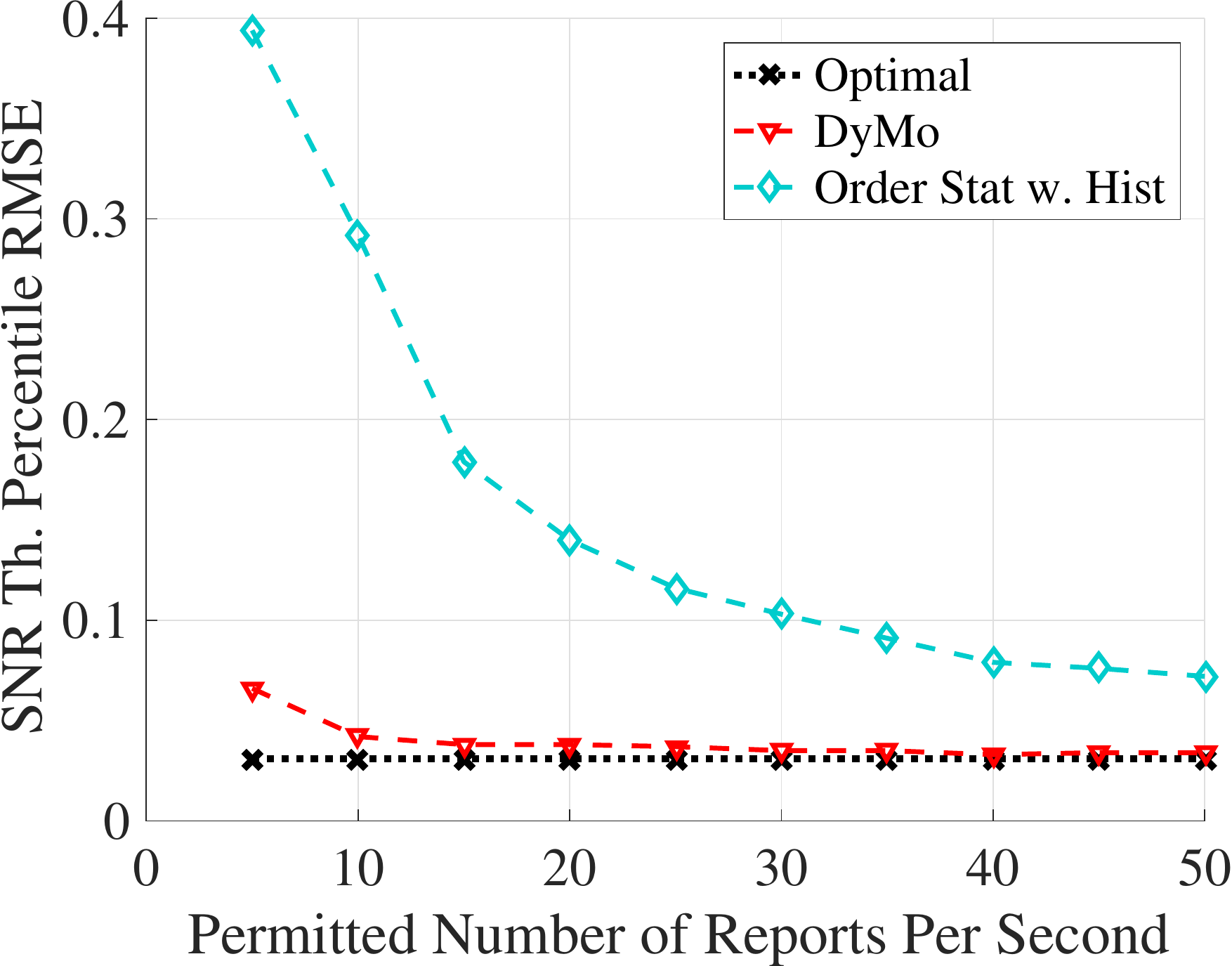}
\label{FIG:UnifSNRthpvsReports} %% -(b)
}%
\\
\subfigure[]{
\includegraphics[trim=10mm 0mm 3mm 5mm, width=0.27\textwidth]{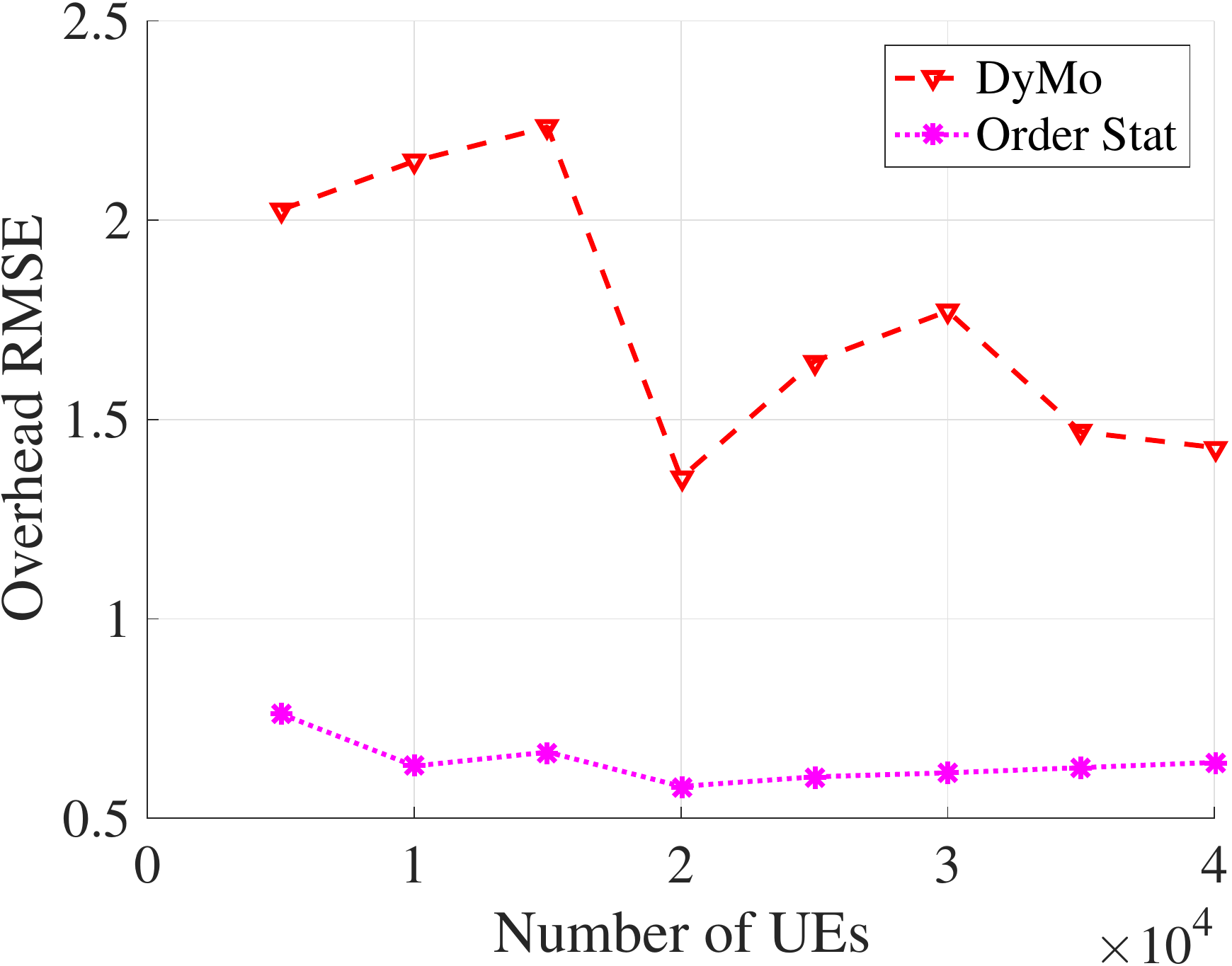}
\label{FIG:UnifOverheadvsUE} %% -(b)
}%
\subfigure[]{
\includegraphics[trim=5mm 0mm 3mm 5mm, width=0.27\textwidth]{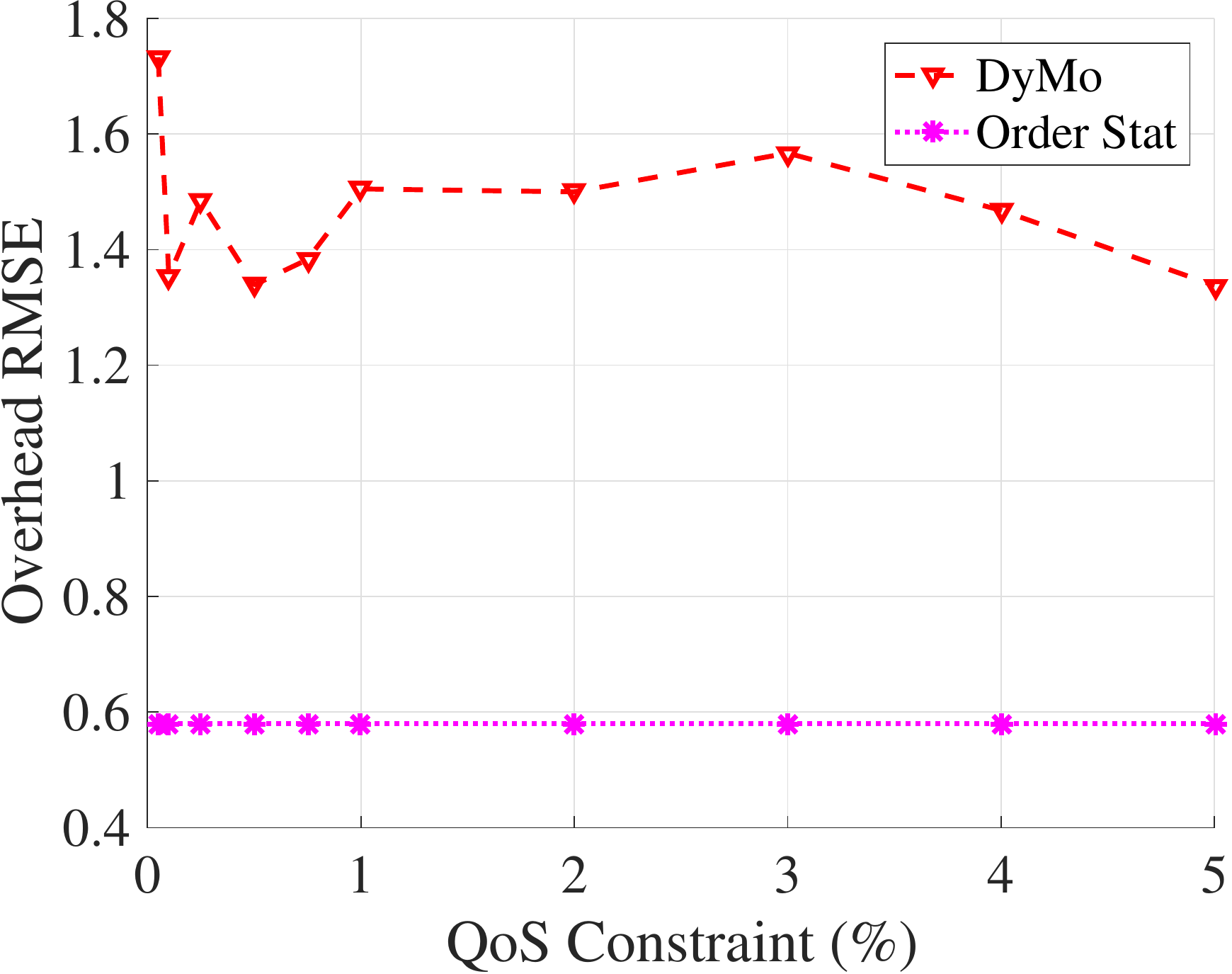}
\label{FIG:UnifOverheadvsp} %% -(b)
}%
\subfigure[]{
\includegraphics[trim=5mm 0mm 3mm 5mm, width=0.27\textwidth]{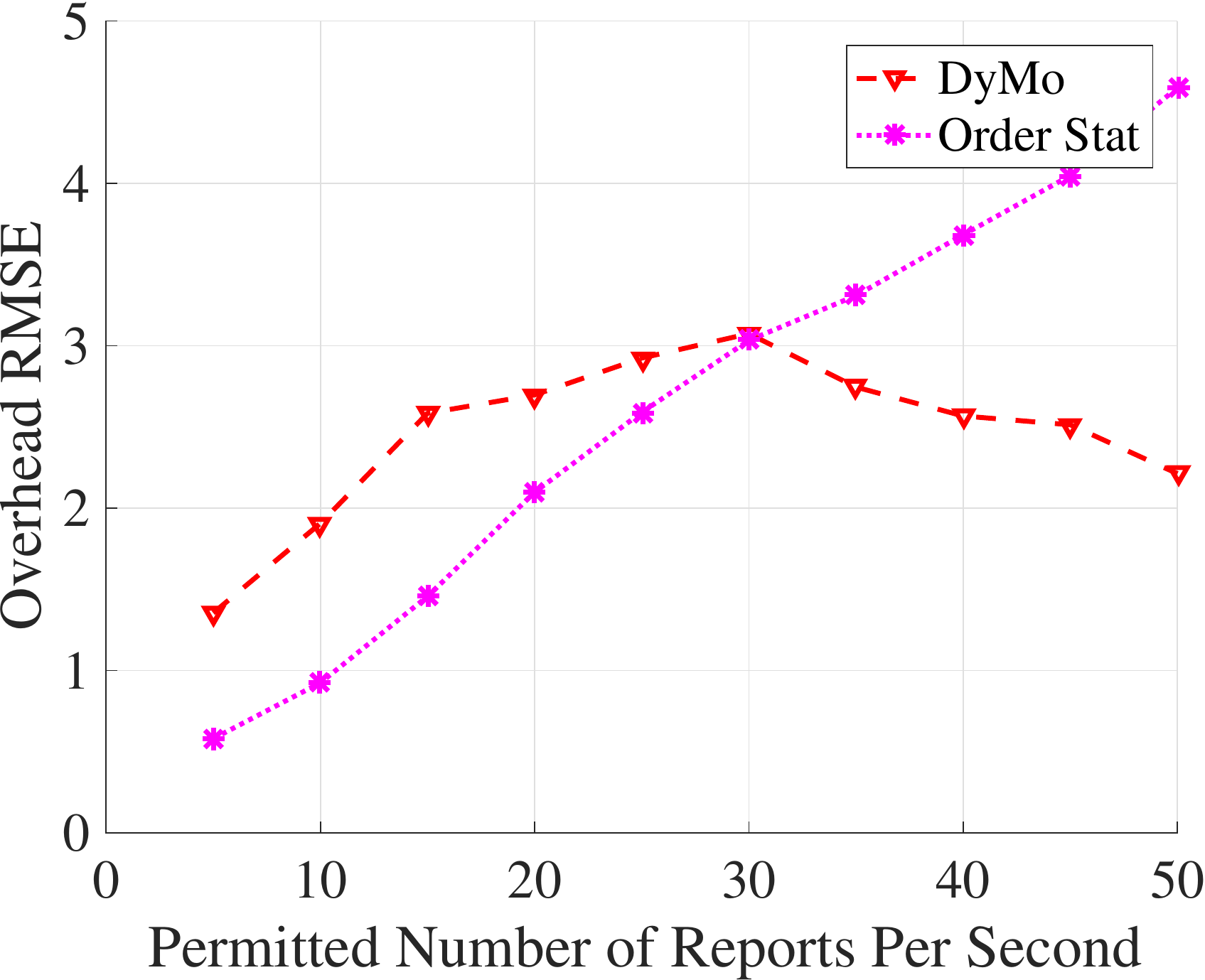}
\label{FIG:UnifOverheadvsReports} %% -(b)
}
\caption[Optional caption for list of figures]{The Root Mean Square Error (RMSE) of different parameters averaged over 5 different simulation instances lasting for $30$mins each in \HOMOGENEOUS scenario with different SNR characteristics and UE mobility patterns. (a) SNR Threshold percentile RMSE vs. the total number of UEs in the system, (b) SNR Threshold percentile RMSE vs. the QoS Constraint $p$, (c) SNR Threshold percentile RMSE vs. the number of permitted reports , (d) Overhead RMSE vs. the number of UEs, (e) Overhead RMSE vs. the QoS constraint $p$, and (f) Overhead RMSE vs. the number of permitted reports.}
\vspace*{-0.3cm}
\label{FIG:agg0}
\end{figure*}

\begin{figure*}[t]
\centering
\subfigure[]{
\includegraphics[trim=10mm 0mm 3mm 5mm, width=0.27\textwidth]{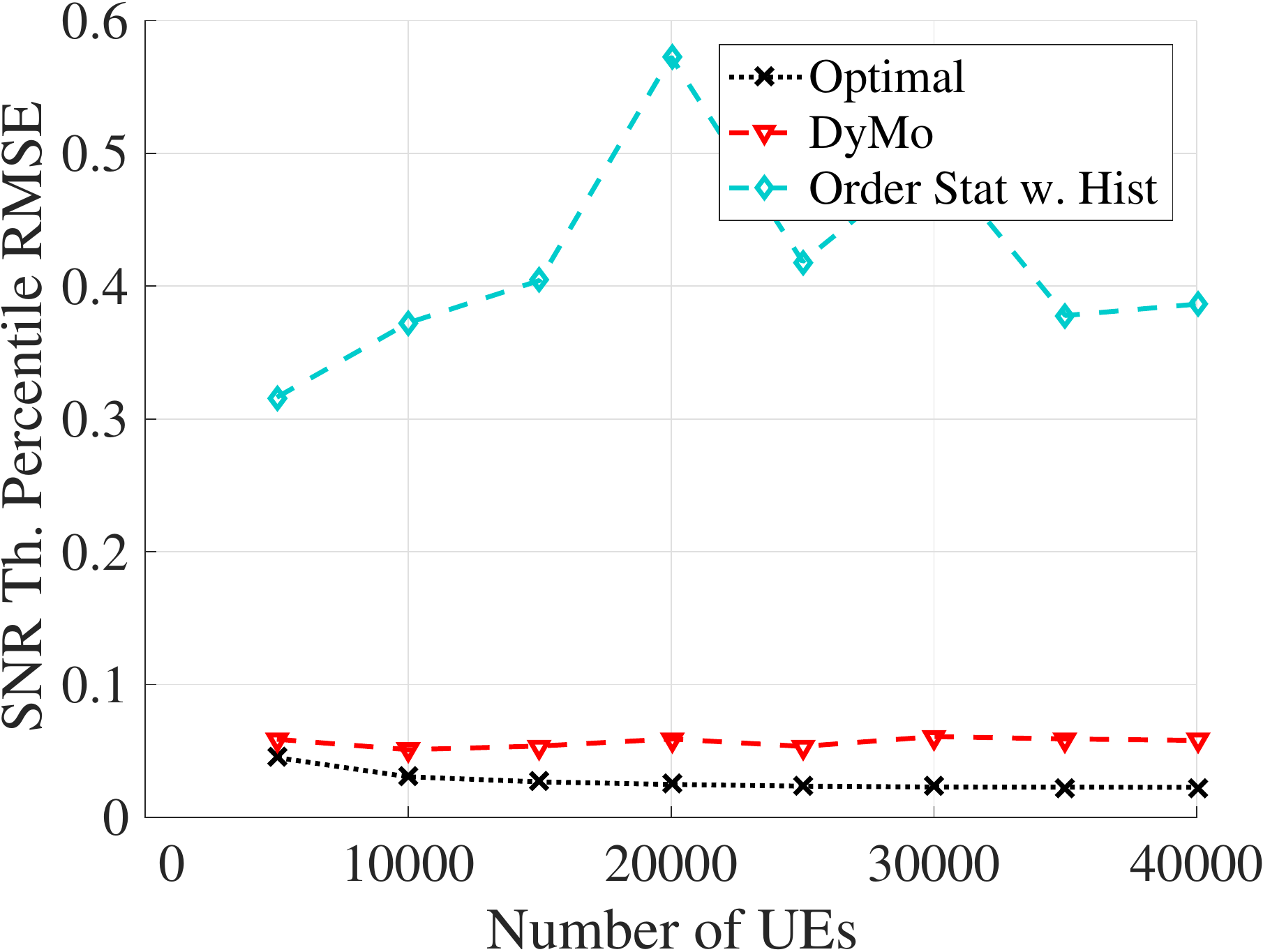}
\label{FIG:SNRthpvsUE} %% -(a)
}%
\subfigure[]{
\includegraphics[trim=5mm 0mm 3mm 5mm, width=0.27\textwidth]{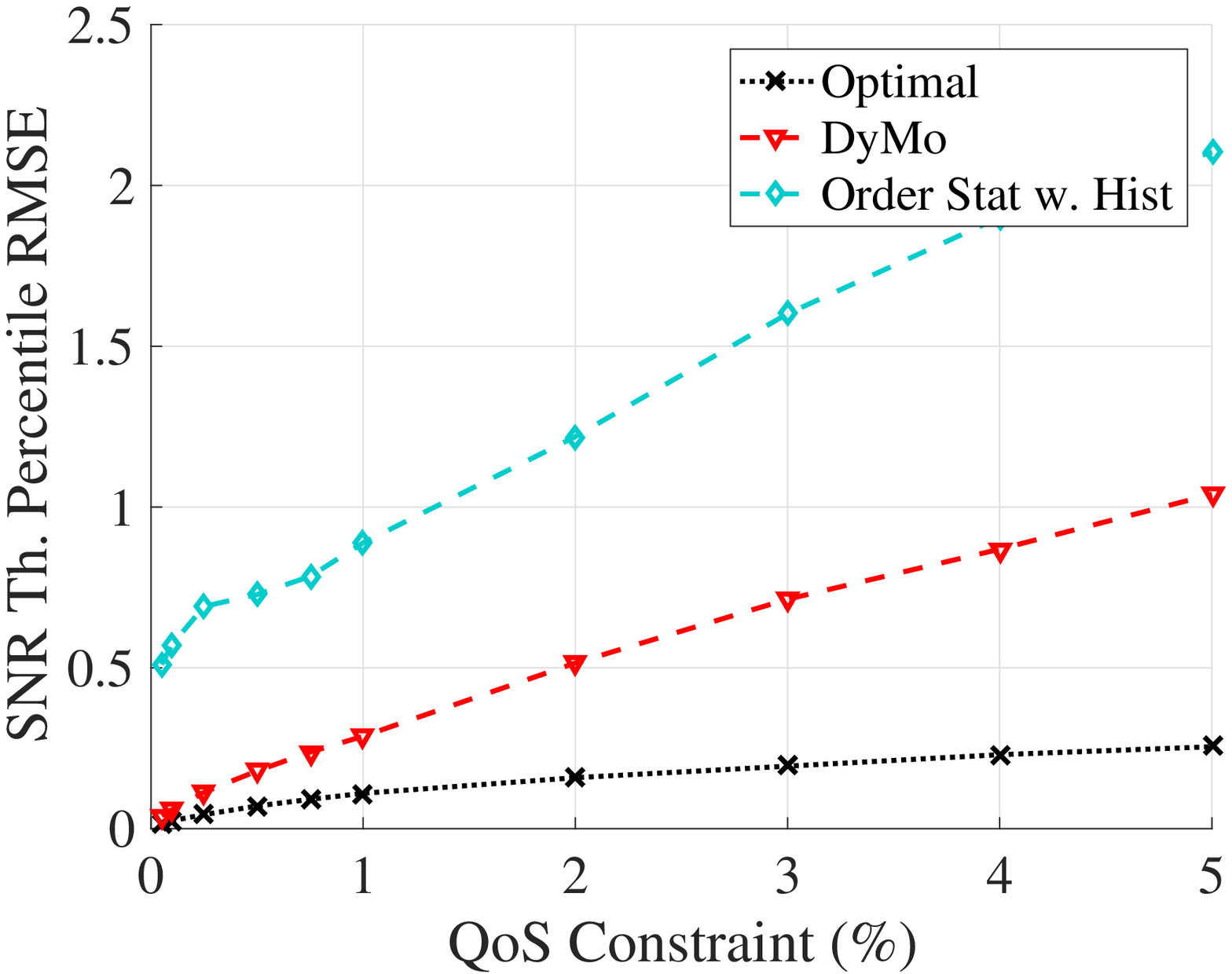}
\label{FIG:SNRthpvsP} %% -(a)
}%
\subfigure[]{
\includegraphics[trim=5mm 0mm 3mm 5mm, width=0.27\textwidth]{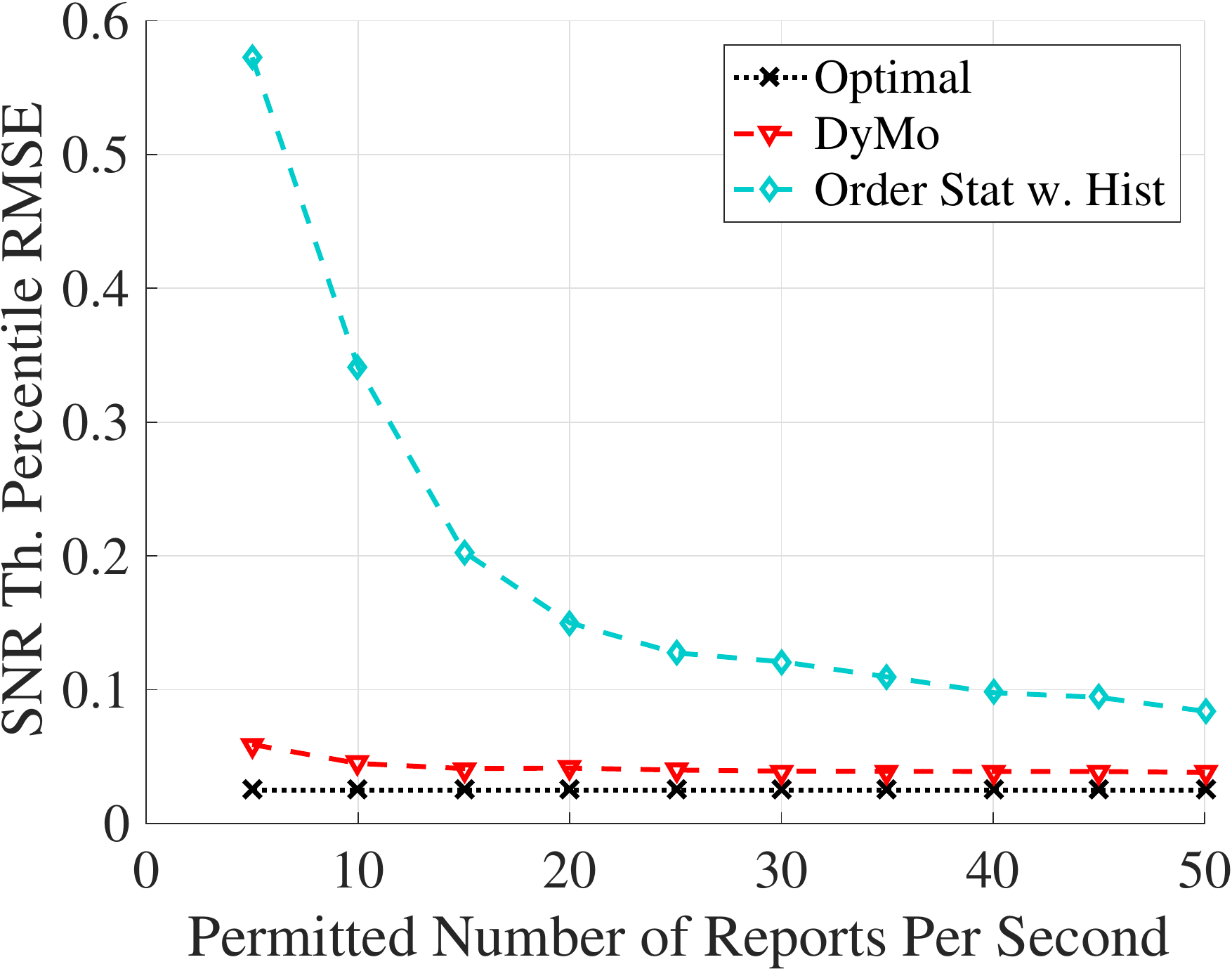}
\label{FIG:SNRthpvsReports} %% -(b)
}%
\\
\subfigure[]{
\includegraphics[trim=10mm 0mm 3mm 5mm, width=0.27\textwidth]{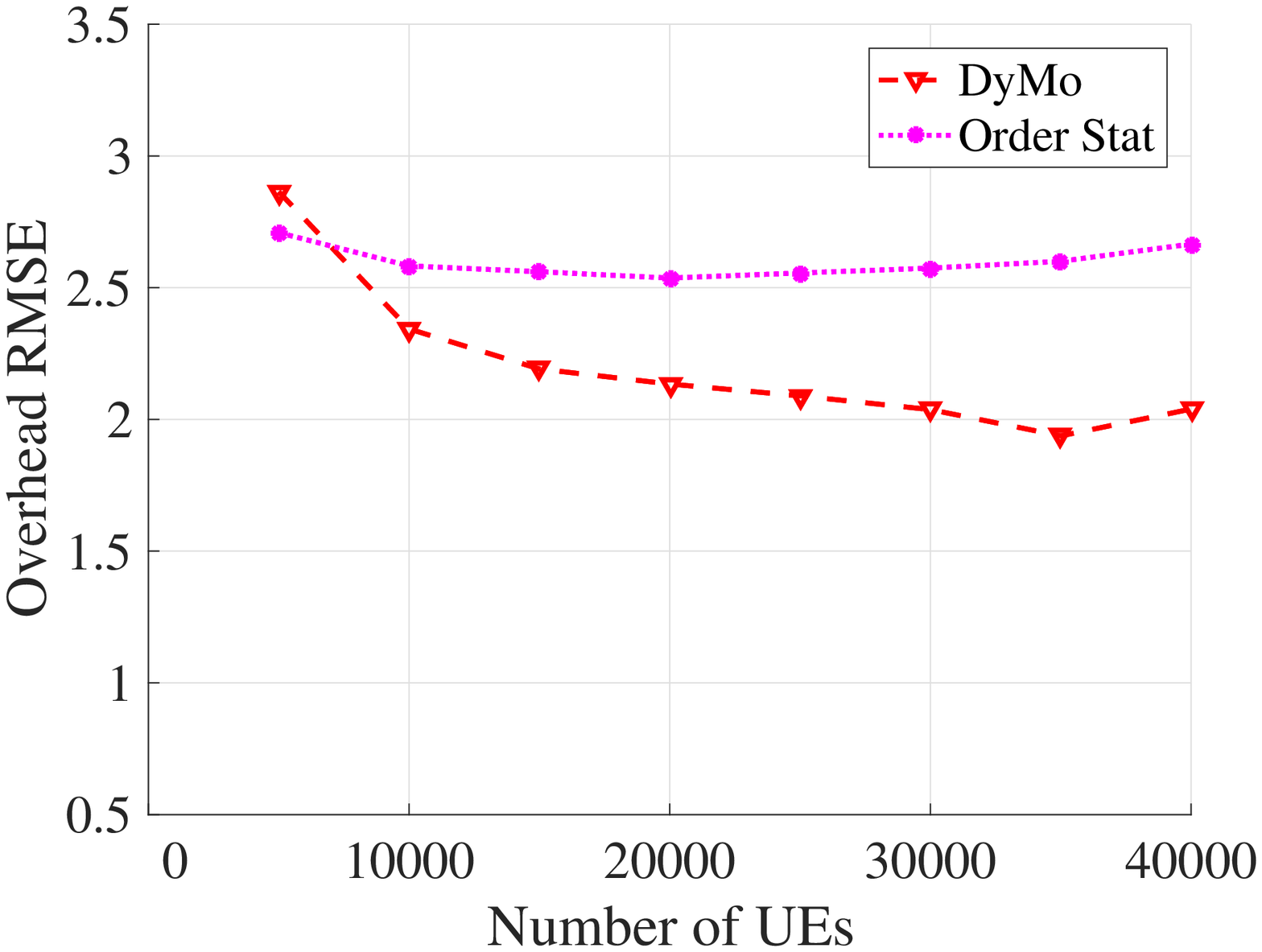}
\label{FIG:OverheadvsUE} %% -(b)
}%
\subfigure[]{
\includegraphics[trim=5mm 0mm 3mm 5mm, width=0.27\textwidth]{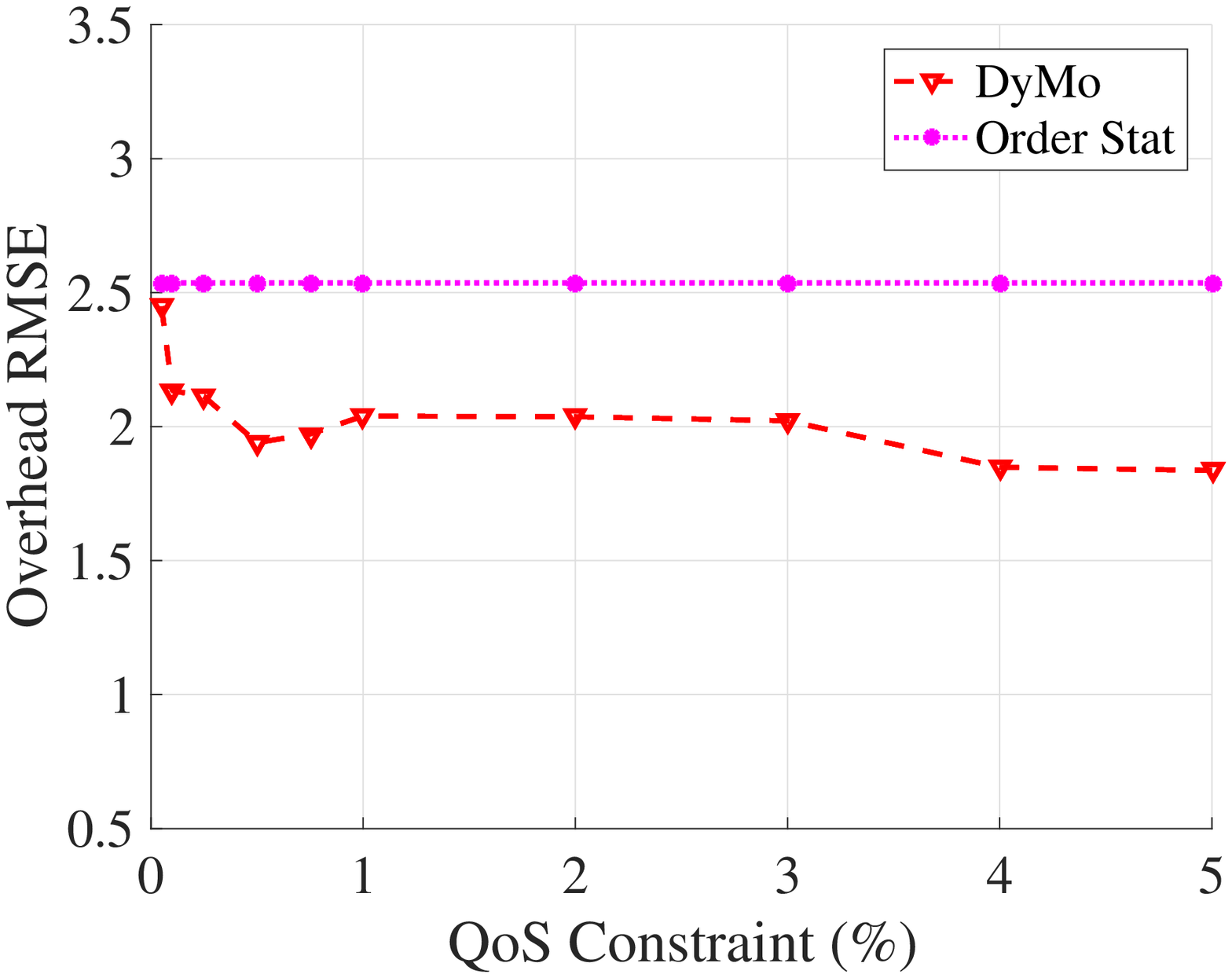}
\label{FIG:OverheadvsP} %% -(b)
}%
\subfigure[]{
\includegraphics[trim=5mm 0mm 3mm 5mm, width=0.27\textwidth]{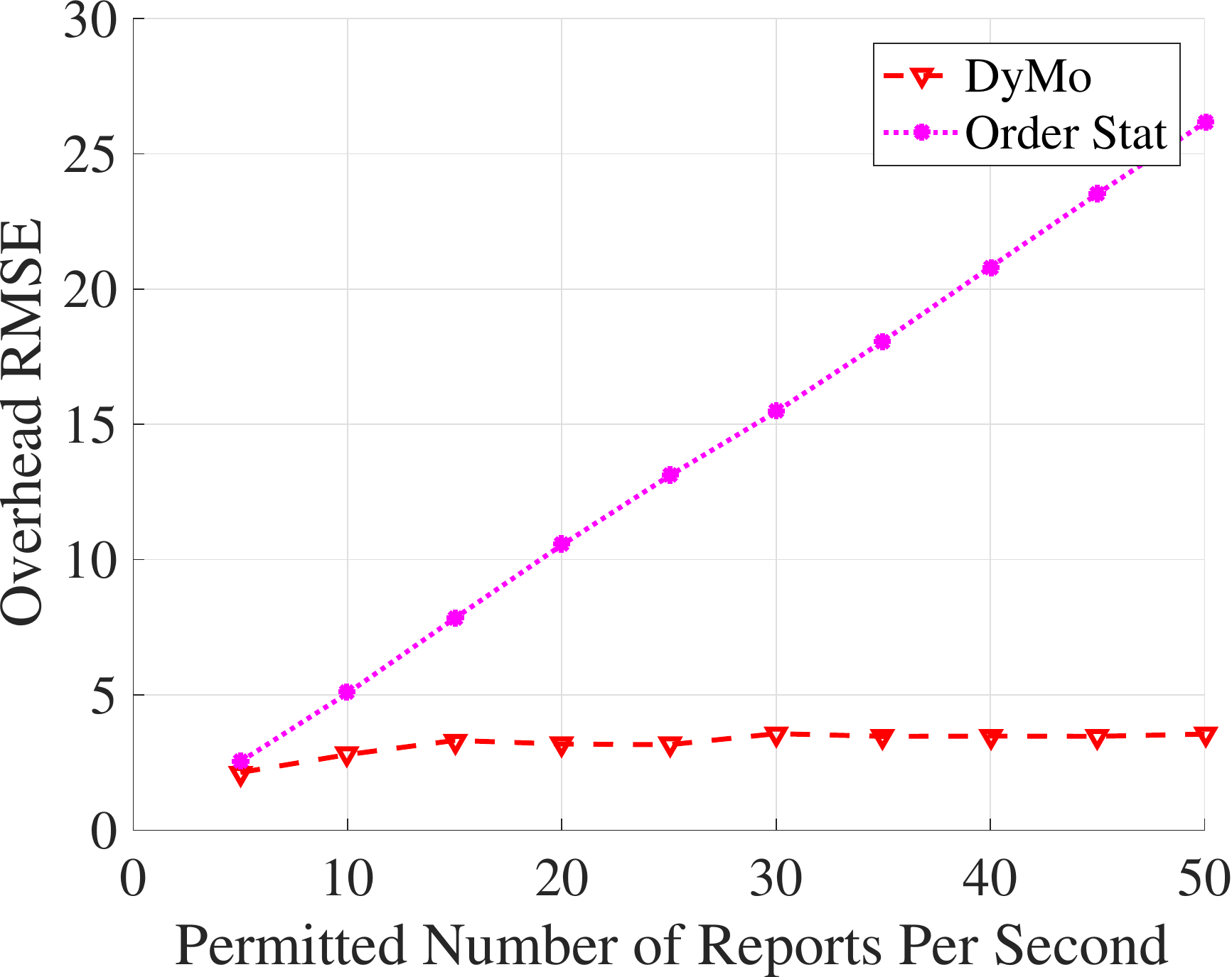}
\label{FIG:OverheadvsReports} %% -(b)
}
\caption[Optional caption for list of figures]{The Root Mean Square Error (RMSE) of different parameters averaged over 5 different simulation instances lasting for $30$mins each in a stadium environment with different SNR characteristics and UE mobility patterns. (a) SNR Threshold percentile RMSE vs. the total number of UEs in the system, (b) SNR Threshold percentile RMSE vs. the QoS Constraint $p$, (c) SNR Threshold percentile RMSE vs. the number of permitted reports, (d) Overhead RMSE vs. the number of UEs, (e) Overhead RMSE vs. the QoS constraint $p$, and (f)  Overhead RMSE vs. the number of permitted reports.}
\vspace*{-0.3cm}
\label{FIG:agg1}
\end{figure*}

\begin{figure*}[t]
\centering
\subfigure[]{
\includegraphics[trim=10mm 0mm 3mm 5mm, width=0.27\textwidth]{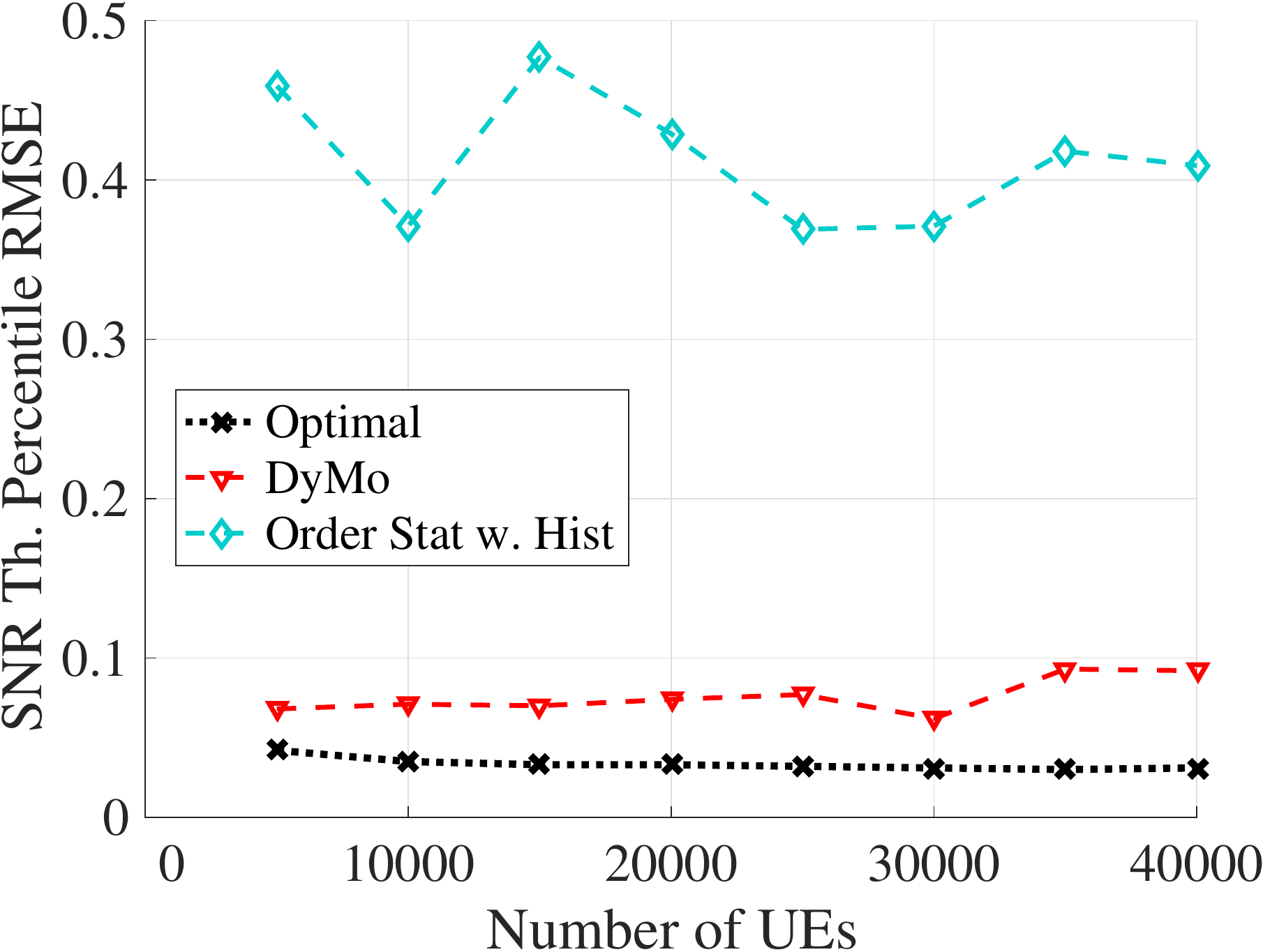}
\label{FIG:FailSNRthpvsUE} %% -(a)
}%
\subfigure[]{
\includegraphics[trim=5mm 0mm 3mm 5mm, width=0.27\textwidth]{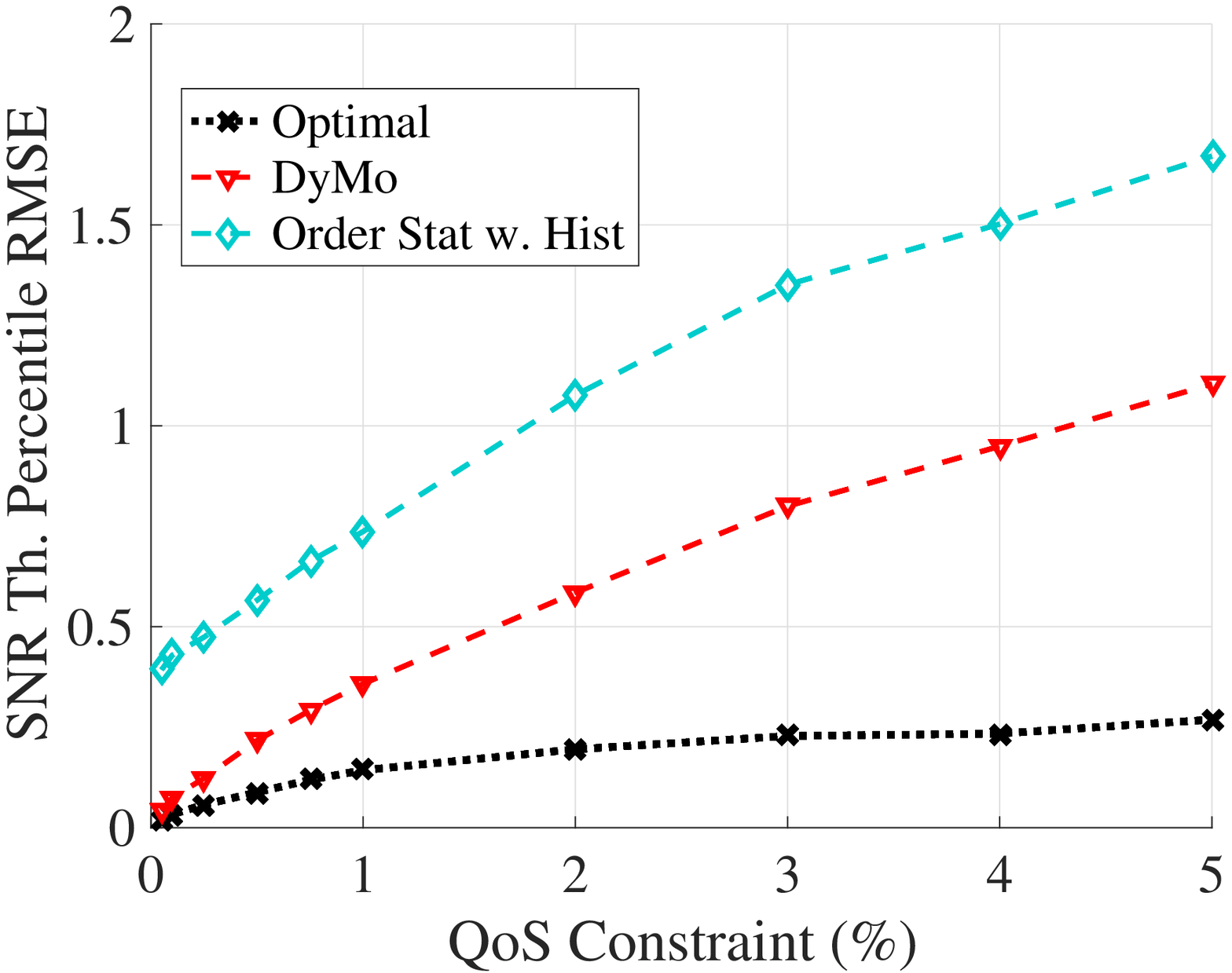}
\label{FIG:FailSNRthpvsP} %% -(a)
}%
\subfigure[]{
\includegraphics[trim=5mm 0mm 3mm 5mm, width=0.27\textwidth]{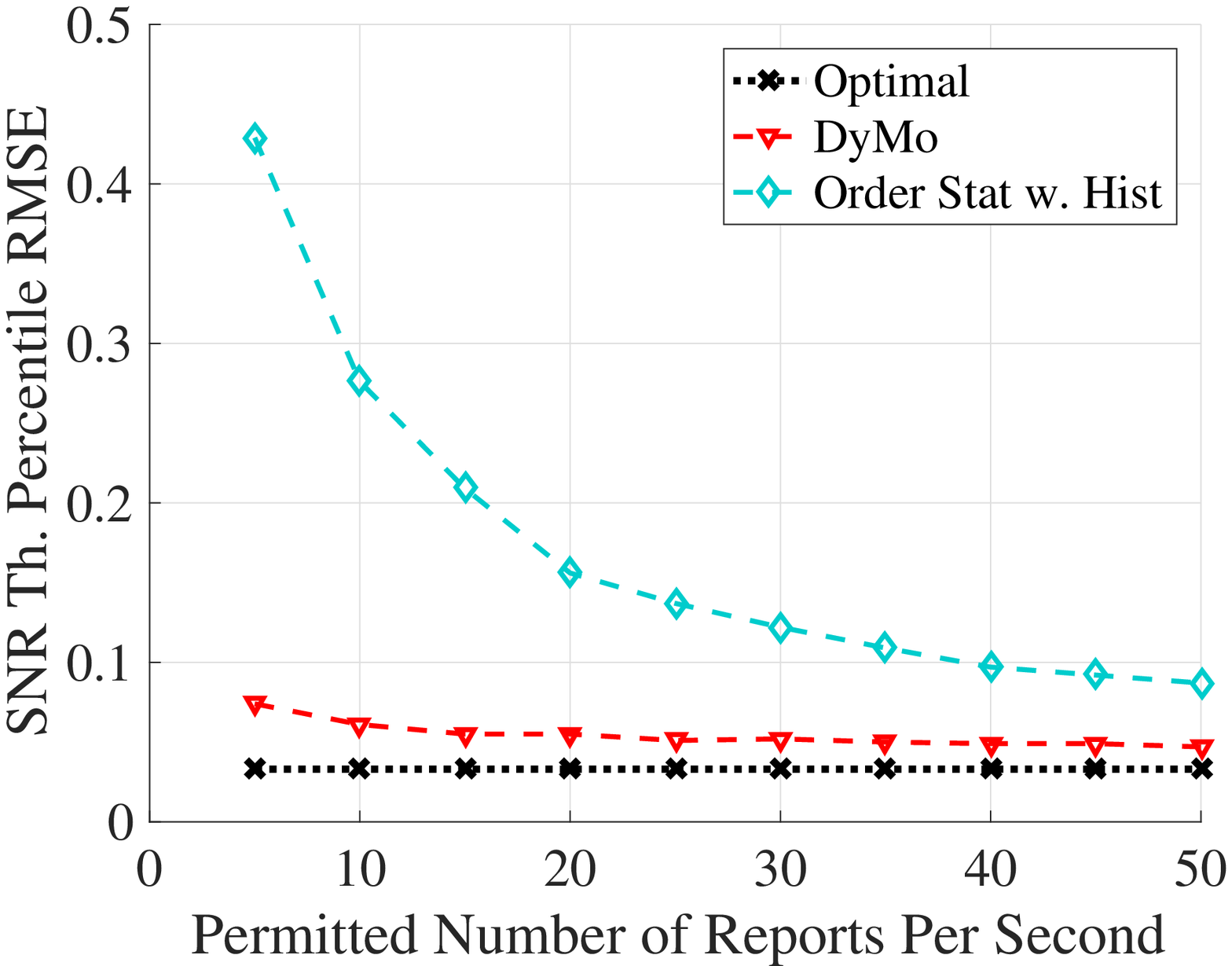}
\label{FIG:FailSNRthpvsReports} %% -(b)
}%
\\
\subfigure[]{
\includegraphics[trim=10mm 0mm 3mm 5mm, width=0.27\textwidth]{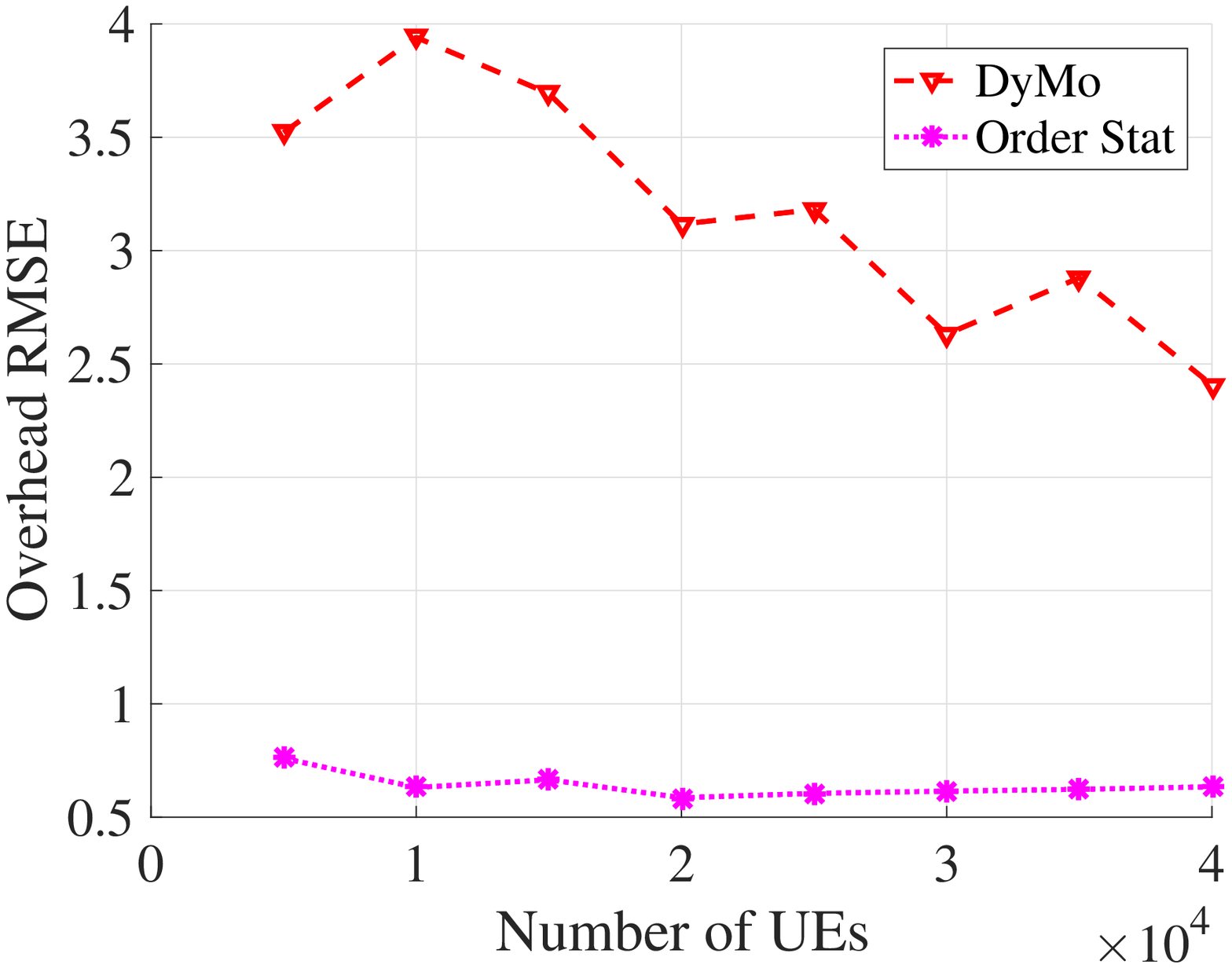}
\label{FIG:FailOverheadvsUE} %% -(b)
}%
\subfigure[]{
\includegraphics[trim=5mm 0mm 3mm 5mm, width=0.27\textwidth]{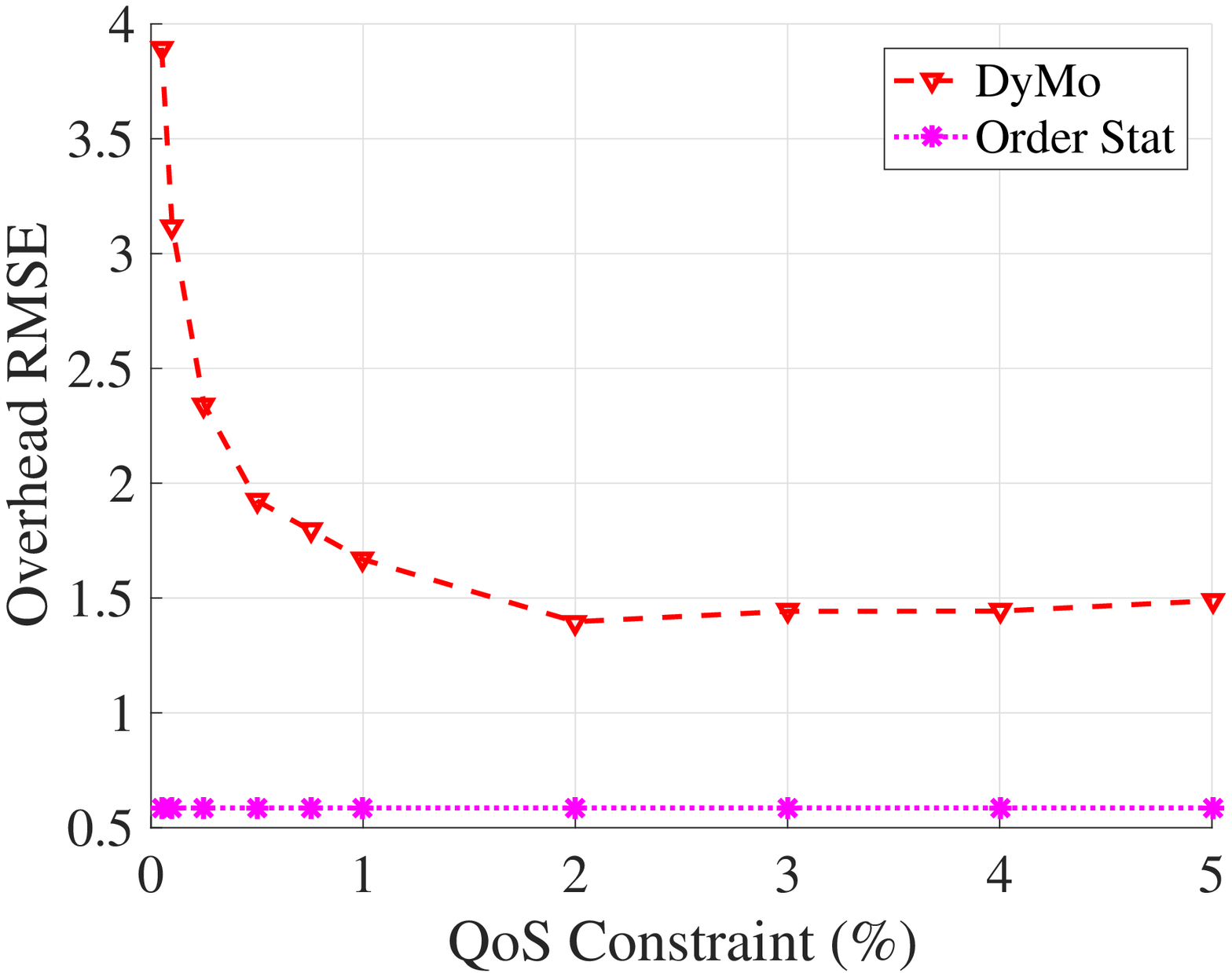}
\label{FIG:FailOverheadvsp} %% -(b)
}%
\subfigure[]{
\includegraphics[trim=5mm 0mm 3mm 5mm, width=0.27\textwidth]{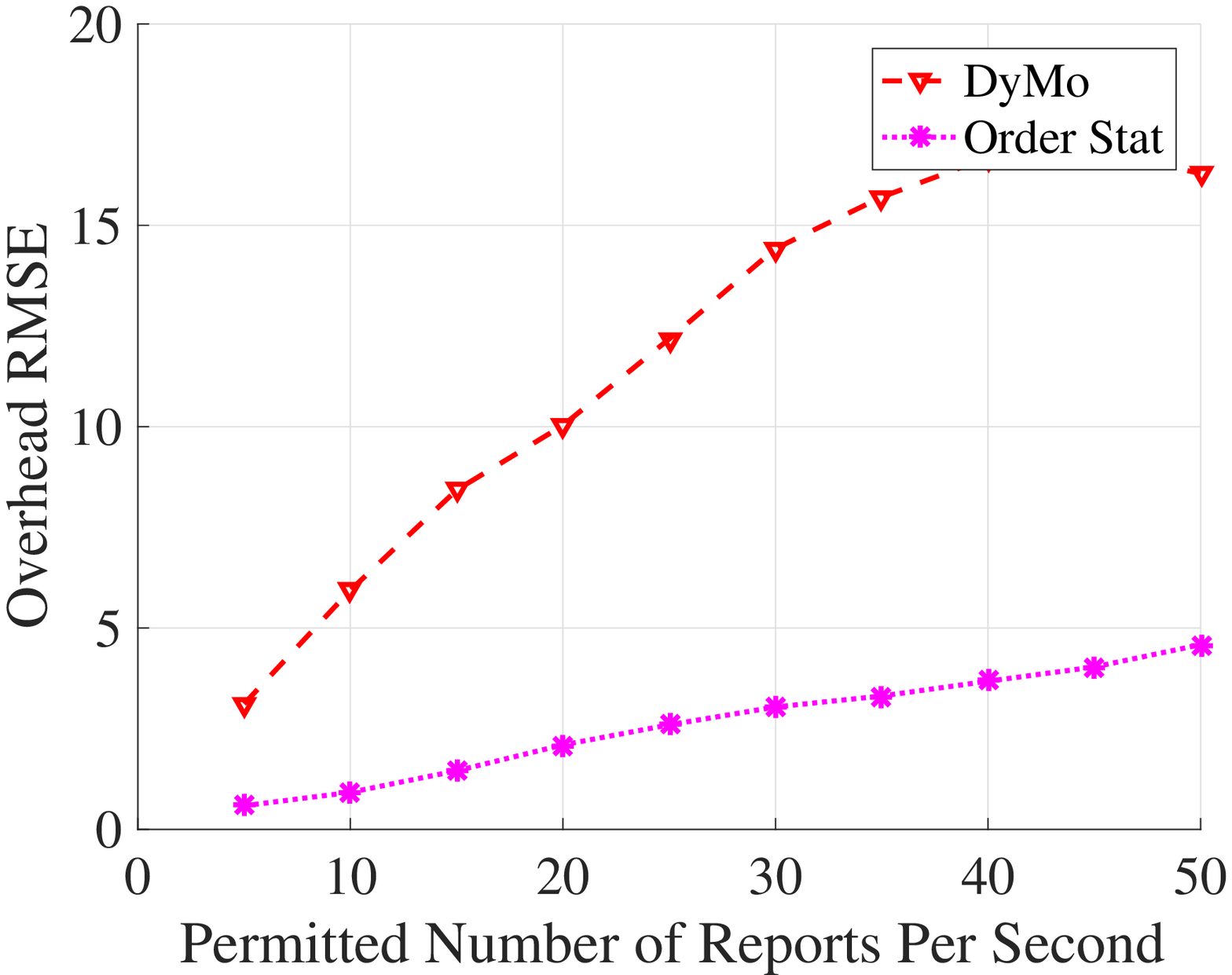}
\label{FIG:FailOverheadvsReports} %% -(b)
}
\caption[Optional caption for list of figures]{The Root Mean Square Error (RMSE) of different parameters averaged over 5 different simulation instances lasting for $30$mins each in failure scenario with different SNR characteristics and UE mobility patterns. (a) SNR Threshold percentile RMSE vs. the total number of UEs in the system, (b) SNR Threshold percentile RMSE vs. the QoS Constraint $p$, (c) SNR Threshold percentile RMSE vs. the number of permitted reports , (d) Overhead RMSE vs. the number of UEs, (e) Overhead RMSE vs. the QoS constraint $p$, and (f)  Overhead RMSE vs. the number of permitted reports.}
\vspace*{-0.3cm}
\label{FIG:agg2}
\end{figure*}

%----------------------------------------------------------------
\subsection{Performance over time}

We first illustrate the performance of the different 
schemes over time for \change{three} given instances, 
\change{a \HOMOGENEOUSNB}, a stadium and a failure scenarios,
with $m=20,000$ UEs, QoS Constraint $p = 0.1\%$, and Overhead constraint $r=5$ reports/sec, i.e., $60$ messages per reporting interval.
\change{The {\em number of permitted outliers depends on 
the number of active UEs at the current reporting interval. 
In the three considered scenarios, it can 
be at most $20$ at any given time}.  } %% End of change
%These values correspond to typical situations in dense 
%eMBMS environments.
The key difference between the \change{different} instances is 
the rate at which the SNR Threshold changes. 
\change{In the \HOMOGENEOUS environment the SNR Threshold is almost fixed with very limited variability.}
In the case of the stadium, the SNR Threshold gradually changes as the UEs change their locations. In the failure scenario, the SNR Threshold is roughly fixed but it drops instantly by $10$dBs for the duration of the failure.
%$s(t)$ is restored to its original value after fixing the failure.

The results of the \change{\HOMOGENEOUSNB}, stadium and failure 
cases are shown in \change{Figs.~\ref{FIG:vstime0},}~ \ref{FIG:vstime1}~and~\ref{FIG:vstime2}, respectively.
Figs.\change{~\ref{FIG:UnifSNRthpDymo},~\ref{FIG:UnifSNRthpOrder},}
~\ref{FIG:SNRthpDymo},~\ref{FIG:SNRthpOrder},~\ref{FIG:FailSNRthpDymo},~and~\ref{FIG:FailSNRthpOrder} show the actual SNR Threshold percentile over time. 
From Figs. \change{\ref{FIG:UnifSNRthpDymo},}~\ref{FIG:SNRthpDymo}~and~\ref{FIG:FailSNRthpDymo}, 
we observe that \AMUSE can accurately infer the SNR Threshold with an estimation error of at most $0.1\%$. Fig.~\ref{FIG:FailSNRthpDymo} shows slightly higher error of $0.25\%$ at the time of the failure (at the $7^{th}$ minute).
The \ORDERSTATS variants suffer from much higher estimation error to the order of a few percentage points, 
as shown by Figs. \change{\ref{FIG:SNRthpOrder}},~\ref{FIG:SNRthpOrder}~and~\ref{FIG:SNRthpOrder}%
\footnote{Notice that \change{the pairs (i) Figs.~\ref{FIG:UnifSNRthpDymo}~and~\ref{FIG:UnifSNRthpOrder}, (ii)} Figs.~\ref{FIG:SNRthpDymo}~and~\ref{FIG:SNRthpOrder} as well as (iii) Figs.~\ref{FIG:FailSNRthpDymo}~and~\ref{FIG:FailSNRthpOrder} use 
different scales for the Y axes.}.
This performance gap results in different estimation 
accuracy of the SNR Threshold for \AMUSE and \ORDERSTATS schemes as illustrated in Figs.\change{~\ref{FIG:UnifSNRth},}~\ref{FIG:SNRth}~and~\ref{FIG:FailSNRth}, respectively.
These figures show that the performance of \AMUSE and \OPT is almost identical.
Even in the event of a failure, \AMUSE reacts immediately 
and detects the SNR Threshold accurately. 
The \ORDERSTATS variants react quickly to a failure but not as accurately as \AMUSENB.
After the recovery, both \AMUSE and \ORDSTwithHIST gradually increase their SNR Threshold estimates,
due to the exponential smoothing process.

The SNR Threshold estimation gap directly impacts the number of outliers as well as the network utilization, i.e., 
the spectral efficiency. 
\change{Figs.~\ref{FIG:UnifOutliersDymo} and~\ref{FIG:UnifOutliersOrder} show the number of outliers of \AMUSE and \ORDERSTATS variants for the \HOMOGENEOUS environment, respectively%
\footnote{Notice that the figure pairs, (i) 
Figs.~\ref{FIG:UnifOutliersDymo} and~\ref{FIG:UnifOutliersOrder}, (ii) Figs.~\ref{FIG:OutliersDymo} and~\ref{FIG:OutliersOrder} as well as (iii) Figs.~\ref{FIG:FailOutliersDymo} and~\ref{FIG:FailOutliersOrder}, 
use different scales for the Y axes.}, 
while Figs.~\ref{FIG:UnifSpectralEfficiencyDymo} and~\ref{FIG:UnifSpectralEfficiencyOrder} show the 
spectral efficiency of the schemes.
Figs.~\ref{FIG:UnifOutliersDymo} and \ref{FIG:UnifSpectralEfficiencyDymo} reveal that after a short adaptation phase \AMUSE converges to the optimal performance, i.e., spectral efficiency, while preserving the QoS constraint.
Fig.~\ref{FIG:UnifSpectralEfficiencyDymo} show that both \OPT and \AMUSE fluctuate between two spectral efficiency levels, $0.29$ and $0.36$ bit/sec/Hz, which results from oscillatation between 
two MCS levels $3$ and $4$. Such oscillations can be easily 
suppressed by enforcing some delay between MCS increase operations. 
The \ORDERSTATS variants over estimate the SNR threshold and 
suffer from higher number of outliers, as shown by 
Fig.~\ref{FIG:UnifOutliersOrder}.
The \HOMOGENEOUS setting represents quasi-static environments with minor variation of the SNR threshold, $s(t)$.
In such settings, the \UNI scheme provides a good estimation%
\footnote{Assuming rigorous field trial measurements.} 
of $s(t)$ and its number of outliers as well as the obtained 
spectral efficiency are comparable to \AMUSENB.
However, this is not the situation when $s(t)$ is time varying.  

The number of outliers of \AMUSE and \ORDERSTATS variants for the stadium environment is shown in Figs.~\ref{FIG:OutliersDymo} and~\ref{FIG:OutliersOrder}, respectively,  
while Figs.~\ref{FIG:FailOutliersDymo} and~\ref{FIG:FailOutliersOrder} 
illustrate the number of outliers of \AMUSE and \ORDERSTATS variants for the failure scenario, in this order.
These figures show that the number of outliers that results 
from the \ORDSTwithHIST and \ORDSTnoHIST  variants are 
occasionally over $200$ and $800$, respectively. 
Whereas, \AMUSE ensures that the number of outliers at 
any time is comparable to \OPT%
and in the worst case it exceeds the permitted 
number by less than a factor of $2$.

Figs.~\ref{FIG:SpectralEfficiencyDymo} and~\ref{FIG:SpectralEfficiencyOrder} show the spectral efficiency for the stadium environment, whereas Figs.~\ref{FIG:FailSpectralEfficiencyDymo} and~\ref{FIG:FailSpectralEfficiencyOrder} show the spectral efficiency for the component failure case. The spectral efficiency for each case is correlated to the SNR Threshold. For the stadium environment, \AMUSE has spectral efficiency close to \OPT while \UNI has the lowest spectral efficiency. In the event of a failure, the spectral efficiency of \AMUSE follows the \OPT as expected from the SNR Threshold estimations.
Since \ORDERSTATS variants typically over estimate the SNR Threshold, they frequently determine MCS and consequently spectral efficiency that exceed the optimal settings. 
Such inaccuracy leads to a high number of outliers.}

Figs.\change{~\ref{FIG:UnifOverhead},}~\ref{FIG:Overhead}
~and~\ref{FIG:FailOverhead} indicate 
only mild violation of the Overhead Constraint by both the \AMUSE and 
\ORDERSTATS variants. 
We observe that accurate SNR Threshold estimation allows \AMUSE to achieve near optimal spectral efficiency with negligible violation of the QoS Constraint.
The other schemes suffer from sub-optimal spectral efficiency, excessive number of outliers, or both. 
Given that the permitted number of outliers is at most $20$,
the \ORDSTwithHIST and \ORDSTnoHIST schemes exceed this value sometimes 
by a factor of $10$ and $40$, respectively.
Among these two alternatives, \ORDSTwithHIST leads to lower number of outliers.
\change{While \UNI provides accurate estimation of $s(t)$ for the 
\HOMOGENEOUS environment, we observe that it yields 
a very conservative eMBMS MCS setting in the stadium example, 
which causes low network utilization.}
In the failure scenario, the conservative eMBMS MCS of \UNI is not sufficient to cope with the low SNR Threshold and it 
leads to excessive number of outliers.

\subsection{Impact of Various Parameters}

We now turn to evaluate the quality of the SNR Threshold estimation and the schemes’ ability to preserve the QoS and Overhead Constraints under 
various settings.
%reliability of not exceeding the overhead constraint for different 
%schemes for various settings. 
We use the same configuration of $m=20,000$ UEs, $p=0.1\%$ and $r=5$ reports/sec and we evaluate the impact of changing the values of one of the parameters. The results for the \change{\HOMOGENEOUSNB,} stadium and failure scenarios are shown in 
\change{Figs.~\ref{FIG:agg0},~\ref{FIG:agg1}~and~\ref{FIG:agg2}, respectively.}  
Each point in the figures is the average of 5 different %\change{stadium/failure} 
simulation instances of $30$mins 
each with different SNR characteristics and UE mobility patterns. 
The error bars are small and not shown. 
In these examples, we compare \AMUSE only with \OPT and 
\ORDSTwithHIST which is the best performing alternative. 
\change{We omit the \UNI scheme since it does not adapt to variation of $s(t)$.}

\change{First, we consider the impact of changing these parameters on the accuracy of the SNR Threshold estimation.
Figs.~\ref{FIG:UnifSNRthpvsUE},~\ref{FIG:SNRthpvsUE}, and \ref{FIG:FailSNRthpvsUE} show} the Root Mean Square Error (RMSE) in SNR Threshold percentile estimation vs.\ $m$, \change{for \HOMOGENEOUSNB, stadium and failure scenarios, respectively.}
The non-zero values of RMSE in \OPT are due to quantization of SNR reports. The RMSE in the SNR Threshold estimation of \AMUSE is close to that of \OPT regardless of the number of UEs, 
\change{while \ORDSTwithHIST suffers from order of magnitude 
higher RMSE.} 

\change{Figs.~\ref{FIG:UnifSNRthpvsP}, \ref{FIG:SNRthpvsP}, and \ref{FIG:FailSNRthpvsP} show}
the RMSE in SNR Threshold estimation as the QoS Constraint $p$ changes, \change{for \HOMOGENEOUSNB, stadium and failure scenarios}.
\AMUSE outperforms the alternative schemes as $p$ increases. 
As $p$ increases, we observe an increasing quantization error, which impacts the RMSE of all the schemes including the \OPTNB.
% Yigal- explnation below - To put back for full version 
\change{Recall that the SNR distribution is represented 
by a histogram where each bar has a width of $0.1dB$. 
As $p$ increases, the number UEs in the bar that contains 
the $p$ percentile UE increases as well. 
Since $s(t)$ should be below the SNR value of this bar,
we notice a higher quantization error.}

\change{Figs.~\ref{FIG:UnifSNRthpvsReports}, \ref{FIG:SNRthpvsReports}, and \ref{FIG:FailSNRthpvsReports} illustrate}
the SNR Threshold percentile RMSE as the Overhead Constraint is relaxed, \change{for \HOMOGENEOUSNB, stadium and failure cases, respectively.} 
The SNR Threshold percentile RMSE of \AMUSE is $0.05\%$ even with Overhead Constraint of $5$ reports/sec, while \OPT RMSE due to quantization is $0.025\%$.
\AMUSE error slightly reduces by relaxing the Overhead Constraint
(\OPT error stays $0.25\%$).
Even with $10$ times higher reporting rate, \AMUSE significantly outperforms the \ORDERSTATS alternatives. 
The RMSE in SNR Threshold percentile for \ORDERSTATS is in the order of the required average value of $0.1$ even 
with a permitted overhead of $50$ reports/sec, i.e,. $3000$ reports per reporting interval.
This is a very high overhead on the unicast traffic,
since in LTE networks \change{the number of simultaneously open unicast connections is limited, i.e., several hundreds per base station} and each connection lasts several hundred msecs even for sending a short update.
Unlike the downlink, uplink resources are not reserved for eMBMS systems and utilize the unicast resources.
The RMSE of number of outliers is qualitatively similar to the SNR Threshold percentile results.
%% Yigal - removed for the RR.
%%  and we exclude the results due to space limitations.

\change{We also compute the overhead RMSE for different UE population sizes, $m$, QoS Constraint $p$, and Overhead Constraints $r$.
The results are shown is sub-figures (d), (e) and (f) of Figs.~\ref{FIG:agg0},~\ref{FIG:agg1}~and~\ref{FIG:agg2}, respectively.
In most cases, the overhead RMSE of \AMUSE is between $1-4$ even when the system parameters change.
We observe an increase in the overhead RMSE only in failure scenarios when the permitted overhead is relaxed,
as shown in Fig.~\ref{FIG:FailOverheadvsReports}.
This is expected immediately after a failure because many more UEs suffer from poor service than \AMUSE estimated.
Thus, as the permitted overhead increases also the spike 
in the number of reports during the first reporting interval
after the failure also increases, which results in a gradual increase of the Overhead RMSE%
\footnote{Notice that the RMSE metric is sensitive to sporadic but very high error.}.

Figs.~\ref{FIG:agg0}~and~\ref{FIG:agg2} show that the \ORDERSTATS variants experience very low violation of the Overhead Constraint in the \HOMOGENEOUS and Failure scenarios.  
This is not surprising, since in these scenarios the variation in the number of active eMBMS receivers is very small and this number is 
roughly $\E[m(t)]$ (the expected number of active eMBMS receivers).
As mentioned in Section~\ref{SSC:EvalMethodology}, this 
observation is misleading, since we assume that $\E[m(t)]$ is known and we ignore the overhead of configuring the UEs with the proper reporting rate.
Obviously, {\em the exact number of active receivers, $\E[m(t)]$, is unknown in practice.}
Furthermore, Fig.~\ref{FIG:OverheadvsReports} shows that in scenarios with high variation in the number of active receivers, $m(t)$, (like the case in the stadium simulations) the violation of the Overhead Constraint is high and it is amplified as the permitted number of reports, $r$, increases.
This is due to the static reporting rate of \ORDERSTATS despite dynamic changes of the number of active eMBMS receivers.
Fig.~\ref{FIG:OverheadvsReports}
confirms that the overhead violation of \ORDERSTATS is very 
sensitive to the estimation of $\E[m(t)]$ and its variance.

{\em Given that the number of active eMBMS receivers, $m(t)$, 
is unknown and may change significantly over time, \ORDERSTATS 
cannot practically preserve the Overhead Constraint without keeping 
track of the active UEs and sending individual messages to a subset of the active UEs.}
However, keeping track of $m(t)$ requires each UE to report when it starts and stops receiving eMBMS services,
which may incur much higher overhead than permitted.
For instance, in our simulations with $m=20,000$ UEs, even if such switching occurs at most once (start and stop) by each UE, 
the total number of reports is $40,000$. When dividing this number by the simulation duration of $30$ minutes ($1,800$ sec) 
we get $22$ messages/second, which is much higher than the permitted overhead.
} %% End of \change 

\iffalse %%%%%
We also notice another interesting case when the permitted overhead is allowed to increase as shown in Fig.~\ref{FIG:OverheadvsReports}. 
While the \AMUSE RMSE is consistently small, the RMSE of \ORDERSTATS scales linearly with the permitted overhead.
This is due to the static reporting rate of \ORDERSTATS despite changing number of active UEs.

This phenomena is significantly milder in the \HOMOGENEOUS and failure instances, as shown 
in Figs.~\ref{FIG:UnifOverheadvsReports} ~and~\ref{FIG:FailOverheadvsReports}, 
since the number of active eMBMS receivers is near constant.
Fig.~\ref{FIG:OverheadvsReports}
demonstrates that the overhead violation of \ORDERSTATS is very sensitive to the estimation of $\E[m(t)]$ 
(the expected number of active eMBMS receivers) 
and its variance. 
Given that the number of active eMBMS receivers, $m(t)$, 
is unknown and may change significantly over time, it is practically 
impossible to preserve the Overhead Constraint without keeping 
track of the active UEs by sending individual messages to a subset of the active UEs. However, as we observe in our evaluation, keeping track of $(m(t)$ may incur higher overhead than permitted.
\fi %%%%%%%%%%%%%%%%%

\noindent
{\bf Summary:} Our simulations show that \AMUSE achieves accurate, close to optimal, estimation of the SNR Threshold even when the number of active eMBMS receivers is unknown. It can improve the spectral efficiency for eMBMS operation, while adding a very low reporting overhead.
\AMUSE can predict the SNR Threshold with lower errors than other alternatives under a wide range of the SNR Threshold requirement $p$ and reporting Overhead Constraint $r$. 
These observations show that \AMUSE exceeds the expectations of our analysis in Section~\ref{SC:Alg}.

%%%%%%%%%%%%%%%%%%%%%%%%%%%%%%%%%%%%%%%%%%%%%

%%%%%%%%%%%%%%%%%%%%%%%%%%%%%%%%%%%%%%%%%%%%%%%%%%%%%%%
% File name:  demo.tex
% Changes (date, author, description):
%%%%%%%%%%%%%%%%%%%%%%%%%%%%%%%%%%%%%%%%%%%%%%%%%%%%%%%

\section{Conclusion}
\label{SC:conclusion} 

\change{This paper presents a {\em Dynamic Monitoring} (\AMUSENB) system for large scale monitoring of eMBMS services, 
based on the concept of  {\em Stochastic Group Instructions}. 
Our extensive simulations show that \AMUSE achieves accurate, close to optimal, estimation of the SNR Threshold 
even when the number of active UEs is unknown. It can improve the spectral efficiency for eMBMS operation while adding 
a low reporting overhead. }
\change{
\section{Acknowledgment}
\label{SC:ack} 

This work was supported in part by NSF grants CNS-16-50669 and CNS-14-23105.
%\bf We would like to thank our parents from their technical support since the moment that we were born (replacing our diapers) until we graduated from college. We love you very much!  Next time we would like you to pay also for our houses and axillary cars.   I am running out of gibberish to write so I assume that this is enough space for your Acknowledgment. 
}

%\fi %%%%%%%%%%%%%%%%%%%%%%%% End of file %%%%%%%%%%%%%%%%%%%%%%

%\pagebreak
\footnotesize{
\bibliographystyle{ieeetr}
\bibliography{AMuSe-ref_V4}  
}
\appendix
%%%%%%%%%%%%%%%%%%%%%%%%%%%%%%%%%%%%%%%%%%%%%%%%%%%%%%%
% File name:  appendix-V1.tex
% Authors: Chun-Nam Yu
% Version: 1.1
%------------------------------------------------------
% Changes (date, author, description):
%%%%%%%%%%%%%%%%%%%%%%%%%%%%%%%%%%%%%%%%%%%%%%%%%%%%%%%
%\comment{GZ: according to IEEE guidelines there is no need to write eq. (x) when reffering to an equation just write (x) (unless its a beginning of a sentence}

%\section{Algorithm Analysis}
%\label{SC:analysis}
 
\subsection{Analysis of the Two-Step Estimation Algorithm}
\label{SC:2SAnalysis}

We now extend the analysis of the  \twostep algorithm given in Section~\ref{SSC:twostep}.
We show that the optimal settings for minimizing 
the error $\epsilon$ of Equation~(\ref{eq:2Serror}) 
is obtained by taking
\[
p_1 \!=\! p_2 \!=\! \sqrt{p}~~~and~~~r_1 \!=\! r_2 \!=\! r/2
\]
Notice that the settings should satisfy the following two constraints:
\begin{equation}
p=p_1\cdot p_2
\label{eq:pcon}
\end{equation}
and
\begin{equation}
r=r_1+r_2
\label{eq:rcon}
\end{equation}
From Equation~(\ref{eq:2Svar}) and by taking $3$ times the 
standard deviation, we get that the errors $\epsilon_1$ and $\epsilon_2$ are
\[
\epsilon_1 = 3 \sqrt{\frac{p_1(1-p_1)}{r_1}}~~~and~~~
\epsilon_2 = 3 \sqrt{\frac{p_2(1-p_2)}{r_2}}
\]
By combining with Equation~(\ref{eq:2Serror}), we get
\begin{equation}
%\begin{split}
  \epsilon  = %\ & \epsilon_1 p_2 + \epsilon_2 p_1\\
                    %=   \ & 
                       3\ \sqrt{\frac{p_1(1-p_1)}{r_1}}\ p_2 +
                       3\ \sqrt{\frac{p_2(1-p_2)}{r_2}}\ p_1
%\end{split}
%\]} %% End of Change
\label{eq:2Serr2}
\end{equation}
By using the two constraints~(\ref{eq:pcon})~and~(\ref{eq:rcon}), 
we assign $p_2=p/p_1$ and $r_2=r-r_1$. Consequently,
\begin{equation}
\begin{split}
  \epsilon  =\ & 
     3\ \sqrt{\frac{p_1(1-p_1)}{r_1}}\ \frac{p}{p_1} +
     3\ \sqrt{\frac{(p/p_1)(1-p/p_1)}{r-r_1}}\ p_1\\
   =\ &
     3\ p \left[\sqrt{\left(\frac{1}{p_1}-1\right)\ \frac{1}{r_1}}\ +
       \sqrt{\left(\frac{p_1}{p}-1\right)\ \frac{1}{r-r_1}}\right]
\end{split}
\label{eq:2Serr3}
\end{equation}
By taking the partial derivative $\DRV{\epsilon}{p_1}$ we get,
\begin{equation}
\begin{split}
\DRV{\epsilon}{p_1} = 
   3\ p & \left[ 
       - \left(2\ p_1^2\ \sqrt{\left(\frac{1}{p_1}-1\right)
             \ \frac{1}{r_1}}\right)^{-1} + \right. \\
      & \ \ \left.\left(2\ p\ \sqrt{\left(\frac{p_1}{p}-1\right)\ 
             \frac{1}{r-r_1}}\right)^{-1}\right]
\end{split}
\label{eq:2SDrvp1}
\end{equation}
For minimizing the error we calculate $\DRV{\epsilon}{p_1} = 
0$ and get that
\begin{equation}
p_1^2\ \sqrt{\left(\frac{1}{p_1}-1\right) \frac{1}{r_1}} = 
p\ \sqrt{\left(\frac{p_1}{p}-1\right)\ \frac{1}{r-r_1}}
\label{eq:2SDrvp1eq}
\end{equation}
By simple mathematical manipulations we get
\begin{equation}
p_1^4\ \left(\frac{1}{p_1}-1\right) \left(r-r_1\right) = 
p^2\ \left(\frac{p_1}{p}-1\right)\ r_1
\label{eq:2SDrvp1eq2}
\end{equation}
Similarly, from the partial derivative $\DRV{\epsilon}{r_1}$ we get
\begin{equation}
\DRV{\epsilon}{r_1} = 
   3\ p \left[ 
       \sqrt{\left(\frac{1}{p_1}-1\right)} \ \frac{-1}{2\ r_1^{3/2}}
       + \sqrt{\left(\frac{p_1}{p}-1\right)}\ 
                   \frac{1}{2\ (r-r_1)^{3/2}} \right]
\label{eq:2SDrvr1}
\end{equation}
For minimizing the error we calculate $\DRV{\epsilon}{r_1} = 
0$ and get that
\begin{equation}
       \sqrt{\left(\frac{1}{p_1}-1\right)} \ \frac{1}{2\ r_1^{3/2}} = 
       \sqrt{\left(\frac{p_1}{p}-1\right)}\ 
                   \frac{1}{2\ (r-r_1)^{3/2}} 
\label{eq:2SDrvr1eq}
\end{equation}
By simple mathematical manipulations we get
\begin{equation}
       \left(\frac{1}{p_1}-1\right)\ \left(r-r_1\right)^3 = 
       \left(\frac{p_1}{p}-1\right)\ r_1^3
\label{eq:2SDrvr1eq2}
\end{equation}
Noticed that Equations~(\ref{eq:2SDrvp1eq2})~and~(\ref{eq:2SDrvr1eq2}) together provide two simple conditions to optimize $p_1$ and $r_1$.
By dividing Equation~(\ref{eq:2SDrvp1eq2}) by Equation~(\ref{eq:2SDrvr1eq2}) we obtain,
\begin{equation}
       \left(r-r_1\right) =  \frac{r_1\ p_1^2}{p}
\label{eq:2Srr1}
\end{equation}
Using Equation~(\ref{eq:2Srr1}) in
Equation~(\ref{eq:2SDrvr1eq2}) results that
\begin{equation}
\begin{split}
       \left(\frac{1}{p_1}-1\right)\ 
       \left(\frac{r_1\ p_1^2}{p} \right)^3 = &
       \left(\frac{p_1}{p}-1\right)\ r_1^3 \\
       \left(\frac{1}{p_1}-1\right)\ \frac{p_1^6}{p^2}  = &
       \left(\frac{p_1}{p}-1\right)
\end{split}
\label{eq:2SDrvr1eq3}
\end{equation}
From this we get the following relation
\begin{equation}
p_1^6-p_1^5+p_1\ p^2 - p^3 = 0
\label{eq:2Sp1relation}
\end{equation}
The only real solutions are $p_1=\pm\sqrt{p}$.
Since $p_1$ must be positive we get that $p=\sqrt{p}$. 
From this solution and Equation~(\ref{eq:2Srr1}), it is implies that the optimal setting is
\[
p_1=p_2=\sqrt{p},~~and ~~r_1=r_2=r/2
\]
Consequently, the errors of the two steps are 
\[
\epsilon_1 \!=\! \epsilon_2 \!=\! 3\sqrt{\sqrt{p}(1-\sqrt{p})/(r/2)}
\]
From this we obtain Proposition~\ref{PR:1} and a bound on the error of,
\[
 6\sqrt{2} \sqrt{\frac{p\sqrt{p}(1-\sqrt{p})}{r}}
\]

\noindent
This concludes our analysis of the \twostep algorithm.

%%%%%%%%%%%%%%%%% End of File %%%%%%%%%%%%%%%%%%%

\end{document}